\documentclass[a4paper,11pt]{article}
\pdfoutput=1 % if your are submitting a pdflatex (i.e. if you have
\usepackage{jheppub} % for details on the use of the package, please
\usepackage[T1]{fontenc} % if needed
\usepackage[utf8]{inputenc}
\usepackage{tikz}
\usetikzlibrary{decorations}

\usepackage{caption}
\usepackage{subcaption}
\usepackage{shuffle}
\usepackage{amssymb,amsmath}
\usepackage{bbm}
\usepackage{bm}
\usepackage{mathtools}
\usepackage{graphicx}
\usepackage{epsfig}
\usepackage{epstopdf}
\usepackage{setspace}
\usepackage{dsfont}
\usepackage{latexsym}
\usepackage{tikz}
\usepackage{psfrag}
\usepackage{booktabs}
\usepackage{bbm}
\usepackage[toc]{appendix}
\usepackage{color}
\usepackage{datetime}
\usepackage{tensor}
\usepackage{lscape}
\usepackage{lipsum}

\usepackage{tikz}
\usetikzlibrary{arrows,shapes}
\usetikzlibrary{trees}
\usetikzlibrary{patterns}
\usetikzlibrary{matrix,arrows} 				% For commutative diagram
\usetikzlibrary{positioning}				% For "above of=" commands
\usetikzlibrary{calc,through}				% For coordinates
\usetikzlibrary{decorations.pathreplacing}  % For curly braces
\usepackage{pgffor}							% For repeating patterns
\usetikzlibrary{decorations.pathmorphing}	% For Feynman Diagrams
\usetikzlibrary{decorations.markings}
\tikzset{
    vector/.style={decorate, decoration={snake}, draw},
	provector/.style={decorate, decoration={snake,amplitude=2.5pt}, draw},
	antivector/.style={decorate, decoration={snake,amplitude=-2.5pt}, draw},
        smallvector/.style={decorate, decoration={snake,amplitude=1.5pt,post length=0.5mm}, draw},
    fermion/.style={draw=black, postaction={decorate},
        decoration={markings,mark=at position .55 with {\arrow[draw=black]{>}}}},
    fermionbar/.style={draw=black, postaction={decorate},
        decoration={markings,mark=at position .55 with {\arrow[draw=black]{<}}}},
    fermionnoarrow/.style={draw=black},
    gluon/.style={decorate, draw=black,
        decoration={coil,amplitude=4pt, segment length=5pt}},
    scalar/.style={dashed,draw=black, postaction={decorate},
        decoration={markings,mark=at position .55 with {\arrow[draw=black]{>}}}},
    scalarbar/.style={dashed,draw=black, postaction={decorate},
        decoration={markings,mark=at position .55 with {\arrow[draw=black]{<}}}},
    scalarnoarrow/.style={dashed,draw=black},
    electron/.style={draw=black, postaction={decorate},
        decoration={markings,mark=at position .55 with {\arrow[draw=black]{>}}}},
    bigvector/.style={decorate, decoration={snake,amplitude=4pt}, draw},
    arrow/.style={draw=black, postaction={decorate},
        decoration={markings,mark=at position 1 with {\arrow[draw=black]{>}}}},
}

\tikzstyle{block} = [draw, rectangle, 
    minimum height=3em, minimum width=6earticlem]

\definecolor{darkblue}{rgb}{0.2, 0, 0.8}

\numberwithin{equation}{section}

\renewcommand{\r}{\rho}

\tikzset{vertex/.style={inner sep=0,minimum size=3pt,circle,fill}}

\usepackage{comment}

\newcommand{\pD}{{\mathcal{D}}}

\newcommand{\pP}{{\mathcal{P}}}

\newcommand{\pA}{\mathcal{A}}

\newcommand{\reef}[1]{(\ref{#1})}

\usepackage{overpic}

\def\be{\begin{equation}}
\def\ee{\end{equation}}
\def\bea{\begin{eqnarray}}
\def\eea{\end{eqnarray}}
\def\ba{\begin{array}}
\def\ea{\end{array}}
\def\bd{\begin{displaymath}}
\def\ed{\end{displaymath}}

\def\Tr{{\rm Tr}}

\def\ra{\rangle}

\def\h{\eta}

\def\r{\rho}                                     %     \varrho
\def\x{\xi}

\def\>{\rangle} %right angle
\def\<{\langle} %left angle
\def\Dsl{D \hskip-.6em \raise1pt\hbox{$ / $ } }

\def\ads{\textrm{AdS}}

\title{BCJ Amplitude Relations for Anti-de Sitter Boundary Correlators in Embedding Space}

\author[a]{Pranav Diwakar,}
\author[b]{Aidan Herderschee,}
\author[a]{Radu Roiban}
\author[a]{and Fei Teng}

\affiliation[a]{Department of Physics, Pennsylvania State University, University Park, PA 16802, USA}
\affiliation[b]{Leinweber Center for Theoretical Physics, Randall Laboratory of Physics\\ The University of Michigan, Ann Arbor, MI 48109-1040, USA}

\emailAdd{pranavd@psu.edu}
\emailAdd{aidanh@umich.edu}
\emailAdd{radu@phys.psu.edu}
\emailAdd{fei.teng@psu.edu}

\abstract{
We generalize the color/kinematics duality of flat-space scattering amplitudes to the embedding space formulation of AdS boundary correlators.
Kinematic numerators and propagators are replaced with differential operators acting 
on a scalar contact diagram that is the AdS generalization of the momentum conserving delta function of flat space scattering amplitudes. 
We show that color/kinematics duality implies differential relations 
among AdS boundary correlators that naturally generalize the flat space BCJ amplitude relations
and verify them for the correlators of Yang-Mills theory and of the 
Nonlinear Sigma Model through four- and  six-points, respectively. For 
the latter we also find representations of the four- and six-point correlator that manifest the duality.
Possible double-copy procedures in AdS space are also discussed. }

\begin{document} 
\begin{flushright}
{\tt LCTP-21-15} \\
\end{flushright}
\maketitle
\flushbottom

%%%%%%%%%%%%

%%%%%%%%%%%%%%%%%%%%%%%%%%%%%%
\section{Introduction}
\label{sec:intro}

Observables specified by boundary data, such as the S matrix in flat space and boundary
correlation functions in anti-de-Sitter space, are at the core of modern developments 
in high energy physics. For example, the latter often correspond to correlation functions 
of gauge-invariant operators in unitary CFTs  \cite{Maldacena:1997re, Gubser:1998bc, 
Witten:1998qj, Penedones:2010ue}, providing a concrete realization of the holographic 
principle \cite{Susskind:1994vu,tHooft:1993dmi}. 
In flat space, significant progress has been made in both understanding the 
underlying structure of scattering amplitudes and developing new computation methods. While 
scattering amplitudes and AdS boundary correlators are different in many respects, their 
shared properties have led to significant synergy between the scattering 
amplitudes program and the study of holographic correlators. For example, the AdS 
generalization of the Froissart-Gribov formula led to the famous OPE inversion  
formula~\cite{Caron-Huot:2017vep,Simmons-Duffin:2017nub}.

In this paper, we discuss the AdS generalizations of color/kinematics duality and BCJ relations of flat space scattering amplitudes, focusing on Yang-Mills (YM) theory 
and the nonlinear sigma model (NLSM). YM in AdS has been studied largely from the 
perspective of holography, as there are a variety of holographic models that 
include a bulk, non-abelian gauge theory as a closed subsector \cite{Harlow:2018tng,Alday:2021odx}.\footnote{A well-known example is that of the $SO(6)$ vector fields of 
maximal supergravity on AdS$_5\times$S$^5$. More involved ones exhibit 
D-branes or M-branes near singularities such that the bulk theory 
has a singular locus of the form $\ads_{d}\times S^{3}$, see e.g. \cite{Fayyazuddin:1998fb,Aharony:1998xz,Seiberg:1996bd,Ganor:1996mu,Seiberg:1996vs}. Alternatively, one could consider adding probe branes which wrap an $S^{3}$ inside the bulk compactified dimension \cite{Karch:2002sh,Hohenegger:2009as,Hikida:2009tp,Gaiotto:2009tk}.} Weak coupling aspects of color/kinematics duality for boundary correlation functions and form factors were discussed in refs.~\cite{Boels:2012ew, Engelund:2012re}. 
NLSMs targeted in a $G/H$ coset manifold, where $G$ is the $U$-duality group and $H$ is the R-symmetry group, are 
the standard description of scalar fields of matter-coupled supergravity theories. Thus, 
NLSMs in AdS space are an integral part of AdS supergravity theories. More recently, 
certain NLSM in AdS, such as the $O(N)$ model, have gained attention 
as tractable models to study QFT in curved space, independent of any high energy completion \cite{Carmi:2018qzm,Marino:2019fvu,Giombi:2020rmc}.

Color/kinematics duality~\cite{Bern:2008qj}, identifying the algebraic properties of color and kinematic factors of amplitudes, and the double copy construction~\cite{Bern:2010ue} 
have been extensively studied with gauge theories in flat space. An impressive array of remarkable results 
show that there is a veritable web of quantum theories connected by these properties, see ref.~\cite{Bern:2019prr} for a review. Originally formulated for maximally-supersymmetric 
gauge and gravity theories, the duality has also been identified in wider classes of theories~\cite{Bern:2010yg, Bern:2010tq, Carrasco:2011mn, Bern:2011rj, BoucherVeronneau:2011qv, Bern:2012uf, Du:2012mt, Oxburgh:2012zr, Yuan:2012rg, Boels:2012ew, Boels:2013bi, Bjerrum-Bohr:2013iza, Bern:2013yya, Ochirov:2013xba, Mafra:2015mja, Mogull:2015adi, He:2015wgf, Yang:2016ear, Bern:2017yxu, Bern:2017ucb, Boels:2017skl, He:2017spx}, including theories with fields in representations other than the adjoint \cite{Johansson:2014zca, Chiodaroli:2013upa, Johansson:2015oia,Bargheer:2012gv,Anastasiou:2016csv,Huang:2013kca,Chiodaroli:2015wal,Anastasiou:2017nsz,Chiodaroli:2018dbu}, the NLSM and Born-Infeld theories \cite{Cheung:2017ems,Cachazo:2014xea,Cheung:2017yef,Elvang:2020kuj,Chen:2013fya, Carrasco:2016ygv, Mizera:2018jbh}. 

Color/kinematics duality and the double copy have been instrumental in higher-loop computations in both 
colored and uncolored theories. For the latter, the double copy expresses the integrands of amplitudes in terms of building blocks extracted from colored
amplitudes. These in turn are constrained by color/kinematics duality, so only a small number of terms need to be computed. At tree-level, a sufficient but not necessary condition for color/kinematics duality is the existence of BCJ relations among partial amplitudes in a minimal color basis~\cite{Bern:2008qj}. For example, the four-gluon color-ordered tree-level 
amplitudes are related by
\begin{equation}\label{orgbcjrel}
s_{12}A_{\textrm{flat}}(1,2,3,4)=s_{13}A_{\textrm{flat}}(1,3,2,4) \ ,   
\end{equation}
where $s_{ij}=(p_i+p_j)^2$ are Mandelstam invariants, implying that a single four-point 
partial amplitude determines the complete color-dressed amplitude at tree-level.  

For theories with fields in the adjoint representation and  couplings governed by the antisymmetric structure constants, the BCJ relations imply that only $(n-3)!$ tree-level partial amplitudes are independent at $n$-point. 
Color/kinematics duality, however, does not always imply amplitude relations. For example, theories with fields transforming in the (anti-)fundamental representation 
exhibit color/kinematics duality, but fewer relations among partial amplitudes \cite{Johansson:2014zca,Johansson:2015oia,delaCruz:2015dpa,Brown:2018wss,Brown:2016hck,
Naculich:2015coa}. It is natural to ask whether or not in AdS space color/kinematics duality, if present, implies nontrivial relations between partial correlation functions.\footnote{Although we restrict ourselves to massless external states in this paper, we note that the double copy also applies when external states are massive \cite{Johnson:2020pny,Plefka:2019wyg,Johansson:2019dnu,Naculich:2014naa,Carrasco:2020ywq,Momeni:2020vvr,Haddad:2020tvs,Momeni:2020hmc}. 
If the mass can be interpreted as momentum in compactified dimensions, and therefore obeys momentum conservation, the massless double copy generalizes without (many) subtleties. Notably, holographic theories in AdS generically include higher Kaluza-Klein modes whose ``mass'' can be interpreted as momentum in a compactified space. Therefore, although these higher-dimensional modes are massive and do not obey any analog of gauge invariance, they should also obey color/kinematics duality in AdS.}

Despite steady effort and significant progress in several directions, a systematic formulation of 
color/kinematics duality and of the double copy in curved space remains elusive. 
Tree-level AdS boundary correlators are a natural starting point to study color/kinematics duality.
Their computation using Witten diagrams in position space is, however, cumbersome. This prompted the development of other representations such as AdS momentum space \cite{Farrow:2018yni,Albayrak:2018tam,Bzowski:2019kwd,Albayrak:2019yve,Lipstein:2019mpu,Albayrak:2020isk} and position-Mellin space \cite{Mack:2009gy,Mack:2009mi,Penedones:2010ue,Paulos:2011ie,Fitzpatrick:2011ia,Kharel:2013mka,Rastelli:2016nze,Penedones:2019tng} and momentum-Mellin space \cite{Sleight:2021iix, Sleight:2019hfp}.
AdS momentum space might be expected to be the most natural representation when searching for generalizations of amplitude relations as AdS boundary correlation functions in momentum space 
contain flat-space scattering amplitudes \cite{Albayrak:2018tam}. However, imposing color/kinematics duality 
on integrated, color-ordered momentum space correlators does not seem to yield BCJ relations \cite{Albayrak:2020fyp,Armstrong:2020woi}. 
In contrast to AdS momentum space, scalar Witten diagrams in Mellin space are simple and 
yield correlation functions that exhibit colour-kinematics duality \cite{Alday:2021odx,Zhou:2021gnu}.
However, with some notable exceptions, see refs. \cite{Sleight:2021iix,Sleight:2018epi}, current 
state-of-the-art techniques in Mellin space often rely on using supersymmetry to relate 
scalar correlators to those of spin-1 and spin-2 states~\cite{Rastelli:2016nze,Zhou:2017zaw,Rastelli:2017udc,Alday:2020dtb,Alday:2020lbp}. Furthermore, while the Mellin space formalism does
lead to a form of color/kinematics duality and double copy~\cite{Alday:2021odx,Zhou:2021gnu} for (maximally) supersymmetric theories, it 
does not immediately lead to additional relations between color-ordered correlators. 
Our results, demonstrating that such relations exist in a position representation of correlation functions, imply that they should also have momentum- and Mellin-space realizations.

In this paper, we use the embedding space formalism \cite{10.2307/1968455,Ferrara:1973yt,deAlfaro:1976vlx,Weinberg:2010fx,Costa:2011mg} 
as a means to generalize color/kinematics duality and amplitude relations to AdS space. In embedding space, the action of conformal representations is 
linear and conformal symmetry takes on a role analogous to that of momentum conservation in flat space. 
Using this formalism, we uncover novel relations for tree-level correlators in AdS space. We conjecture that certain AdS boundary correlators can generically be expressed as 
(nonlocal) differential operators acting on a single contact diagram, thus giving a \textit{differential representation} of the correlator and quantifying in what sense
this differential operator exhibits color/kinematics duality.\footnote{Amplitude representations that are similar in spirit are used in the double copy of 
celestial amplitudes in flat space~\cite{Casali:2020vuy,Casali:2020uvr,Kalyanapuram:2020aya}.} 
We are partially motivated by the recent generalization of the ambitwistor string to AdS${}_{3}\times S^{3}$ \cite{Roehrig:2020kck}, which provides 
an explicit example of color/kinematics duality in an AdS space.\footnote{In flat space, particular BCJ representations of a variety of theories can be derived from ambitwistor string models~\cite{Mason:2013sva,Berkovits:2013xba,Adamo:2013tsa,Casali:2015vta}.}
Our conjectures are the natural generalization of the results in ref.~\cite{Roehrig:2020kck} 
to higher dimensions and the natural extension of the results of ref.~\cite{Eberhardt:2020ewh} to non-scalar theories.

Using our conjectural AdS generalization of color/kinematics duality, we derive novel relations for AdS boundary correlators, which are schematically similar 
to flat space BCJ relations with the suitable replacement of Mandelstam invariants with combinations of conformal generators $D_{i}^{AB}$, given in 
section~\ref{amplitudesinembeddingspace}:
\begin{equation}\label{coinjecturrepla}
s_{I}\rightarrow \Big(\sum_{i\in I}D_{i}^{AB}\Big)^{2} \ .    
\end{equation}
We construct the explicit color/kinematics-satisfying representation of the NLSM AdS boundary correlators at four- and six-points and show that these correlators 
indeed obey the AdS generalization of the BCJ amplitude relations. We also construct the four-point gluon correlator of YM theory in AdS of general dimension and verify 
that, for a four-dimensional boundary, it obeys these relations
\begin{equation}\label{YMBCJrel}
(D_{1}^{AB}+D_{2}^{AB})^{2}A(1,2,3,4)=(D_{1}^{AB}+D_{3}^{AB})^{2}A(1,3,2,4) \ .    
\end{equation}
We expect (but do not prove) that they are obeyed for general dimensions.
The existence of the relation~\eqref{YMBCJrel} supports the existence of a representation for the AdS boundary correlators that manifests the color/kinematics duality therein , although there is no proof for that yet.
Nevertheless, in section~\ref{AdS_AmplitudesRelationsGeneral}, we propose one possible realization of the color/kinematics duality for AdS boundary correlators in terms of differential operators. We then explicitly show that up to six points NLSM correlators can be arranged into this form.

A natural next step is to extend our proposal for color/kinematics duality to a double copy relation for gauge theories. We consider a variety of proposals that appear to be valid in different limits for three-point correlators. We first study a differential double copy procedure that naturally generalizes our proposal for color/kinematics duality in AdS, which seems to yield self-consistent results in AdS${}_{3}$. For higher dimensional AdS spacetimes, we find that the differential double copy leads to a current conserving three-point graviton correlator only if we supplement the YM with specific higher dimensional operators.
%
%even at three-points. The inconsistency of this procedure for $d>2$ is interesting given that color/kinematics duality seems to hold for $d>2$. 
We also consider other double copy procedures in position space and Mellin space. We reproduce the result of ref.~\cite{Mizeraetal}, giving a double copy-like relation for the three-point 
Mellin amplitude without supersymmetry in the limit $d\rightarrow \infty$. Furthermore, we compare our results with the Mellin space double copy construction of ref.~\cite{Zhou:2021gnu}, which gives super-graviton AdS boundary correlators on AdS${}_{5}\times S^{5}$ in terms of super-gluon AdS boundary correlators on AdS${}_{5}\times S^{3}$. We conclude with a heuristic discussion of double copy procedures for various formulations of AdS boundary correlators in the high energy limit. 

The paper is organized as follows. 
In section~\ref{amplitudesinembeddingspace} we review properties of 
AdS boundary correlators in embedding space.
In section~\ref{AdS_AmplitudesRelationsGeneral} we motivate and conjecture a color/kinematics duality for 
tree level correlators in AdS, generalizing the BCJ representation of ref.~\cite{Bern:2008qj}. 
We use our conjectural BCJ representation of correlators to find the AdS generalization of BCJ amplitude relations.
In section~\ref{AdSrulesEmbeddingSpace} we review the AdS Feynman rules in embedding space for NLSM and YM.
In section~\ref{NSLMsection} we show that NLSM AdS boundary correlators obey color/kinematics duality, verify the 
amplitudes relations and construct the manifest color/kinematic-satisfying  representation of 
the four- and six-point correlators.
In section~\ref{sec:YMsec} we construct a differential representation of the on-shell and off-shell YM three-point correlators; using them we construct the four-point correlator and confirm that it obeys AdS BCJ relations. 
section~\ref{towardsadsdoublecopy} gives a brief discussion of the double copy in AdS space and section~\ref{conclusions} contains our conclusions.
Appendices \ref{app_embedding_space} and \ref{contactdiagramsinads} include a short review of the embedding space formalism and a review of $D$-functions and of the relations they obey. \\ 

\paragraph{Notation:} We use $\mathcal{A}$ and $A(\alpha)$ to denote color-dressed AdS boundary correlators and color-ordered partial AdS boundary correlators, respectively; $\mathcal{M}$ refers to the AdS boundary correlators of uncolored theories, such as gravity. We will use the subscript ``flat'' when referring to flat-space scattering amplitudes.

%%%%%%%%%%%%%%%%%%%%%%%%%%%%%%

\section{AdS boundary correlators in the embedding space}\label{amplitudesinembeddingspace}

In this section, we review certain general properties of boundary correlators on $\ads_{d+1}$ background. We are particularly interested in their embedding space form, since they exhibit interesting properties that are analogous to those of flat space scattering amplitudes in momentum space. 

We write AdS boundary correlators as $\mathcal{A}(P_i,Z_i)$, where $P_i$ is a point on the conformal boundary $\partial\ads_{d+1}$ and $Z_{i}$ is a polarization vector. In the following, all the quantities we discuss are given in the embedding space. If the external particles have spin, then the AdS boundary correlator $\mathcal{A}$ is also a multilinear function in the polarization vector $Z_i$. The complete definitions of $P_i$ and $Z_i$, which are not necessary for the discussion here, are given in appendix~\ref{app_embedding_space}. 
A boundary correlator $\mathcal{A}$ is a homogeneous function in both $P_i$ and $Z_i$,
\begin{align}
    \mathcal{A}(\lambda P_i,Z_i)=\lambda^{-\Delta_i}\mathcal{A}(P_i,Z_i)\,,\qquad \mathcal{A}( P_i,\lambda Z_i)=\lambda^{l_i}\mathcal{A}(P_i,Z_i)\,,
\end{align}
where $\Delta_i$ is the conformal weight of particle $i$ and $l_i$ is its spin. Note that the conformal weight is defined as the negative of the scaling dimension. AdS boundary correlators are expected to be scalar quantities invariant under the action of the conformal group $\text{SO}(d{+}1,1)$, which is isomorphic to the Lorentz group of the embedding space. The conformal generator acting on the $i$-th particle is
\begin{align}\label{eq:confgen}
    D_{i}^{AB}&=P_{i}^{A}\frac{\partial}{\partial P_{i,B}}-P_{i}^{B}\frac{\partial}{\partial P_{i,A}}+Z_{i}^{A}\frac{\partial}{\partial Z_{i,B}}-Z_{i}^{B}\frac{\partial}{\partial Z_{i,A}}\,.
\end{align}
Because the embedding space realizes the conformal transformations linearly, the \emph{conformal Ward identity (CWI)} capturing the conformal invariance of $\mathcal{A}$ is
\begin{align}\label{eq:CWIn}
    \sum_{i=1}^n D_i^{AB}\mathcal{A}=0\,.
\end{align}
It resembles the momentum conservation of flat space amplitudes. For an external 
spinning particle, we can peel off a polarization vector, such that
\begin{align}
    \mathcal{A}=Z_{i,M}\mathcal{A}^{M}\,.
\end{align}
Written as an embedding space vector, $\mathcal{A}^M$ is transverse to the conformal boundary $\partial\ads_{d+1}$ if and only if
\begin{align}
    P_{i,M}\mathcal{A}^{M}=0\,.
\end{align}
See appendix~\ref{app_embedding_space} for more details. Therefore, for $\mathcal{A}(P_i,Z_i)$ to be an AdS boundary correlator, it has to satisfy the \emph{transversality condition},
\begin{align}\label{eq:transverse}
    \mathcal{A}(P_i,Z_i)\Big|_{Z_i\rightarrow P_i}=0\,,
\end{align}
which is analogous to the linearized gauge invariance of flat space amplitudes.

In flat space, momentum conservation leads to relations between Mandelstam variables, for example, $s+t+u=0$ for four-point massless kinematics. We now show how similar relations among conformal generators arise for AdS boundary correlators. We first define for convenience the inner product of conformal generators as
\begin{align}\label{conformalgener}
    D_{ij}^{2}\equiv D_i\cdot D_j=\eta_{AC}\eta_{BD}D_{i}^{AB}D_{j}^{CD}\,,\qquad D_{i}^{2}\equiv D_i\cdot D_i\,.
\end{align}
Clearly, $D_{ij}^2=D_{ji}^2$ since $D_i$ commutes with $D_j$ as they act on different variables. $D_i^2$ is proportional to the quadratic Casimir operator of particle $i$,
\begin{align}\label{eq:casimir}
    -\frac{1}{2}D_i^2 &= \left(P_i\cdot\frac{\partial}{\partial P_i}\right)\left(d+P_i\cdot\frac{\partial}{\partial P_i}\right)+ \left(Z_i\cdot\frac{\partial}{\partial Z_i}\right)\left(d-2+Z_i\cdot\frac{\partial}{\partial Z_i}\right) \nonumber\\
    &\quad +2\left(Z_i\cdot\frac{\partial}{\partial P_i}\right)\left(P_i\cdot\frac{\partial}{\partial Z_i}\right)\\
    &\cong\Delta_i(\Delta_i-d)+l_i(l_i+d-2)\,.\nonumber
\end{align}
When acting on an AdS boundary correlator, or more generally a conformal partial wave, the second line of the above equation does not contribute due to transversality~\eqref{eq:transverse}. We use $\cong$ to denote ``equivalent when acting on a conformal partial wave''. Therefore, we can use the eigenvalue of $D_i^2$ to define the on-shell mass of particle $i$. A scalar particle is massless if $\Delta_i=d$ while a vector particle is massless if $\Delta_i=d-1$ (see section~\ref{onshelloffshell} for further comments on this definition). Thus, massless scalar and vector correlators satisfy
\begin{align}\label{onshellcondition}
    D_i^2\mathcal{A}=0\quad\text{for all }i\,.
\end{align}
Massless vector correlators further satisfy the current conservation~\cite{Costa:2011mg},
\begin{align}\label{currentconservationform}
    \frac{\partial}{\partial P_{i,M}}\left[\left(\frac{d}{2}-1+Z_i\cdot\frac{\partial}{\partial Z_i}\right)\frac{\partial}{\partial Z_i^M}-\frac{1}{2}Z_{i,M}\frac{\partial^2}{\partial Z_i\cdot\partial Z_i}\right]\mathcal{A}=0\,.
\end{align}
Eq. (\ref{currentconservationform}) assumes that the $i$-th particle has conformal weight $\Delta_{i}=d-2+l_{i}$. Notably, current conservation for graviton boundary correlators requires $\Delta_i=d$, which leads to $-\frac{1}{2}D_i^2\cong 2d$. This is an exception to the naive definition of masslessness described above. 

The massless condition and the CWI~\eqref{eq:CWIn} together give rise to very simple relations among conformal generators. As a simple example, we consider the four-point CWI,
\begin{equation}\label{eq:cwi4}
(D_{1}^{AB}+D_{2}^{AB}+D_{3}^{AB}+D_{4}^{AB}) \mathcal{A}=0\,,
\end{equation}
from which we can derive
\begin{align}\label{eq:D12D34}
    (D_1+D_2)^2\mathcal{A}=-(D_1+D_2)\cdot (D_3+D_4)\mathcal{A}=(D_3+D_4)^2\mathcal{A}\,.
\end{align}
For massless AdS boundary correlators, we thus get
\begin{align}
    D_{12}^2\mathcal{A}=D_{34}^2\mathcal{A}\,,\quad\text{or}\quad D_{12}^2\cong D_{34}^2\,.
\end{align}
Similarly, we can derive that
\begin{align}\label{eq:stuAdS}
    D_{12}^2+D_{13}^2+D_{23}^2\cong 0\,,
\end{align}
which is the AdS incarnation of the flat space relation $s+t+u=0$. For correlators of 
higher multiplicity, we define
\begin{equation}
D_{I}^{2}\equiv \frac{1}{2}\Big(\sum_{a\in I}D_{a}\Big)^{2} \,.
\end{equation}
Using a slight generalization of eq.~\eqref{eq:D12D34}, we can show that 
\begin{align}\label{eq:compident}
    D_I^2 \cong D_{\bar{I}}^2\,,
\end{align}
where $\bar{I}$ is the complement of set $I$ in the set of labels of all external particles.
Furthermore, we can show that the relations between massless on-shell Mandelstam variables can all be realized as relations between various $D_I^2$ when acting on a conformal partial wave. One can also prove that~\cite{Eberhardt:2020ewh}
\begin{align}\label{eq:DIcommute}
    [D_{I}^{2},D_{I'}^{2}]&= 0 \quad \textrm{ if } \quad (I\cap I'=\emptyset) \textrm{ or } (I\subset I') \textrm{ or } (I'\subset I) \ .   
\end{align}
We will often be interested in understanding how the inverse of $D_I^2$ acts on conformal correlators. We will show in the next section that the inverse of $D_I^2$ acting on a contact diagram can be related to a bulk-bulk propagator in Witten diagram computations. However, one can understand how $(D_{I}^{2})^{-1}$ acts on more generic conformal correlators by decomposing the conformal correlator into conformal partial waves \cite{Dolan:2003hv}. Conformal partial waves are, by construction, eigenfunctions of $D_{I}^{2}$. Therefore, since any conformal correlator can be expanded as a linear combination of conformal partial waves, one can use such a conformal partial wave decomposition to systematically understand how $D_{I}^{2}$ acts on any conformal correlator. Conformal partial waves beyond four-point were recently considered in refs.~\cite{Buric:2021ywo,Buric:2020dyz}.

Gauge invariance and on-shell kinematics are crucial for flat space amplitudes to have additional structures, like color/kinematics duality, BCJ amplitude relations, and double copy. Due to the properties listed above,  we therefore intuitively expect that embedding space is a promising stage to explore such hidden structures in AdS boundary correlators.

\section{BCJ relations for AdS boundary correlators}
\label{AdS_AmplitudesRelationsGeneral}

In this section, we begin by discussing the cubic bi-adjoint scalar (BAS) 
theory in AdS space.  We then use the results we obtain to motivate a 
generalization of color/kinematics duality and of the 
BCJ amplitudes relations for certain AdS boundary correlators. 

\subsection{Cubic bi-adjoint scalar in AdS}\label{cubicbasreview}

To motivate the BCJ amplitude relations in an AdS setup, we first consider the simplest theory in AdS that could exhibit color/kinematics duality -- cubic BAS, defined by the Lagrangian 
\begin{equation}
\mathcal{L}=\frac{1}{2}(\nabla\phi)^{2}-\frac{g}{6}f^{abc}\tilde{f}^{a'b'c'}\phi^{aa'}\phi^{bb'}\phi^{cc'} \ .  
\end{equation}
The scalars transform in the bi-adjoint representation of $\text{SU}(N)\times\text{SU}(N')$. Perturbative computations of amplitudes and of AdS boundary correlators involve summing cubic Feynman and Witten diagrams respectively, dressed with appropriate color factors. We will review a particular representation of the cubic BAS boundary correlators given in ref.~\cite{Roehrig:2020kck} for $d=2$ and in ref.~\cite{Eberhardt:2020ewh} for general $d$. We will first consider the four-point formula before generalizing to the $n$-point case. 

First, we posit and justify later that the four-point BAS AdS boundary correlator 
can be represented as
\begin{equation}\label{eq:4pointformula}
\mathcal{A}^{\textrm{BAS}}=\Bigg( \frac{C_s\tilde{C}_s}{D_{12}^{2}}+\frac{C_t\tilde{C}_t}{D_{23}^{2}}+\frac{C_u\tilde{C}_u}{D_{13}^{2}}\Bigg ) \Bigg (\frac{\Gamma(d)}{2\pi^{d/2} \Gamma(d/2+1)} \Bigg )^{4} D_{d,d,d,d}\,,
\end{equation}
where the color factors are $C_s=f^{a_{1}a_{2}x}f^{xa_{3}a_{4}}$, $C_t=f^{a_{2}a_{3}x}f^{xa_{1}a_{4}}$, $C_u=f^{a_{3}a_{1}x}f^{xa_{2}a_{4}}$ and similarly for $\tilde{C}_{s,t,u}$. We define $\frac{1}{D_{ij}^2}$ as the inverse of the operator $D_{ij}^2$. They act on $D_{d,d,d,d}$, which is defined as
\begin{equation}
    D_{d,d,d,d}=\int_{\ads} dX\prod_{i=1}^4\frac{1}{(-2P_i\cdot X)^d}\, ,
\end{equation}
and corresponds to the four-point scalar contact diagram.
Since we are working with external scalars, we can ignore the $Z$ component of $D_{i}^{AB}$. To justify the structure described above, we now show that the first term in eq.~(\ref{eq:4pointformula}) is equivalent to an $s$-channel Witten diagram. We first write the action of $\frac{1}{D_{12}^2}$ on $D_{d,d,d,d}$ as
\begin{align}
\label{AdSBBpropagator}
\frac{1}{D_{12}^{2}}&\int_{\ads} dX \prod_{i=1}^{4}\frac{1}{(-2 P_{i}\cdot X)^{d}}\nonumber\\
& =\int_{\ads} dX \frac{1}{(-2P_{3}\cdot X)^{d}}\frac{1}{(-2 P_{4}\cdot X)^{d}}\frac{1}{\square_{X}}\Bigg[\frac{1}{(-2 P_{1}\cdot X)^{d}}\frac{1}{(-2 P_{2}\cdot X)^{d}}\Bigg] \ ,   
\end{align}
\noindent where $\square_X\equiv\frac{1}{2}D_{X}^{2}$ can be identified with the AdS Laplacian,
\begin{align}
    \square_X = \frac{1}{2}D_X^2 = - \nabla^2_{\ads}\,,
\end{align}
and the measure is $\int_{\ads}dX=\int \frac{dzd^dx}{z^{d+1}}$ in the Poincar\'e coordinates. See appendix~\ref{app_embedding_space} for more details. Eq.~\eqref{AdSBBpropagator} can be proved by noticing that the action of $D_X^{AB}$ and $D_i^{AB}$ on the bulk-boundary propagator $\frac{1}{(-2P_i\cdot X)^d}$ are the same. Getting the Witten diagram representation 
of the right-hand side of eq.~\eqref{AdSBBpropagator}
now only involves some straightforward algebra:
\begingroup
\allowdisplaybreaks
\begin{align}\label{eq:lastalgebra}
&\quad\int_{\ads} dX \frac{1}{(-2 P_{3}\cdot X)^{d}}\frac{1}{(-2 P_{4}\cdot X)^{d}}\frac{1}{\square_{X}}\Bigg[\frac{1}{(-2 P_{1}\cdot X)^{d}}\frac{1}{(-2 P_{2}\cdot X)^{d}}\Bigg]\nonumber \\
& =\int_{\ads} dX_{1}dX_{2}\frac{1}{(-2 P_{1}\!\cdot\!X_1)^{d}}\frac{1}{(-2 P_{2}\!\cdot\!X_1)^{d}}\frac{1}{(-2 P_{3}\!\cdot\!X_2)^{d}}\frac{1}{(-2P_{4}\!\cdot\!X_2)^{d}} \left [\frac{1}{\square_{X_{1}}}\delta^{d+1}(X_{1}-X_{2}) \right ] \nonumber\\ 
& =\int_{\ads} dX_{1}dX_{2}\frac{1}{(-2 P_{1}\cdot X_1)^{d}}\frac{1}{(-2 P_{2}\cdot X_1)^{d}}G(X_{1},X_{2})\frac{1}{(-2 P_{3}\cdot X_2)^{d}}\frac{1}{(-2 P_{4}\cdot X_2)^{d}} \ .   
\end{align}
\endgroup
\noindent We have used the fact that the bulk-bulk propagator can be viewed as the inverse of the Laplacian,
\begin{equation}
\square_{X_{1}}G(X_{1},X_{2})=\delta^{d+1}(X_{1}-X_{2})\quad  \Rightarrow \quad  G(X_{1},X_{2})=\frac{1}{\square_{X_{1}}}\delta^{d+1}(X_{1}-X_{2}) \ ,   
\end{equation}
\noindent to obtain the last line, which is nothing but an $s$-channel Witten diagram.
Analogous manipulations show that the other two terms in 
eq.~(\ref{eq:4pointformula}) are equivalent to $t$-channel 
and $u$-channel Witten diagrams. 

The above calculation can be easily generalized to higher points. For example, the following expression corresponds to a five-point Witten diagram~\cite{Eberhardt:2020ewh},
\begin{equation}\label{eq:5pexp}
\pgfmathsetmacro{\r}{1.2}
\begin{tikzpicture}[baseline={([yshift=-.5ex]current bounding box.center)},every node/.style={font=\scriptsize}]
\draw [] (0,0) circle (\r cm);
\filldraw (\r,0) circle (1pt) node[right=0pt]{$5$};
\filldraw (72:\r) circle (1pt) node[above=0pt]{$4$};
\filldraw (144:\r) circle (1pt) node[above=0pt]{$3$};
\filldraw (-144:\r) circle (1pt) node[left=0pt]{$2$};
\filldraw (-72:\r) circle (1pt) node[below=0pt]{$1$};
\filldraw (\r/2,0) circle (1pt) (108:\r/2) circle (1pt) (-108:\r/2) circle (1pt);
\draw [thick] (\r,0) -- (\r/2,0) (72:\r) -- (108:\r/2) -- (144:\r) (-72:\r) -- (-108:\r/2) -- (-144:\r);
\draw [thick] (108:\r/2) -- (\r/2,0) node[pos=0.5,above right=-5pt]{$\frac{1}{D_{34}^2}$};
\draw [thick] (-108:\r/2) -- (\r/2,0) node[pos=0.5,below right=-5pt]{$\frac{1}{D_{12}^2}$};
\end{tikzpicture}
=\left (\frac{\Gamma(d)}{2\pi^{d/2} \Gamma(d/2+1)} \right )^{5}\frac{1}{D_{12}^{2}D_{34}^{2}}D_{d,d,d,d,d}\ . 
\end{equation}
\noindent The $n$-point color-dressed BAS correlator can then be written as 
\begin{equation}\label{eq:colordressformula}
\mathcal{A}^{\textrm{BAS}}=\left (\frac{\Gamma(d)}{2\pi^{d/2} \Gamma(d/2+1)} \right )^{n} \sum_{\textrm{cubic }g}C(g|\alpha_{g})\tilde{C}(g|\alpha_{g})\prod_{I\in g}\frac{1}{D_{I}^{2}}D_{\underbrace{\scriptstyle d,d,\ldots,d}_n}\,,
\end{equation}
\noindent where the sum runs over all cubic graphs, and $C(g|\alpha_{g})$ and $\tilde{C}(g|\alpha_{g})$ are the color factors associated with the cubic graph $g$. 
They are simply contractions of structure constants,
\begin{equation}\label{eq:colorfactor}
C(g|\alpha)=\prod_{v}f^{a_{v}b_{v}c_{v}}\,,    
\end{equation}
and the index contraction is implicitly assumed. We introduce an ordering $\alpha_g$ for each graph $g$ to specify the orderings of the adjoint indices in the product in eq.~\eqref{eq:colorfactor}. This fixed a sign ambiguity in the definition. We have to choose the same $\alpha_g$ for both $C$ and $\tilde{C}$. Of course, the boundary correlator does not depend on this choice. We will use eq.~(\ref{eq:colordressformula}) to motivate the AdS analogs of many flat-space formulas.

\subsection{Color/kinematics duality for flat space amplitudes}

\begin{figure}
    \centering
    \begin{tikzpicture}[dl/.style={line width=1pt,dash pattern=on 2pt off 2pt}]
    \pgfmathsetmacro{\r}{0.4}
    \pgfmathsetmacro{\x}{1.2}
    \pgfmathsetmacro{\y}{1}
    \begin{scope}[xshift=0]
    \draw [thick] (-\x,-\y) -- (-0.5,0) -- (-\x,\y);
    \draw [thick] (\x,-\y) -- (0.5,0) -- (\x,\y);
    \draw [thick] (-0.5,0) -- (0.5,0) node[pos=0.5,below=0]{$s_{I_1I_2}$};
    \draw [dl,black,fill=white] (-\x,-\y) circle (\r);
    \draw [dl,red,fill=white] (-\x,\y) circle (\r);
    \draw [dl,blue,fill=white] (\x,\y) circle (\r);
    \draw [dl,green,fill=white] (\x,-\y) circle (\r);
    \node at (-\x,-\y) {$I_1$};
    \node at (-\x,\y) {$I_2$};
    \node at (\x,\y) {$I_3$};
    \node at (\x,-\y) {$I_4$};
    \node at (0,-1.5) {$g_s$};
    \end{scope}
    \begin{scope}[xshift=4.5cm]
    \draw [thick] (-\x,-\y) -- (-0.5,0) -- (-\x,\y);
    \draw [thick] (\x,-\y) -- (0.5,0) -- (\x,\y);
    \draw [thick] (-0.5,0) -- (0.5,0) node[pos=0.5,below=0]{$s_{I_1I_4}$};
    \draw [dl,black,fill=white] (-\x,-\y) circle (\r);
    \draw [dl,green,fill=white] (-\x,\y) circle (\r);
    \draw [dl,red,fill=white] (\x,\y) circle (\r);
    \draw [dl,blue,fill=white] (\x,-\y) circle (\r);
    \node at (-\x,-\y) {$I_1$};
    \node at (-\x,\y) {$I_4$};
    \node at (\x,\y) {$I_2$};
    \node at (\x,-\y) {$I_3$};
    \node at (0,-1.5) {$g_t$};
    \end{scope}
    \begin{scope}[xshift=9cm]
    \draw [thick] (-\x,-\y) -- (-0.5,0) -- (-\x,\y);
    \draw [thick] (\x,-\y) -- (0.5,0) -- (\x,\y);
    \draw [thick] (-0.5,0) -- (0.5,0) node[pos=0.5,below=0]{$s_{I_1I_3}$};
    \draw [dl,black,fill=white] (-\x,-\y) circle (\r);
    \draw [dl,blue,fill=white] (-\x,\y) circle (\r);
    \draw [dl,green,fill=white] (\x,\y) circle (\r);
    \draw [dl,red,fill=white] (\x,-\y) circle (\r);
    \node at (-\x,-\y) {$I_1$};
    \node at (-\x,\y) {$I_3$};
    \node at (\x,\y) {$I_4$};
    \node at (\x,-\y) {$I_2$};
    \node at (0,-1.5) {$g_u$};
    \end{scope}
    \end{tikzpicture}
    \caption{A triplet of three cubic tree graphs that differ by one propagator.}
    \label{figjacobi}
\end{figure}
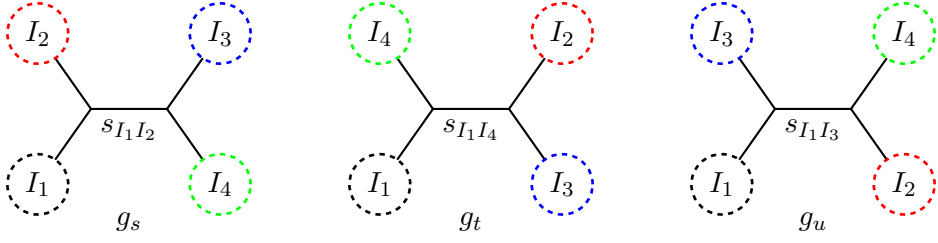

We now take a slight detour and review the derivation of the flat-space 
BCJ amplitudes relations before generalizing them to AdS boundary correlators. 
In flat space, amplitudes that satisfy color/kinematics duality can be 
written as a sum over cubic graphs~\cite{Bern:2008qj},
\begin{equation}\label{eq:bcjrepflatspace}
\mathcal{A}_{\textrm{flat}}=\sum_{\textrm{cubic }g}C(g|\alpha_{g})N(g|\alpha_{g})\prod_{I\in g}\frac{1}{s_{I}}  \ . 
\end{equation}
\noindent where both the kinematic numerators (sometimes referred to as BCJ numerators) 
and color factors obey Jacobi-like relations corresponding to triplets of cubic graphs as shown in figure~\ref{figjacobi},
\begin{equation}\label{eq:newkinematicnumerator}
\begin{split} 
N(g_{s}|I_{1}I_{2}I_{3}I_{4})+N(g_{t}|I_{1}I_{4}I_{2}I_{3})+N(g_{u}|I_{1}I_{3}I_{4}I_{2})&=0 \ , \\
C(g_{s}|I_{1}I_{2}I_{3}I_{4})+C(g_{t}|I_{1}I_{4}I_{2}I_{3})+C(g_{u}|I_{1}I_{3}I_{4}I_{2})&=0 \ .
\end{split}   
\end{equation}
\noindent For cubic BAS, the numerator $N(g|\alpha_g)$ is simply another the color factor $\tilde{C}(g|\alpha_g)$ so that the Jacobi identity trivially holds. We can expand the color factors in eq.~\eqref{eq:bcjrepflatspace} into the $(n-2)!$ dimensional Del Duca-Dixon-Maltoni (DDM) basis, consisting of the color factors of the half ladder graphs~\cite{DelDuca:1999rs},
\begin{align}\label{eq:halfladder}
    C_{1,\alpha(2,3,\ldots,n-1),n}&\equiv C\bigg(
    \begin{tikzpicture}[baseline={([yshift=-.5ex]current bounding box.center)},every node/.style={font=\scriptsize}]
    \pgfmathsetmacro{\ra}{0.3}
    \pgfmathsetmacro{\rb}{0.7}
    \pgfmathsetmacro{\h}{0.4}
    \draw (0,0) node[left=0]{$1$} -- (3,0) node[right=0]{$n$};
    \draw (\ra,0) -- ++(0,\h) node[above=0]{$\alpha(2)$};
    \draw (\ra+\rb,0) -- ++(0,\h) node[above=0]{$\alpha(3)$};
    \draw (3-\ra,0) -- ++(0,\h) node[above=0]{$\alpha(n{-}1)$};
    \node at (1.85,0.2) {$\cdots$};
    \end{tikzpicture}\bigg|1,\alpha,n
    \bigg)\nonumber\\
    &=f^{a_1 a_{\alpha(2)}x_2}f^{x_2 a_{\alpha(3)}x_3}\cdots f^{x_{n-2}a_{\alpha(n-1)}a_n}\,.
\end{align}
As an example, we give the decomposition of the following five-point color factor explicitly,
\begin{align}
    C\Bigg(\begin{tikzpicture}[baseline={([yshift=-.5ex]current bounding box.center)},every node/.style={font=\scriptsize}]
    \draw (0,0) node[left=0]{$1$} -- (1,0) node[right=0]{$5$};
    \draw (0.5,0) -- ++ (0,0.25);
    \draw (0.5,0.25) -- ++(135:0.5) node[left=0]{$2$};
    \draw (0.5,0.25) -- ++(45:0.5) node[right=0]{$4$};
    \draw (0.5,0.25) ++(45:0.25) -- ++(135:0.25) node[above=0pt]{$3$};
    \end{tikzpicture}\Bigg|1,2,3,4,5\Bigg)&=f^{a_1 x a_5}f^{x a_2 y}f^{y a_3a_4}\nonumber\\
    &=C_{1,2,3,4,5}-C_{1,2,4,3,5}-C_{1,3,4,2,5}+C_{1,4,3,2,5}\,.
\end{align}
The decomposition of both color factors in the BAS amplitude leads to the double (color-ordered) partial amplitude denoted as $m$,
\begin{align}
    \mathcal{A}^{\text{BAS}}_{\text{flat}}=\sum_{\alpha,\beta\in S_{n-2}}C_{1,\alpha(2,3,\ldots,n-1),n}m(1,\alpha,n|1,\beta,n)\tilde{C}_{1,\beta(2,3,\ldots,n-1),n}\,.
\end{align}
In general, $m(\alpha|\beta)$ receives contributions from all cubic Feynman diagrams that are planar for both permutations $\alpha$ and $\beta$. Here we give a few low multiplicity examples,
\begingroup
\allowdisplaybreaks
\begin{align}\label{eq:mexample}
    m(1,2,3,4|1,2,3,4)&=\frac{1}{s_{12}}+\frac{1}{s_{23}}\,,\qquad\quad m(1,3,2,4|1,2,3,4)=-\frac{1}{s_{23}}\,, \nonumber\\
    m(1,2,3,4,5|1,2,3,4,5)&=\frac{1}{s_{12}s_{34}}+\frac{1}{s_{12}s_{45}}+\frac{1}{s_{23}s_{45}}+\frac{1}{s_{15}s_{23}}+\frac{1}{s_{15}s_{34}}\,,\\
    m(1,3,2,4,5|1,2,3,4,5)&=-\frac{1}{s_{23}}\left(\frac{1}{s_{45}}+\frac{1}{s_{15}}\right),\qquad m(1,3,4,2,5|1,2,3,4,5)=-\frac{1}{s_{15}s_{34}}\,,\nonumber
\end{align}
\endgroup
and the full definition of $m(\alpha|\beta)$ can be found in ref.~\cite{Cachazo:2013iea}. For a generic amplitude that exhibits color/kinematics duality, the DDM basis decomposition gives
\begin{align}
    \mathcal{A}_{\text{flat}}=\sum_{\alpha\in S_{n-2}}C_{1,\alpha(2,3,\ldots,n-1),n}A_{\text{flat}}(1,\alpha,n)\,,
\end{align}
where $A_{\text{flat}}(1,\alpha,n)$ is the (color-ordered) partial amplitude given by
\begin{align}
    A_{\text{flat}}(1,\alpha,n)=\sum_{\beta\in S_{n-2}}m(1,\alpha,n|1,\beta,n)N_{1,\beta(2,3,\ldots,n-1),n} \ .
\end{align}
Here $N_{1,\beta,n}$ are DDM-basis numerators associated with half ladder graphs as in eq.~\eqref{eq:halfladder}. They also form a basis for all the BCJ numerators. As a consequence of the color structure, we can use the Kleiss-Kuijf relation~\cite{Kleiss:1988ne}
\begin{align}\label{eq:KKrel}
A_{\text{flat}}(1,\alpha,n,\beta)&=(-1)^{|\beta|}\sum_{\sigma \in \alpha\shuffle\beta^{T} }A_{\text{flat}}(1,\sigma,n)
\end{align}
and cyclicity to expand any partial amplitudes in terms of the DDM basis ones.

We note that the DDM basis is minimal for color factors, but over-complete for partial amplitudes on the support of on-shell massless kinematics. This is reflected by the fact that the rank of $m(\alpha|\beta)$, as a matrix in the DDM basis, is only $(n-3)!$. Crucially, the null vectors of $m(\alpha|\beta)$ translate to BCJ amplitude relations for partial amplitudes,
\begin{align}\label{eq:nvflat}
    \sum_{\beta\in S_{n-2}}v(\beta)m(1,\beta,n|\alpha)=0\quad \longrightarrow \quad \sum_{\beta\in S_{n-2}}v(\beta)A_{\text{flat}}(1,\beta,n)=0\,.
\end{align}
For example, at four points the $m(\alpha|\beta)$ matrix, 
\begin{align}
    \begin{bmatrix}
m(1,2,3,4|1,2,3,4) & &m(1,2,3,4|1,3,2,4) \\ 
m(1,3,2,4|1,2,3,4) & &m(1,3,2,4|1,3,2,4)
\end{bmatrix}=\begin{bmatrix}
\frac{1}{s_{12}}+\frac{1}{s_{23}} & &-\frac{1}{s_{23}} \\ 
-\frac{1}{s_{23}} & &\frac{1}{s_{13}}+\frac{1}{s_{23}}
\end{bmatrix},
\end{align}
has a null vector $v=[s_{12},\,-s_{13}]$, which leads to the BCJ amplitude relation
\begin{align}
    s_{12}A_{\text{flat}}(1,2,3,4)=s_{13}A_{\text{flat}}(1,3,2,4)\,.
\end{align}
More generally, the fundamental BCJ relations can be written as~\cite{Bern:2008qj}
\begin{equation}\label{eq:fBCJ}
0=s_{12}A_{\text{flat}}(1,2,\ldots,n)+\sum_{j=3}^{n-1}\Big(s_{12}+\sum_{k=3}^{j}s_{2j}\Big)A_{\text{flat}}(1,3,\ldots,j,2,j+1,\ldots,n)\,.  
\end{equation}

\subsection{Color/kinematics duality for AdS boundary correlators}

To define an extension of color/kinematics duality to field theories in AdS space,  
we need to first assume a suitably general form for their boundary correlators. 
Motivated by eq.~(\ref{eq:colordressformula}), a natural generalization of eq.~(\ref{eq:bcjrepflatspace}) is that an AdS boundary correlator $\mathcal{A}$ 
can be cast into the form
\begin{equation}\label{eq:bcjrepads}
\mathcal{A}=\sum_{\textrm{cubic }g}C(g|\alpha_{g}) \left ( \prod_{I\in g}\frac{1}{D_{I}^{2}} \right )  \hat{N}(g|\alpha_{g}) D_{\underbrace{\scriptstyle d,d,\ldots,d}_n}\,,
\end{equation}
\noindent where the numerators $\hat{N}(g|\alpha_g)$ are now differential operator-valued, act directly on the $D$-functions and absorb the normalization factors of the bulk-boundary propagators. Note that we have placed the product of $(D_{I}^{2})^{-1}$ to the left of the kinematic numerators in eq.~(\ref{eq:bcjrepads}) for reasons that will be clarified shortly.
A more general form of the boundary correlator replaces the factor $\hat{N}(g|\alpha_{g}) D_{\scriptstyle d,d,\ldots, d}$ by some more general structure ${N}(g|\alpha_{g})$ which may perhaps be written as a linear combination of differential operators acting on $D$ functions. 

With the definition \eqref{eq:bcjrepads}, the kinematic Jacobi relations are 
taken to  be the operator relations,
\begin{equation}\label{eq:kinematicnumerator}
\hat{N}(g_{s}|I_{1}I_{2}I_{3}I_{4})+\hat{N}(g_{t}|I_{1}I_{4}I_{2}I_{3})
+\hat{N}(g_{u}|I_{1}I_{3}I_{4}I_{2})=0 \ .   
\end{equation}
With the more general form of correlators, the kinematic Jacobi relations are 
functional relations, in close similarity with flat space scattering amplitudes.
We will comment briefly on its consequences at the end of section~\ref{sec:4pBCJrep}.

Kinematic numerators as differential operators have already appeared in the study of celestial amplitudes in flat space~\cite{Casali:2020vuy,Casali:2020uvr,
Kalyanapuram:2020aya}, so it should not be surprising that it may also happen in AdS space. 
In this paper, the representation~\eqref{eq:bcjrepads} is realized manifestly 
for the NLSM at four and six points. 

The kinematic Jacobi relation \eqref{eq:kinematicnumerator} can be relaxed so that the combination
of numerator factors on the right-hand side is required to vanish only when acting on functions
of the type $D_{\Delta_1, \Delta_2, \dots}$, as it is the case when the numerator factors are assembled into a correlator. While we have not explicitly verified, it is natural to expect that only this weaker relation is required by the gauge invariance of eq.~\eqref{eq:bcjrepads}.
There is no analog of this weaker relation for tree-level flat space amplitudes in momentum
space; at loop level, however, this is analogous to the requirement that the kinematic Jacobi relations hold only up to total derivatives.

The leap from eq.~(\ref{eq:colordressformula}) to eq.~(\ref{eq:bcjrepads}) is partially motivated by a recent generalization of ambitwistor string models to $\ads_{3}\times S^{3}$ \cite{Roehrig:2020kck}. These models can be interpreted as taking the infinite tension limit of a WZW model with $\ads_{3}\times S^{3}$ target space. For a non-abelian spin-1 theory on $\ads_{3}\times S^{3}$, the ambitwistor model in ref.~\cite{Roehrig:2020kck} provides a CHY-like formula for the differential representation of $A(\alpha)$ in a YM-Chern-Simons theory. We expect that these formulas for the differential correlator simplify to eq.~(\ref{eq:bcjrepads}), just as in flat space. Furthermore, we tentatively expect that the $\ads_{3}\times S^{3}$ ambitwistor model generalizes to higher dimensions, at least for the YM sector.  Proving these expectations, however, is beyond the scope of this paper. Therefore, we simply take the $\ads_{3}\times S^{3}$ computation as inspiration and conjecture that eqs.~(\ref{eq:newkinematicnumerator}) and (\ref{eq:bcjrepads}) hold for certain single-colored theories in higher dimensional AdS. 

Before we proceed, let us note that our discussion has been restricted to colored theories in AdS. There is a natural generalization of eq.~(\ref{eq:bcjrepads}) to gravitational theories which will be discussed in section \ref{towardsadsdoublecopy}. 

\subsection{BCJ relations for AdS boundary correlators}\label{sec:4pBCJrep}

We now demonstrate that the color/kinematics dual form~\eqref{eq:bcjrepads} of the AdS boundary correlators naturally lead to additional relations among the AdS partial correlators. First, we use the color Jacobi identity~(\ref{eq:newkinematicnumerator}) to expand the AdS correlator in the DDM basis,
\begin{equation}\label{eq:conjecturedadsBCJrep}
\mathcal{A}=\sum_{\alpha \in S_{n-2}} C_{1,\alpha(2,3,\ldots,n-1),n}A(1,\alpha,n) \ ,
\end{equation}
where $A(1,\beta,n)$ are the AdS partial correlators. We then perform the same expansion for the kinematic numerators, $\hat{N}(g|\alpha)$, now finding 
\begin{equation}\label{eq:ddmexpressionamp}
A(1,\alpha,n)=\sum_{\beta\in S_{n-2}}\hat{m}(1,\alpha,n|1,\beta,n)\hat{N}_{1,\beta(2,3,\ldots,n-1),n}D_{\underbrace{\scriptstyle d,d,\ldots,d}_n}\,,
\end{equation}
where $\hat{m}(\alpha|\beta)$ is the double partial correlator of BAS obtained by simply replace $s_I$ by $D_I^2$ in the flat space amplitude $m(\alpha|\beta)$. We note that $D_I^2$ and $D_{I'}^2$ always commute if they belong to the same Feynman diagram since $I$ and $I'$ always satisfy the condition in eq.~\eqref{eq:DIcommute}. The DDM basis partial correlators form a basis for all partial correlators due to the Kleiss-Kuijf relation~\eqref{eq:KKrel} and cyclicity, which depends on color Lie algebra only.

Similar to flat space amplitudes, eq.~(\ref{eq:ddmexpressionamp}) yields  relations among the partial correlators, since the null vectors of $\hat{m}(\alpha|\beta)$ are orthogonal to the vector 
of partial correlators,
\begin{align}\label{eq:nullvectorcond}
    \sum_{\beta\in S_{n-2}}\hat{v}(\beta)\hat{m}(1,\beta,n|\alpha)=0\quad \longrightarrow \quad \sum_{\beta\in S_{n-2}}\hat{v}(\beta)A(1,\beta,n)=0\,,
\end{align}
\emph{cf.} eq.~\eqref{eq:nvflat}. The null vectors $\hat{v}(\beta)$ in general are themselves differential operators. If $v(\beta)$ is a null vector of $m(\alpha|\beta)$ that is first order in Mandelstam variables, then it is not difficult to see that $\hat{v}(\beta)$ is still a null vector of $\hat{m}(\alpha|\beta)$ after the replacement $s_I\rightarrow D_I^2$.\footnote{The reasoning goes as follows. In flat space, proving BCJ relations requires using on-shell identities to cancel certain numerators with propagators. Now for AdS correlators, CWI works the same as on-shell identites, and cancellation between numerators and denominators will not be affected by non-commutativity since for each term every $D_{I}^2$ in the denominator commutes and there is only a single term in the numerator. Of course, finding more generic null vectors is difficult.} We can then conjecture that the rank of $\hat{m}(\alpha|\beta)$ is still $(n-3)!$ on the support of CWI~\eqref{eq:CWIn}. In particular, it leads to the conclusion that the partial correlators of the form~\eqref{eq:ddmexpressionamp} satisfy the fundamental BCJ relations
\begin{equation}\label{eq:BCJADS}
0=D_{12}^{2}A(1,2,\ldots,n)+\sum_{j=3}^{n-1}\Big(D_{12}^2+\sum_{k=3}^{j}D_{2j}^{2}\Big)A(1,3,\ldots,j,2,j+1,\ldots,n)\,,  
\end{equation}
which may formally be obtained from eq.~\eqref{eq:fBCJ} through the replacement 
$s_I\rightarrow D_I^2$.

To better understand the above statement, we now consider some explicit examples. From eq.~\eqref{eq:ddmexpressionamp}, the four-point DDM basis partial correlators are given by
\begin{equation}\label{eq:4pexplciitex}
\begin{bmatrix}
A(1,2,3,4)\\ 
A(1,3,2,4)
\end{bmatrix}=\underbrace{\begin{bmatrix}
\frac{1}{D_{12}^{2}}+\frac{1}{D_{23}^{2}} & -\frac{1}{D_{23}^{2}} \\ 
-\frac{1}{D_{23}^{2}} & \frac{1}{D_{13}^{2}}+\frac{1}{D_{23}^{2}}
\end{bmatrix}}_{\hat{m}}\begin{bmatrix}
\hat{N}_{1,2,3,4}\\ 
\hat{N}_{1,3,2,4}
\end{bmatrix}D_{d,d,d,d}\,.
\end{equation}
\noindent We would like to show that 
\begin{equation}\label{eq:testnullvector}
\hat{v}=\begin{bmatrix}
D_{12}^{2}\\ 
-D_{13}^{2}
\end{bmatrix} 
\end{equation}
annihilates the vector of partial correlators when acted from the left.
To this end, it is sufficient to show that $\hat{v}$ annihilates the $\hat{m}$,
\begin{equation}\label{nullvector1}
%\forall \alpha\in \{(1,2,3,4),\ (1,3,2,4) \}: \quad  
D^{2}_{12}\hat{m}(1,2,3,4|\alpha)-D^{2}_{13}\hat{m}(1,3,2,4|\alpha)\cong 0 \quad\text{for }\alpha=(1,2,3,4)\text{ and }(1,3,2,4)\,, 
\end{equation}
\noindent which can be checked explicitly. For example, fixing $\alpha=(1,2,3,4)$, we get
\begin{align}
&\quad D_{12}^{2}\hat{m}(1,2,3,4|1,2,3,4)-D^{2}_{13}\hat{m}(1,3,2,4|1,2,3,4)\nonumber\\
&\cong D_{12}^{2}\left(\frac{1}{D_{12}^2}+\frac{1}{D_{23}^2}\right)-(D_{23}^{2}+D_{12}^{2})\frac{1}{D_{23}^2} = 1+D_{12}^2\frac{1}{D_{23}^2}-1-D_{12}^2\frac{1}{D_{13}^2}=0\,. 
\end{align}
The expressions for $\hat{m}$ are obtained from eq.~\eqref{eq:mexample} through the replacement $s_{ij}\rightarrow D_{ij}^2$, and using the ``momentum conservation'' identity~\eqref{eq:stuAdS}. One can repeat the above exercise to show that eq.~(\ref{nullvector1}) holds for $\alpha=(1,3,2,4)$. Therefore, taking the dot product of eq.~(\ref{eq:testnullvector}) and eq.~(\ref{eq:4pexplciitex}) yields
\begin{equation}\label{eq:BCJ4pooint}
D_{12}^{2}A(1,2,3,4)=D_{13}^{2}A(1,3,2,4) \ .     
\end{equation}
As one of the main results of this paper, we will show that the four-point partial correlators of NLSM and of YM theory satisfy this relation.

For our second example we consider the five-point BCJ relation
\begin{align}\label{expamplebcj5}
0&=D_{12}^{2}A(1,2,3,4,5)+(D_{12}^{2}+D_{23}^{2})A(1,3,2,4,5)\,.\nonumber\\
&\quad +(D_{12}^{2}+D_{23}^{2}+D_{24}^{2})A(1,3,4,2,5)\,.    
\end{align}
According to eq.~\eqref{eq:ddmexpressionamp}, it suffices to prove that
\begin{equation}\label{eq:5pBCJex}
\begin{split}
0&\cong D_{12}^{2}\hat{m}(1,2,3,4,5|1,\alpha,5)+(D_{12}^{2}+D_{23}^{2})\hat{m}(1,3,2,4,5|1,\alpha,5) \\
&\quad +(D_{12}^{2}+D_{23}^{2}+D_{24}^{2})\hat{m}(1,3,4,2,5|1,\alpha,5)
\end{split}
\end{equation}
for all $\alpha\in S_3$. Here we choose $\alpha=(2,3,4)$ such that the relevant double partial correlators are all given in eq.~\eqref{eq:mexample}. It is now straightforward to show that
\begin{align}\label{eq:5pBCJexfullexp}
&\quad D_{12}^{2}\hat{m}(1,2,3,4,5|1,2,3,4,5)+(D_{12}^{2}+D_{23}^{2})\hat{m}(1,3,2,4,5|1,2,3,4,5) \nonumber\\
&\quad +(D_{12}^{2}+D_{23}^{2}+D_{24}^{2})\hat{m}(1,3,4,2,5|1,2,3,4,5) \\
&\cong D_{12}^{2}\left ( \frac{1}{D_{12}^{2}D_{34}^{2}}+\frac{1}{D_{12}^{2}D_{45}^{2}}+\frac{1}{D_{45}^{2}D_{23}^{2}}+\frac{1}{D_{23}^{2}D_{51}^{2}}+\frac{1}{D_{15}^{2}D_{34}^{2}} \right ) \nonumber\\
&\quad -(D_{12}^{2}+D_{23}^{2})\frac{1}{D_{23}^{2}}\left(\frac{1}{D_{45}^{2}}+\frac{1}{D_{15}^{2}} \right )-(D_{12}^{2}+D_{15}^{2}-D_{34}^{2})\frac{1}{D_{15}^{2}D_{34}^{2}}=0\,,\nonumber
\end{align}
where we have also used $D_{23}^2+D_{24}^2+D_{34}^2\cong D_{234}^2\cong D_{15}^2$ for the conformal generators on the second line.

Before proceeding, we note that while the AdS boundary correlators of the form~\eqref{eq:bcjrepads} naturally give rise to the BCJ relations~\eqref{eq:BCJADS}, the inverse does not hold. In other words, color/kinematics duality in the AdS boundary correlators might have a different manifestation than eq.~\eqref{eq:bcjrepads}. For example, at four-points, the following correlator,
\begin{equation}\label{BCJnondifferential}
\mathcal{A}=\frac{f^{a_{1}a_{2}x}f^{a_{3}a_{4}x}}{D_{12}^{2}}N_{s}(Z_{i},P_{i})+\frac{f^{a_{1}a_{4}x}f^{a_{2}a_{3}x}}{D_{23}^{2}}N_{t}(Z_{i},P_{i})+\frac{f^{a_{1}a_{3}x}f^{a_{4}a_{2}x}}{D_{13}^{2}}N_{u}(Z_{i},P_{i}) \,,  
\end{equation}
where
\begin{equation}
N_{s}(Z_{i},P_{i})+N_{t}(Z_{i},P_{i})+N_{u}(Z_{i},P_{i})=0 \,,     
\end{equation}
still leads to eq.~(\ref{eq:BCJ4pooint}). However, eq.~(\ref{BCJnondifferential}) is not 
equivalent to eq.~(\ref{eq:bcjrepads}) as $N_{s,t,u}$ need not necessarily 
be written in the form $\hat{N}_{s,t,u}D_{d,d,d,d}$. In practice, it is 
easier to verify relations like eq.~\eqref{eq:BCJ4pooint} than to directly 
construct kinematic numerators. While we have argued for the 
form~\eqref{eq:bcjrepads}, it is nevertheless important to keep an open mind.

\section{Vertex rules from embedding space action}\label{AdSrulesEmbeddingSpace}

By manifesting all of its symmetries, the embedding space formalism has proven an important
framework for organizing the results of calculations in AdS spaces of various dimensions.
While the translation of contact term Feynman graphs between AdS and the embedding space 
is straightforward, it becomes less so for exchange diagrams of vector and tensor fields.

In this section, we discuss an action-based approach to the embedding space Feynman rules. 
We will construct actions for the NLSM and YM theory in embedding space from which 
vertices can be extracted in the usual way and used for Witten-diagram calculations. 
While we do not spell it out, Einstein's gravity has a similar (though slightly more involved) presentation.
In addition, we will review the split-representation of the propagators and an algorithm for computing $n$-point AdS boundary correlators as explicit polynomials of $Z_{i}$, $P_{i}$ and $D$-functions. 

Those already familiar with the embedding space formalism, and its subtleties, may skip this section. 

\subsection{Nonlinear sigma model}

The action for the nonlinear sigma model in a curved space with metric $g$ is
\begin{equation}
{\cal L}_\text{NLSM}= -\Tr{(U^{-1}\nabla_\mu U)(U^{-1}\nabla^\mu U)}
\end{equation}
with $U$ an element in some group $G$ and $\nabla$ is (formally) 
the gravitational covariant derivative. 
In AdS embedding space, the action remains essentially unchanged, with the 
exception of the metric. While the embedding space is flat, to ensure that 
the metric reduces to the desired one, it is necessary to choose
\begin{equation}
G^{AB} = \eta^{AB} - \frac{X^A X^B}{X^2} \ ,
\label{projector}
\end{equation}
see ref.~\cite{Costa:2016hju, Penedones:2016voo, Giombi:2017hpr} 
and appendix~\ref{app_embedding_space}. 
The Feynman rules depend on the parametrization of the group element $U$;
with the standard exponential parametrization, $U = \exp(i \phi_a t^a)$ with real fields $\phi$, and generators $t^a$ obeying $[t^a, t^b]=i f^{abc} t^c$ and $\Tr[t^a t^b] = \frac{1}{2}\delta^{ab}$. The relevant Lagrangian to sixth order in fields 
is 
\begin{align}
{\cal L}_\text{NLSM} &= 
\frac{1}{2} G^{AB} \partial_A\phi_a\partial_B\phi_b\delta^{ab}
+ \frac{1}{24} G^{AB} f^{a_1a_2x}f^{a_3a_4x} \phi_{a_1}(\partial_A\phi_{a_2})(\partial_B\phi_{a_3})\phi_{a_4}\nonumber\\
&\quad + \frac{1}{720}G^{AB}f^{a_1a_2x}f^{xa_3y}f^{a_4a_5z}f^{za_6y}\phi_{a_1}(\partial_{A}\phi_{a_2})\phi_{a_3}\phi_{a_4}(\partial_{B}\phi_{a_5})\phi_{a_6}+\mathcal{O}(\phi^8)\,,
\label{expanded_Lagrangian}
\end{align}
For this choice of fields, the Lagrangian contains no terms with an odd number of fields, so all odd-point amplitudes vanish identically.\footnote{This holds, of course, also in flat space where all odd-point amplitudes also vanish identically.
In flat space we may choose nonvanishing color-kinematics-satisfying numerators \cite{Carrasco:2016ldy}. While we will not discuss this here, we expect that the same is true in AdS space. }
The four- and six-point vertices that enter the Feynman rules can be read off from eq.~(\ref{expanded_Lagrangian}). In addition to the vertices, the bulk-boundary and bulk-bulk propagators are necessary to calculate generic correlators. In terms of the embedding space coordinates, the scalar equation of motion on the AdS background $X^2=-1$ is given by
\begin{align}\label{eq:eomEmbedding}
    \partial_A(G^{AB}\partial_B\phi)-\Delta(\Delta-d)\phi =J\,,
\end{align}
where $J$ corresponds to scalar source terms. From eq.~(\ref{eq:eomEmbedding}), the 
scalar bulk-boundary propagator is\footnote{We follow here the normalization in \cite{Fitzpatrick:2011ia, Paulos:2011ie}, which is slightly different from that of \cite{Freedman:1998tz}, for the scalar bulk-boundary propagator. 
Together with the normalization for the vector-field bulk-boundary propagator in
eq.~\eqref{spin1btobound}, they are convenient to simplify certain overall factors 
for $d\ne 2$ in later sections.}
\begin{align}
\label{Kusual}
E_\Delta(P_{k}, X)
&=\frac{\mathcal{N}_{\Delta}}{(-2P_k\cdot X)^\Delta}\,,\qquad \mathcal{N}_{\Delta}=\frac{\Gamma(\Delta)}{2\pi^{d/2} \Gamma(\Delta-d/2+1)}\,,
\end{align}
The assumption that $z\ge 0$ implies $X\cdot P\le 0$. Another solution is the bulk-bulk propagator, which we write using the split representation 
\begin{equation}\label{scalarbulktobulk}
G_{\Delta}(X,Y)=\int_{-i\infty}^{i\infty}\frac{dc}{2\pi i} f_{\Delta}(c)\Omega_{c}(X,Y)
\end{equation}
where 
\begin{equation}
\begin{split}
\Omega_{c}(X,Y)&=-2c^2\int_{\partial \textrm{AdS}}dQ E_{d/2+c}(Q, X)E_{d/2-c}(Q, Y) \\
f_{\Delta}&=\frac{1}{(\Delta-d/2)^{2}-c^2}  \ .   
\end{split}
\end{equation}
The bulk-bulk propagator is normalized such that 
\begin{equation}
\big[\, \partial_A (G^{AB}\partial_B) - \Delta(\Delta-d)\big]G_{\Delta}(X,Y)=-\delta^{d+1}(X,Y) \ .     
\end{equation}
Physically, the split representation corresponds to a decomposition of the AdS bulk-bulk propagator in terms of AdS harmonic functions, $\Omega_{c}(X,Y)$, which are eigenfunctions of the AdS Laplacian that are divergence free. It is easy to check that eq.~(\ref{scalarbulktobulk}) is the bulk propagator by using that eq.~(\ref{Kusual}) is the bulk-boundary propagator and an identity that decomposes the AdS delta function into AdS harmonic functions. The crucial insight of the split representation is that the AdS harmonic functions can be represented as products of bulk-boundary propagators integrated over the boundary. Therefore, the split representation allows us to sew three-point correlators together in a manner reminiscent of BCFW recursions in flat space.

\subsection{Yang-Mills}\label{YMfeynamnrules}

We now turn to Yang-Mills, a theory of massless spin-1 states in AdS. The Lagrangian is given by
\begin{align}\label{adsYMlagranf}
{\cal L}_\text{YM}= -\frac{1}{4}F_{\mu\nu}^{a}F^{a,\mu\nu}
\end{align}
where the indices are contracted with the AdS metric, 
$F_{\mu\nu}^a=\nabla_\mu A_\nu^a - \nabla_\nu A_\mu^a - g f^{abc}A_\mu^b A_\nu^c$ and, as before, $\nabla$ is the gravitational covariant derivative. Similiar to the NLSM, the AdS embedding of eq.~(\ref{adsYMlagranf}) is essentially unchanged. Under the Lorentz gauge, it is given by
\begin{align}\label{eq:YMlagrangian}
\mathcal{L}&=-\frac{1}{2}G^{AB}G^{CD}\partial_{A}A^{a}_{C}\partial_{B}A^{a}_{D}+gf^{abc}G^{AB}G^{CD}(\partial_{A}A_{C}^{a})A_{B}^{b}A_{D}^{c} \\
&\quad -\frac{g^2}{4}G^{AB}G^{CD}f^{abx}f^{xcd}A_{A}^{a}A_{C}^{b}A_{B}^{c}A_{D}^{d}\,,  
\end{align}
from which we can read off the three-point and four-point vertices in the embedding space. The equations of motion for the spin-1 state are
\begin{equation}\label{equationsofmotion}
(\nabla^{2}-\Delta(\Delta-d)+1)A^{a,A}=J^{a,A}
\end{equation}
where $J^{a,A}$ corresponds to vector source terms. From eq.~(\ref{equationsofmotion}), the bulk-boundary propagator is 
\begin{equation}\label{spin1btobound}
E^{MA}_{\Delta}(P,X)=\left( \eta^{MA}-\frac{X^{M}P^{A}}{P\cdot X}\right)\frac{\mathcal{N}_{\Delta,1}}{(-2P\cdot X)^{\Delta}}\,,\qquad \mathcal{N}_{\Delta,1}=\frac{\Delta}{\Delta-1}\mathcal{N}_{\Delta}\,,
\end{equation}
which is well defined on the AdS hypersurface because $E^{MA}_{\Delta}X_{A}=0$ and $P_{M}E^{MA}_{\Delta}=0$. Crucially, we can write eq.~(\ref{spin1btobound}) in terms of the scalar bulk-boundary propagator using a differential operator, $\pD^{MA}$:
\begin{equation}\label{bulktoboundYMN}
E^{MA}_{\Delta}=%\frac{\Delta}{\Delta-1}
\frac{\Delta}{\Delta-1}\pD^{MA}E_{\Delta}, \quad \textrm{where} \quad \pD^{MA}_{\Delta}=\eta^{MA}+\frac{1}{\Delta}P^{A}\frac{\partial}{\partial P_{M}} \ .
\end{equation}
Another solution to the equations of motion is the bulk-bulk propagator, which we again write using the split-representation,
\begin{equation}\label{spin1bulk}
G^{AB}_{\Delta}(X,Y)=\int_{-i\infty}^{i\infty}\frac{dc}{2\pi i} f_{\Delta}(c)\Omega_{c}^{AB}(X,Y)   
\end{equation}
where $f_\Delta$ is the same as for a scalar field and
\begin{equation}
\begin{split}
\Omega^{AB}_{c}(X,Y)&=-2c^{2}\int_{\partial\textrm{AdS}}dQ\,\eta_{MN}E^{MA}_{d/2+c}(Q,X)E^{NB}_{d/2-c}(Q,Y) \\
\end{split}
\end{equation}
Similar to the scalar split representation, the vector split representation also corresponds to a decomposition of the bulk-bulk propagator in terms of spin-1 AdS harmonic functions, $\Omega^{AB}_{c}(X,Y)$, which is well defined in the bulk embedding because $X_{A}\Omega^{AB}_{c}(X,Y)=0$ and $\Omega^{AB}_{c}(X,Y)Y_{B}=0$. The same property is also satisfied by $G^{AB}_{\Delta}$. Using eqs. (\ref{bulktoboundYMN}) and (\ref{spin1bulk}) in the evaluation of position space correlators prevents the appearance of uncontracted bulk integration variables in the AdS boundary correlators and will allow us to write the correlator manifestly in terms of $Z_{i}$, $P_{i}$ and $D$-functions.

\subsection{On-shell and off-shell correlators}\label{onshelloffshell}

As it is well-known, field equations in AdS space generically exhibit two solutions with distinct asymptotics near the boundary,
\begin{eqnarray}
\phi(z, x) = \phi_0(z^{\Delta_0}+\dots)+\phi_1(z^{\Delta_1} + \dots) \ ,
\end{eqnarray}
where $\Delta_0$ and $\Delta_1$ are the smaller and larger solutions to a second-order equation which relates the $\text{SO}(d+1,1)$ quantum 
numbers of the field and its AdS mass term, respectively.
They are distinguished by the fact that a solution with $\phi_0$ asymptotics is not 
normalizable near the boundary while a solution with $\phi_1$ asymptotics is normalizable, $\int_{\ads} d^dx \int_{0} dz \sqrt{g} |\phi|^2<\infty$.\footnote{Technically, if $\Delta$ lies in the range $(d-2)/2<\Delta<d/2$, either $\phi_{0}$ or $\phi_{1}$ can correspond to the source term. The choice between $\phi_{0}$ and $\phi_{1}$ as the source term simply corresponds to how there are two different quantizations of the bulk scalar field \cite{Klebanov:1999tb}.}

For scalar fields, the traditional definition of $\ads$ mass is related to the conformal weight by the formula 
\begin{equation}
\Delta(\Delta-d) = m^2 \ .
\end{equation}
Therefore, for a massless scalar, such as the fields of the NLSM, we get
\begin{equation}
\Delta_1 = \Delta = d \ ,
\qquad
\Delta_0 = d-\Delta = 0 \ .
\end{equation}
A similar consideration for vectors and gravitons in the bulk yields $\Delta=d-1$ and $\Delta=d$ respectively \cite{Witten:1998qj,Gubser:1998bc}.
It is worth mentioning that there is no invariant meaning to the AdS mass because fields with the same properties and belonging to 
the same multiplet have different AdS energies \cite{Gunaydin:1984vz}.  A possible definition of massless fields in AdS is that they occur in the tensor 
product of  two doubleton multiplets,  which correspond to massless conformal fields on the boundary~\cite{Gunaydin:1998sw}; the mass can 
then be interpreted as a suitable shift of the corresponding quadratic Casimir  of $\text{SO}(d+1,1)$.\footnote{We thank Murat G\"unaydin for discussion on this point.}
For ${\cal N}=8$ supergravity in $\ads_5\times S^5$, the corresponding operators are conserved currents in ${\cal N}=4$ sYM theory, belonging to the stress tensor multiplet.

The leading field asymptotics on a surface parallel to the boundary at $z=\epsilon$ serves as a source for gauge-invariant operators of dimension $\Delta$, as
\begin{equation}
S_\text{boundary} \sim \int_{\ads} d^d x \sqrt{-\gamma_\epsilon} \phi(\epsilon, x) {\cal O}(\epsilon, x)= \int_{\ads} d^d x \phi_0(x) \epsilon^{-\Delta} {\cal O}(\epsilon, x) \ ,
\end{equation}
and ${\cal O}(\epsilon, x)=\epsilon^\Delta {\cal O}(x)$ render this term independent of $\epsilon$. Thus, by differentiating the effective action with respect to $\phi_0$, one 
evaluates~\cite{Witten:1998qj} correlation functions of gauge-invariant operators in the boundary theory. From the perspective of the bulk theory, they can be interpreted as 
correlation functions of the fields with these prescribed asymptotics; we shall refer to them as {\em on-shell} correlation functions. By analogy with the case of flat space correlation functions with external states not obeying the free equations of motion, we will refer to bulk correlation functions
whose asymptotics are not $\phi_0$ as {\em off-shell}. 

In general, off-shell correlation functions do not have an immediate boundary interpretation for specific values of the conformal weight. However, they feature prominently in the split representation of the bulk-bulk propagator.  In that form, the propagator is written as a sum of products of 
bulk-boundary propagators, see e.g. eq.~\eqref{scalarbulktobulk}, and thus higher-point correlators are written as sums of products of lower-point correlators which have at least one leg off-shell in the sense defined above. 
For scalar fields, one bulk-boundary propagator factor corresponds to the conformal weight $\Delta$ of the scalar field while the other corresponds to a weight $d-\Delta$ 
associated to the normalizable mode of the same scalar field. While the former is a non-normalizable mode and thus leads to an on-shell field, the latter is a normalizable 
mode of the same  field; its usual interpretation is that it defines a particular state in the boundary theory so the correlator with one such insertion may be interpreted as 
a term in the perturbative expansion of an on-shell correlator in that state. A similar interpretation should hold for correlators with more than one off-shell leg, except that the 
relevant state corresponds to turning on normalizable modes of several fields.
The bulk-bulk propagators of higher-spin fields have support on bulk-boundary propagators 
with AdS energies beyond those corresponding to the normalizable and non-normalizable modes. 
They have a less straightforward boundary interpretation, but may perhaps be understood as 
needed to obtain a representation of the $(d+1)$-dimensional rotation group.

In flat space, field redefinitions change the correlation functions of fundamental fields, but these changes are projected out of S-matrix elements by the LSZ reduction. 
It is interesting to ask whether our definition of on-shellness has similar properties. One might expect this to be the case in light of the holographic duality between 
on-shell correlators and gauge theory correlation functions of gauge-invariant operators. Indeed, the Schwinger-Dyson equation\footnote{Here the hat signifies that the field at that 
position is absent from the correlation function and $\phi$ denotes a generic field, not necessarily a scalar.}
\begin{align}
\Big\langle \frac{\delta S}{\delta\phi(x)}\phi(x_1)\dots\phi(x_n)\Big\rangle = \sum_{i=1}^n \delta(x-x_i)\langle \phi(x_1)\dots {\widehat \phi(x_i)}\dots\phi(x_n)\rangle \ ,
\end{align}
holds in AdS space (and more generally in curved space); since the field sources at $x_i$, $i=1,...,n$ are placed on the boundary while the argument of 
$\frac{\delta S}{\delta\phi(x)}$ is a bulk point, they cannot coincide so the right-hand side vanishes identically implying that AdS on-shell correlation functions in the 
sense defined above are invariant under suitable field redefinitions.\footnote{Similar to field redefinitions that leave invariant the 
S matrix of a flat-space field theory, field redefinitions that leave correlators invariant should not change quadratic term of bulk fields and vanish at the boundary. For example, $\phi_i\mapsto \phi'_i = \sum_j a_{ij}\phi_j + \text{nonlinear}$ yields a sum of the original correlation functions weighted by coefficients $a_{ij}$. This may be easily understood by noticing that the boundary operators sourced by the fields $\phi'$ 
are linear combinations of those sourced by the fields $\phi$. More generally, it is not difficult to see that, if a nonlinear field redefinition do not change the boundary conditions of a field, then the nonlinear terms are subleading at the boundary and therefore do not change the on-shell correlators.}

The Schwinger-Dyson equation, however, does not hold for off-shell correlation functions. This is easiest to see by looking at a free field theory for $n=1$; the off-shell 
two-point function of as defined above does not correlate the conformal weight with the AdS Lagrangian mass term while $\frac{\delta S}{\delta\phi(x)}$ depends on the AdS 
Lagrangian mass term, so $\langle \frac{\delta S}{\delta\phi(x)}\phi(x_1) \rangle$ cannot be proportional to  $\delta(x-x_1)$. This is consistent with the earlier observation
that off-shell correlation functions do not have a straightforward gauge theory interpretation. It is interesting that this dependence on the choice of fields cancels
out when off-shell correlators are assembled into on-shell ones. 

It has been recently shown that it is interesting to consider varying the mass of bulk fields. For example, one can extract the proper time from the event horizon to a black 
hole singularity by studying how thermal one point functions vary with the mass of the bulk field \cite{Grinberg:2020fdj}, allowing one to probe the bulk geometry of thermal states beyond the quantum entanglement wedge. Furthermore, analytic continuations in spin and in $\Delta$ are famously connected \cite{Caron-Huot:2017vep, Simmons-Duffin:2017nub} 
to light ray operators and the OPE inversion formula. Further discussion is beyond the scope of this paper. 

\section{NLSM in AdS}
\label{NSLMsection}
It is not difficult to evaluate the four-point boundary correlator of the NLSM fields in AdS 
using the embedding space action discussed in the previous section. We will then verify that
they obey the AdS BCJ relations discussed in section~\ref{sec:4pBCJrep} and put them into 
a form that manifests color/kinematics duality.

As we will see, the $X$ dependence in the $n$-point vertex following from
the presence of the projector~\eqref{projector} in eq.~\eqref{expanded_Lagrangian} does not 
contribute to the on-shell $n$-point contract-term contributions to $n$-point functions. 
It however becomes important in the contributions of $n$-point vertices to 
higher multiplicity correlators or even in their contribution to at the same-multiplicity 
correlators if at least one of the external lines is taken off-shell.

\subsection{NLSM correlators from AdS vertices}

The four-point Witten diagram for the NLSM in AdS is straightforward to evaluate using the four-point vertex from eq.~\eqref{expanded_Lagrangian} and the bulk-boundary propagator in eq. (\ref{Kusual}). The result contains three related color structures,
\begin{align}
\label{WittenDiag4ptNLSM}
\mathcal{A}^{a_1a_2a_3a_4}_{\Delta\Delta\Delta\Delta}
%{\cal A}(1^a, 2^b, 3^c, 4^d) 
&= 
  f^{a_1a_2x}f^{xa_3a_4} \pA_{1,2,3,4}
+ f^{a_2a_3x}f^{xa_1a_4} \pA_{2,3,1,4}
+ f^{a_3a_1x}f^{xa_2a_4} \pA_{3,1,2,4}\,,
\\
\pA_{1,2,3,4}&=-\frac{\mathcal{N}_{\Delta}^4}{6} 
%\left( \frac{\pi^{-d/2} \Gamma[\Delta]}{2\Gamma[1+\Delta-d/2]} \right)^4
\int_{\ads} dX
\prod_{i=1}^4 \frac{1}{(-2 P_i\cdot X)^\Delta} (\eta^{AB} + X^A X^B)\cr
&\quad \times 2\Delta^2
    \left(\frac{P_{1,A}}{(-2 P_1\cdot X)} - \frac{P_{2,A}}{(-2 P_2\cdot X)}\right)
    \left(\frac{P_{3,B}}{(-2 P_3\cdot X)} - \frac{P_{4,B}}{(-2 P_4\cdot X)}\right) \,,
\end{align}
where $\pA_{2,3,1,4}$ and $\pA_{3,1,2,4}$ can be obtained from $\pA_{1,2,3,4}$ by an index relabeling. Here we also assume that all four external scalars have the same weight $\Delta$. The color-ordered partial correlators can then be evaluated in the usual way, for example,
\begin{align}
\label{NLSMorderedamplitudes}
A(1_{\Delta},2_{\Delta},3_{\Delta},4_{\Delta}) &= \pA_{1,2,3,4} - \pA_{2,3,1,4}\,,
\nonumber\\
A(1_{\Delta},3_{\Delta},2_{\Delta},4_{\Delta}) &= \pA_{2,3,1,4} - \pA_{3,1,2,4}\,,
\quad
\text{etc.}
\end{align}
and, as is the case for all theories whose color factor only contains structure constants $f^{abc}$, they obey the Kleiss-Kuijf relation~\eqref{eq:KKrel}.

Using that the $X$-dependent factors in $G^{AB}$ drop out, the integral can be expressed in terms of $D$-functions:
\begin{align}
\pA_{1,2,3,4}=\frac{\Delta^2\mathcal{N}_{\Delta}^4}{6} 
&(P_{13} D_{\Delta+1, \Delta,\Delta+1,\Delta} 
+ P_{24} D_{\Delta, \Delta+1,\Delta,\Delta+1} \nonumber\\
&- P_{23} D_{\Delta, \Delta+1,\Delta+1,\Delta} 
- P_{14} D_{\Delta+1, \Delta,\Delta,\Delta+1})\,,
\end{align}
where $P_{ij}=-2P_i \cdot P_j$. The other two coefficients are obtained by permuting the 
labels of external lines, making sure to appropriately permute the indices of the $D$-functions.
However, a form that is more useful for our purpose of exploring relations between partial correlators, 
\begin{align}
\pA_{1,2,3,4}&=\frac{\mathcal{N}_{\Delta}^4}{24}
(
  D^2_{13} 
+ D^2_{24} 
- D^2_{23}
- D^2_{14}) D_{\Delta, \Delta,\Delta,\Delta} \ ,
\end{align}
may be derived using the integral representation of the $D$-functions as well as the definition~\eqref{conformalgener} of the $D^2_{ij}$ operators.
Using eq. (\ref{eq:compident}), the
position-dependent coefficients of the correlator \eqref{WittenDiag4ptNLSM} become 
\begin{subequations}
\begin{align}
\pA_{1,2,3,4}&=\frac{\mathcal{N}_{\Delta}^4}{12} 
(
  D^2_{13} 
- D^2_{23}
) D_{\Delta, \Delta,\Delta,\Delta} \ ,
\\
\pA_{2,3,1,4}&=\frac{\mathcal{N}_{\Delta}^4}{12} 
(
  D^2_{12} 
- D^2_{13}
) D_{\Delta, \Delta,\Delta,\Delta} \ ,
\\
\pA_{3,1,2,4}&=\frac{\mathcal{N}_{\Delta}^4}{12} 
(
  D^2_{23} 
- D^2_{12}
) D_{\Delta, \Delta,\Delta,\Delta} \ .
\end{align}
\end{subequations}
Interestingly, we notice 
\begin{align}
\pA_{1,2,3,4} + \pA_{3,1,2,4} + \pA_{2,3,1,4} = 0 \,.
\end{align}
While this relation is reminiscent of color/kinematics duality, it is also different 
from its flat-space counterpart as it effectively includes the denominators of diagrams, 
not only their numerators. We will see shortly a form of the four-point NLSM AdS boundary
correlator that obeys color/kinematics duality in a sense closer to that of flat space.

First however, let us explore whether the four-point partial correlator of the NLSM 
obeys the AdS generalization of the BCJ relations, as conjectured in section \ref{AdS_AmplitudesRelationsGeneral}. Using eq.~\eqref{NLSMorderedamplitudes} and identities 
in appendix~\ref{Dfunctionidentities}, we get
\begin{align}
\label{PartialNLSM}
A(1_{\Delta},2_{\Delta},3_{\Delta},4_{\Delta}) &= \frac{\mathcal{N}^4_{\Delta}}{4} \,
D^2_{13}  D_{\Delta,\Delta,\Delta,\Delta}\,,\quad
A(1_{\Delta},3_{\Delta},2_{\Delta},4_{\Delta}) = \frac{\mathcal{N}_{\Delta}^4}{4} \,
D^2_{12}  D_{\Delta,\Delta,\Delta,\Delta} \ ,
\end{align}
with the third partial correlator being determined either by explicit 
calculation as above or through the Kleiss-Kuijf relation~\eqref{eq:KKrel}.

The expected four-point AdS BCJ relation \eqref{eq:BCJ4pooint} is
\begin{align}\label{eq:BCJ4NLSM}
  D^2_{12}A(1_{\Delta},2_{\Delta},3_{\Delta},4_{\Delta})-D^2_{13}A(1_{\Delta},3_{\Delta},2_{\Delta},4_{\Delta})& = \frac{\mathcal{N}_{\Delta}^4}{4} 
  [D^2_{12}, D^2_{13}]  D_{\Delta, \Delta,\Delta,\Delta}
  \\
&  \propto
  f_{AB, CD, EF} D_1^{AB} D_2^{CD} D_3^{EF} D_{\Delta, \Delta,\Delta,\Delta}\,,\nonumber
\end{align}
where $f_{AB, CD, EF}$ are the structure constants of $\text{SO}(d+1,1)$.
While the commutator $[D_{12}^2, D_{13}^2] $ is generically nonvanishing, the fact that
{\em the integrand} of $D_{\Delta, \Delta,\Delta,\Delta}$ is in its kernel,
\begin{align}
\label{vanishing}
 f_{AB, CD, EF} D_1^{AB} D_2^{CD} D_3^{EF} \prod_{i=1}^{n\ge 3} \frac{1}{(-2P_i\cdot X)^\Delta} = 0  \ ,
\end{align}
can be understood by noticing that the variables that can appear, $X\cdot P_i$ and 
$P_i\cdot P_j$, do not allow the construction of a scalar function that is completely antisymmetric in the labels $1, 2$ and $3$. Of course, this statement can be verified by a direct calculation by using the relation~\eqref{eq:D12D13}.

We therefore see that the four-point NLSM partial correlators obey the AdS generalization 
\begin{align}
D^2_{12}A(1_{\Delta},2_{\Delta},3_{\Delta},4_{\Delta})-D^2_{13}A(1_{\Delta},3_{\Delta},2_{\Delta},4_{\Delta})& = 0 \ .
\end{align}
of the flat-space BCJ amplitudes relations ~\cite{Bern:2008qj}.
While this does not require that the external lines be massless, we will see that  
masslessness will be required by the BCJ relation for the six-point amplitude
computed in section~\ref{6pcomputation}.
We will now construct a representation of the 
four-point NLSM AdS boundary correlator that manifests color/kinematics duality. We find that this representation only exists for massless external particles. Therefore, for scalars, we need $\Delta=d$. Thus, for massless AdS boundary correlators, we omit the labels of the conformal weight and write
\begin{align}
    \mathcal{A}^{a_1a_2\cdots a_n}_{\Delta_1\Delta_2\cdots\Delta_n}\xrightarrow{\text{massless}}\mathcal{A}_n^{a_1a_2\ldots a_n}\,.
\end{align}
The same shorthand notation will also be used in later sections for boundary correlators of AdS vector fields, for which the massless limit is $\Delta_i=d-1$.

\subsection{BCJ representation of NLSM correlator}

To construct a color/kinematics-satisfying representation of the massless color-dressed 
correlator found above, we use eq.~\eqref{AdSBBpropagator} to represent the exchange graphs in the $s$, $t$ and $u$ channels. For example,
\begin{align}
\label{scalar_exchange}
E_{ij}\equiv \int_\ads dX
\frac{1}{(-2P_k\cdot X)^d(-2P_l\cdot X)^d}
\frac{1}{\Box_X}
\frac{1}{(-2P_i\cdot X)^d(-2P_j\cdot X)^d}= \frac{1}{D^2_{ij}}D_{d,d,d,d}
\end{align}
with $i\ne j\ne k\ne l = 1,\dots,4$ represents the scalar four-point exchange graph with external legs $i$ and $j$ meeting at a cubic vertex, and legs $k$ and $l$ meeting at the other cubic vertex. The two bulk vertices are connected by a bulk-bulk propagator resulting from the action of $\frac{1}{D_{ij}^2}$ on the contact term $D_{d,d,d,d}$.

Since both vertex factors
and bulk-bulk propagators are represented as non-commuting differential operators in the position-space framework we are employing here, 
the precise definition of kinematic numerators is not a priori clear. 
We will {\em define} the kinematic numerator operators $\hat N_{s, t, u}$ as:
\begin{align}\label{eq:A4NLSM}
{\cal A}^{a_1a_2a_3a_4}_4 = \frac{\mathcal{N}^4_{d}}{4} \left(
  \frac{C_s}{D^2_{12}} \hat N_s  
+ \frac{C_t}{D^2_{23}} \hat N_t  
+ \frac{C_u}{D^2_{13}} \hat N_u \right) D_{d,d,d,d} \ ,
\end{align}
with the color factors $C_{s,t,u}$ corresponding to the three color factors in
eq.~\eqref{WittenDiag4ptNLSM}, respectively. The yet-be-determined kinematic 
numerators are interpreted as operators acting towards the right. Here we pull out a normalization factor in order to simplify the kinematic numerators presented below.

Taking inspiration from the known flat-space 
color/kinematic-satisfying NLSM numerators \cite{Cheung:2017yef,Du:2016tbc,Du:2017kpo,Carrasco:2016ygv} %
\begin{align}
    N_{s,\textrm{flat}} = su \qquad  N_{t,\textrm{flat}} = 0 \qquad  N_{u,\textrm{flat}} = - s u \ ,
\end{align}
and requiring the NLSM partial correlators \eqref{PartialNLSM} be 
reproduced, we find that a suitable set of kinematic numerators exhibiting a
kinematic analog of the color Jacobi relation is
\begin{align}
     \hat N_s = D^2_{12} D^2_{13}
     \,,\quad
     \hat N_u = -D^2_{13}D^2_{12}
     \,,\quad
     \hat N_{t} = - [D^2_{12}, D^2_{13}] \, .
\end{align}
While $\hat N_{t}$ is generically a nonvanishing operator, its action on
$D_{d,d,d,d}$ vanishes according to eq.~\eqref{vanishing}, so the 
partial correlators~\eqref{PartialNLSM} are correctly reproduced. 
The analog of the generalized gauge symmetry allows us to modify these operators as
\begin{align}\label{eq:deltaN}
     \delta {\hat N}_{s} = D^2_{12} {\cal O} + \mathcal{O}_s
     ~,\quad
     \delta {\hat N}_u = D^2_{13}{\cal O} + \mathcal{O}_u
     ~,\quad
     \delta {\hat N}_{t} = D^2_{23}{\cal O} + \mathcal{O}_t \ ,
\end{align}
for some operator ${\cal O}$ and $\mathcal{O}_{s,t,u}$ that satisfy $\mathcal{O}_{s,t,u}D_{d,d,d,d}=0$ and $\mathcal{O}_s+\mathcal{O}_t+\mathcal{O}_u=0$, while maintaining the color/kinematics. It is trivial to see that, similarly to its flat space 
analog, this modification does not alter the partial correlator. For example, the choice 
\begin{gather}
    \mathcal{O}=-\frac{1}{3}D_{23}^2-\frac{2}{3}D_{13}^2\,,\nonumber\\
    \mathcal{O}_s=\frac{1}{3}[D_{12}^2,D_{23}^2]\,,\quad\mathcal{O}_{u}=\frac{1}{3}[D_{13}^2,D_{12}^2]\,,\quad\mathcal{O}_t=[D_{12}^2,D_{13}^2]+\frac{1}{3}[D_{23}^2,D_{13}^2]\,,
\end{gather}
can be used to make the numerators satisfy the symmetry of respective graphs,
\begin{align}
    \hat{N}_s+\delta\hat{N}_s &\cong\frac{1}{3}(D_{12}^2D_{13}^2-D_{23}^2D_{12}^2)\,,\nonumber\\
    \hat{N}_u+\delta\hat{N}_u &\cong\frac{1}{3} (D_{13}^{2}D_{23}^{2}-D_{12}^2D_{13}^2)\,,\\
    \hat{N}_t+\delta\hat{N}_t &\cong \frac{1}{3} (D_{23}^2D_{12}^2-D_{13}^{2}D_{23}^{2})\,.\nonumber
\end{align}
We use $\cong$ here because we have used identities that hold on the support of CWI, for example, eq.~\eqref{eq:stuAdS}, to reach the right hand side.

Next, we argue that a color/kinematics-satisfying representation is not present if we instead try to use the representation in which $1/D_{ij}^2$ is placed to the right of the numerator factors. In this alternative form, the four-point AdS boundary correlator, which is intuitively 
more similar to the flat space one, is
\begin{align}
\mathcal{A}_4^{a_1a_2a_3a_4} = 
  C_s \tilde N_s  E_{12}
+ C_t \tilde N_t  E_{23}
+ C_u \tilde N_u  E_{13} \, ,
\end{align}
with the scalar exchange graphs $E_{ij}$ defined in eq.~\eqref{scalar_exchange}. Similar to eq.~\eqref{eq:deltaN}, we can modify the $\tilde n_i$ by operators that do not change the final AdS boundary correlator,
\begin{gather}
    {\tilde N}_s = (D^2_{13} +{\cal O} + \mathcal{O}_s)D^2_{12}
    \,,\qquad
    {\tilde N}_u = (-D^2_{12} +{\cal O} + \mathcal{O}_u)D^2_{13}
    ~,\nonumber\\
    {\tilde N}_t = \frac{1}{D^2_{23}} ([D^2_{13}, D^2_{12}] 
    +D^2_{23}{\cal O} + D^2_{23}\mathcal{O}_t)D^2_{23}\, .
\end{gather}
The would-be kinematic Jacobi relations reduce to
\begin{align}\label{eq:jacobialt}
 \frac{1}{D^2_{23}}[[D^2_{13}, D^2_{12}], D^2_{23}] 
    +{\cal O}(D^2_{12}+D^2_{13}+D^2_{23})+\mathcal{O}_sD_{12}^2+\mathcal{O}_tD_{23}^2+\mathcal{O}_uD_{13}^2 \stackrel{?}{=} 0\,.
\end{align}
To formally solve for the $\mathcal{O}$'s, one can study the matrix elements of the above combination in the space spanned by the conformal partial waves. However, since only the first term in eq.~\eqref{eq:jacobialt} has nonvanishing matrix elements, imposing the kinematic Jacobi relations leads to a contradiction.

\subsection{The six-point correlator}\label{6pcomputation}
Computing the six-point NLSM boundary correlator requires both the four-point and six-point vertices in the Lagrangian~\eqref{expanded_Lagrangian}. There are four contributing Witten diagrams,
\begin{align}\label{eq:6pNLSM}
\pgfmathsetmacro{\r}{0.75}
\mathcal{A}_6^{a_1a_2\cdots a_6} = \begin{tikzpicture}[baseline={([yshift=-.5ex]current bounding box.center)},every node/.style={font=\scriptsize}]
\draw [] (0,0) circle (\r cm);
\filldraw (\r,0) circle (1pt) node[right=0pt]{$5$};
\filldraw (60:\r) circle (1pt) node[above=0pt]{$4$};
\filldraw (120:\r) circle (1pt) node[above=0pt]{$3$};
\filldraw (-\r,0) circle (1pt) node[left=0pt]{$2$};
\filldraw (-120:\r) circle (1pt) node[below=0pt]{$1$};
\filldraw (-60:\r) circle (1pt) node[below=0pt]{$6$};
\filldraw (-0.25,0) circle (1pt) (0.25,0) circle (1pt);
\draw [thick] (-\r,0) -- (\r,0) (60:\r) -- (0.25,0) (-60:\r) -- (0.25,0) (120:\r) -- (-0.25,0) (-120:\r) -- (-0.25,0);
\end{tikzpicture}
+
\begin{tikzpicture}[baseline={([yshift=-.5ex]current bounding box.center)},every node/.style={font=\scriptsize}]
\draw [] (0,0) circle (\r cm);
\filldraw (\r,0) circle (1pt) node[right=0pt]{$6$};
\filldraw (60:\r) circle (1pt) node[above=0pt]{$5$};
\filldraw (120:\r) circle (1pt) node[above=0pt]{$4$};
\filldraw (-\r,0) circle (1pt) node[left=0pt]{$3$};
\filldraw (-120:\r) circle (1pt) node[below=0pt]{$2$};
\filldraw (-60:\r) circle (1pt) node[below=0pt]{$1$};
\filldraw (-0.25,0) circle (1pt) (0.25,0) circle (1pt);
\draw [thick] (-\r,0) -- (\r,0) (60:\r) -- (0.25,0) (-60:\r) -- (0.25,0) (120:\r) -- (-0.25,0) (-120:\r) -- (-0.25,0);
\end{tikzpicture}
+
\begin{tikzpicture}[baseline={([yshift=-.5ex]current bounding box.center)},every node/.style={font=\scriptsize}]
\draw [] (0,0) circle (\r cm);
\filldraw (\r,0) circle (1pt) node[right=0pt]{$4$};
\filldraw (60:\r) circle (1pt) node[above=0pt]{$3$};
\filldraw (120:\r) circle (1pt) node[above=0pt]{$2$};
\filldraw (-\r,0) circle (1pt) node[left=0pt]{$1$};
\filldraw (-120:\r) circle (1pt) node[below=0pt]{$6$};
\filldraw (-60:\r) circle (1pt) node[below=0pt]{$5$};
\filldraw (-0.25,0) circle (1pt) (0.25,0) circle (1pt);
\draw [thick] (-\r,0) -- (\r,0) (60:\r) -- (0.25,0) (-60:\r) -- (0.25,0) (120:\r) -- (-0.25,0) (-120:\r) -- (-0.25,0);
\end{tikzpicture}
+
\begin{tikzpicture}[baseline={([yshift=-.5ex]current bounding box.center)},every node/.style={font=\scriptsize}]
\draw [] (0,0) circle (\r cm);
\filldraw (\r,0) circle (1pt) node[right=0pt]{$5$};
\filldraw (60:\r) circle (1pt) node[above=0pt]{$4$};
\filldraw (120:\r) circle (1pt) node[above=0pt]{$3$};
\filldraw (-\r,0) circle (1pt) node[left=0pt]{$2$};
\filldraw (-120:\r) circle (1pt) node[below=0pt]{$1$};
\filldraw (-60:\r) circle (1pt) node[below=0pt]{$6$};
\filldraw (0,0) circle (1pt);
\draw [thick] (-\r,0) -- (\r,0) (60:\r) -- (-120:\r) (120:\r) -- (-60:\r);
\end{tikzpicture}\,.
\end{align}
Similar to the four-point computation, it is straightforward to write the six-point contact diagram contribution in terms of derivative operators. We expand the result in terms of the DDM basis color factors $C_{1,\sigma(2,3,4,5),6}$,
\begin{align}\pgfmathsetmacro{\r}{0.75}
\begin{tikzpicture}[baseline={([yshift=-.5ex]current bounding box.center)},every node/.style={font=\scriptsize}]
\draw [] (0,0) circle (\r cm);
\filldraw (\r,0) circle (1pt) node[right=0pt]{$5$};
\filldraw (60:\r) circle (1pt) node[above=0pt]{$4$};
\filldraw (120:\r) circle (1pt) node[above=0pt]{$3$};
\filldraw (-\r,0) circle (1pt) node[left=0pt]{$2$};
\filldraw (-120:\r) circle (1pt) node[below=0pt]{$1$};
\filldraw (-60:\r) circle (1pt) node[below=0pt]{$6$};
\filldraw (0,0) circle (1pt);
\draw [thick] (-\r,0) -- (\r,0) (60:\r) -- (-120:\r) (120:\r) -- (-60:\r);
\end{tikzpicture}
= -\frac{\mathcal{N}_{d}^6}{720}
%\Bigg( \frac{\pi^{-d/2} \Gamma[\Delta]}{2\Gamma[1{+}\Delta{-}d/2]} \Bigg)^6 
\sum_{\sigma\in S_4}C_{1,\sigma(2,3,4,5),6}\,\pA_{1,\sigma(2,3,4,5),6}^{\text{contact}}\,,
\end{align}
where the position dependent piece is given by
\begin{align}
    \pA_{1,2,3,4,5,6}^{\text{contact}}=\Big(D^2_{12}-4D^2_{13}+3D^2_{14}+\text{cyclic}(1,2,3,4,5,6)\Big)D_{d,d,d,d,d,d}\,.
\end{align}
For the first three diagrams in eq.~\eqref{eq:6pNLSM}, we would like to write the bulk-bulk propagators therein by the corresponding derivative operators, for example,
\begin{align}\label{eq:channel123}
    \int_{\ads}dXdY\, G(X,Y)\prod_{i=1}^3\frac{1}{(-2P_i\cdot X)^{d}}\prod_{i=4}^{6}\frac{1}{(-2P_i\cdot Y)^{d}}=\frac{1}{D_{123}^2}D_{d,d,d,d,d,d,d}\,.
\end{align}
In the presence of nontrivial vertex functions, this can be done following different strategies. First, we notice that the above replacement is directly applicable when there are no derivatives acting on the bulk-bulk propagators. We can thus use integration-by-parts to move all derivatives with respect to the bulk point to the bulk-boundary propagators. As an example, we consider the first Witten diagram in eq.~\eqref{eq:6pNLSM}. The scalar current that flows from the bulk point $X$ to $Y$ can then be written as
\begin{align}\pgfmathsetmacro{\r}{0.75}
\begin{tikzpicture}[baseline={([yshift=-.5ex]current bounding box.center)},every node/.style={font=\scriptsize}]
\draw [] (0,0) circle (\r cm);
\filldraw (120:\r) circle (1pt) node[above=0pt]{$3$};
\filldraw (-\r,0) circle (1pt) node[left=0pt]{$2$};
\filldraw (-120:\r) circle (1pt) node[below=0pt]{$1$};
\filldraw (-0.25,0) circle (1pt) (0.25,0) circle (1pt);
\draw [thick] (-\r,0) -- (\r/3,0) (120:\r) -- (-\r/3,0) (-120:\r) -- (-\r/3,0);
\node at (\r/3,0) [right=0pt]{$Y$};
\node at (-\r/3,0) [below right=-3pt]{$X$};
\end{tikzpicture}
&=\frac{\mathcal{N}_{d}^3}{12}
\left[f^{a_1a_2b}f^{ba_3x} \hat{\pA}_{1,2,3}+(2\leftrightarrow 3)\right]\int_{\ads}dX\,G(X,Y)\prod_{i=1}^3\frac{1}{(-2P_i\cdot X)^{d}}\,,
\end{align}
where $\hat{\pA}_{1,2,3}=D_{123}^2-3D_{13}^2$. 
We can obtain the contribution from the full Witten diagram by gluing the above current with the one flowing out of $Y$,
\begin{align}\pgfmathsetmacro{\r}{0.75}
\begin{tikzpicture}[baseline={([yshift=-.5ex]current bounding box.center)},every node/.style={font=\scriptsize}]
\draw [] (0,0) circle (\r cm);
\filldraw (\r,0) circle (1pt) node[right=0pt]{$5$};
\filldraw (60:\r) circle (1pt) node[above=0pt]{$4$};
\filldraw (120:\r) circle (1pt) node[above=0pt]{$3$};
\filldraw (-\r,0) circle (1pt) node[left=0pt]{$2$};
\filldraw (-120:\r) circle (1pt) node[below=0pt]{$1$};
\filldraw (-60:\r) circle (1pt) node[below=0pt]{$6$};
\filldraw (-0.25,0) circle (1pt) (0.25,0) circle (1pt);
\draw [thick] (-\r,0) -- (\r,0) (60:\r) -- (0.25,0) (-60:\r) -- (0.25,0) (120:\r) -- (-0.25,0) (-120:\r) -- (-0.25,0);
\end{tikzpicture}
&=-\frac{\mathcal{N}_{d}^6}{144}
\left[f^{a_1a_2b}f^{ba_3x} \hat{\pA}_{1,2,3}+(2\leftrightarrow 3)\right]\left[f^{a_4a_5c\vphantom{b}}f^{ca_6x} \hat{\pA}_{4,5,6}+(5\leftrightarrow 6)\right]  \nonumber\\[-1.2em]
&\quad\times\int_{\ads}dXdY\, G(X,Y)\prod_{i=1}^3\frac{1}{(-2P_i\cdot X)^{d}}\prod_{i=4}^{6}\frac{1}{(-2P_i\cdot Y)^{d}}\,,
\end{align}
where the last line is now of the form~\eqref{eq:channel123}. The operators in the first square bracket commute with those in the second square bracket, and they all commute with $1/D^2_{123}$ as well. This feature allows us to simplify the final result significantly, which leads to
\begin{align}\pgfmathsetmacro{\r}{0.75}
\begin{tikzpicture}[baseline={([yshift=-.5ex]current bounding box.center)},every node/.style={font=\scriptsize}]
\draw [] (0,0) circle (\r cm);
\filldraw (\r,0) circle (1pt) node[right=0pt]{$5$};
\filldraw (60:\r) circle (1pt) node[above=0pt]{$4$};
\filldraw (120:\r) circle (1pt) node[above=0pt]{$3$};
\filldraw (-\r,0) circle (1pt) node[left=0pt]{$2$};
\filldraw (-120:\r) circle (1pt) node[below=0pt]{$1$};
\filldraw (-60:\r) circle (1pt) node[below=0pt]{$6$};
\filldraw (-0.25,0) circle (1pt) (0.25,0) circle (1pt);
\draw [thick] (-\r,0) -- (\r,0) (60:\r) -- (0.25,0) (-60:\r) -- (0.25,0) (120:\r) -- (-0.25,0) (-120:\r) -- (-0.25,0);
\end{tikzpicture}
=-\frac{\mathcal{N}_{d}^6}{144}
&\Big( C_{1,2,3,4,5,6}\pA_{1,2,3,4,5,6}+C_{1,3,2,4,5,6}\pA_{1,3,2,4,5,6}\nonumber\\[-1.2em]
&+C_{1,2,3,5,4,6}\pA_{1,2,3,5,4,6}+C_{1,3,2,5,4,6}\pA_{1,3,2,5,4,6}\Big)\,,
\end{align}
where $C$'s are DDM basis color factors. In the on-shell limit $\Delta=d$, we have
\begin{align}
    \pA_{1,2,3,4,5,6}&=\left[\frac{9}{D^{2}_{123}}D_{13}^2D_{46}^2-3D_{13}^2-3D_{46}^2+D_{123}^2\right]D_{d,d,d,d,d,d}\,.
\end{align}
The full color-ordered six-point correlator is given by
\begin{align}\label{eq:A6NLSM}
    A(1,2,3,4,5,6)&=-\frac{\mathcal{N}_d^6}{96}
    \Bigg[\frac{3}{D^2_{123}}D_{13}^2D_{46}^2-D^2_{135}+\text{cyclic}(1,2,3,4,5,6)\Bigg]D_{d,d,d,d,d,d}\,.
\end{align}
The derivative operators in each term commute. One can then check that the six-point fundamental BCJ relations does hold, for example,
\begin{align}
    0&=D_{12}^2A(1,2,3,4,5,6)+(D_{12}^2+D_{23}^2)A(1,3,2,4,5,6)\\
    &\quad+(D_{12}^2+D_{23}^2+D_{24}^2)A(1,3,4,2,5,6)+(D_{12}^2+D_{23}^2+D_{24}^2+D_{25}^2)A(1,3,4,5,2,6)\,.\nonumber
\end{align}
The AdS BCJ relations can in principle be generalized to higher Kaluza-Klein modes with $\Delta\neq d$, similar to eq.~\eqref{eq:BCJ4NLSM} at four points. We leave this study to a future work.

We have also constructed the six-point NLSM differential numerators that manifestly obey the Jacobi relations~\eqref{eq:kinematicnumerator}. As at four points, one can again start 
from the flat space DDM-basis numerators~\cite{Cheung:2017yef,Du:2016tbc,Du:2017kpo,
Carrasco:2016ygv} and make the replacement  $s_{I}\rightarrow D_{I}^{2}$. For example, a possible choice is
\begin{align}
    \hat{N}_{1,2,3,4,5,6}=-\frac{\mathcal{N}_d^6}{16}D_{12}^2(D_{13}^2+D_{23}^2)
    D_{56}^2(D_{45}^2+D_{46}^2)\, ;
\end{align}
the other DDM-basis numerators are obtained by relabeling. One can of course further use 
the analog of generalized gauge freedom~\eqref{eq:deltaN} to make this numerator more symmetric. Remarkably, just as the four-point case, these numerators yield the partial correlator~\eqref{eq:A6NLSM} despite the non-commutative nature of its various building blocks. To see an explicit example, consider the following numerator constructed using 
the procedure~\eqref{eq:kinematicnumerator},
\begin{align}\label{eq:NLSMn2}
    \hat{N}\Bigg(\begin{tikzpicture}[baseline={([yshift=-.5ex]current bounding box.center)},every node/.style={font=\scriptsize}]
    \draw (0,0) node[left=0]{$1$} -- (2,0) node[right=0]{$6$};
    \draw (0.5,0) -- ++ (0,0.25);
    \draw (0.5,0.25) -- ++(135:0.5) node[above=0]{$2$};
    \draw (0.5,0.25) -- ++(45:0.5) node[above=0]{$3$};
    \draw (1,0) -- ++ (0,0.25) node[above=0]{$4$};
    \draw (1.5,0) -- ++ (0,0.25) node[above=0]{$5$};
    \end{tikzpicture}\Bigg|1,2,3,4,5,6\Bigg)&=\hat{N}_{1,2,3,4,5,6}-\hat{N}_{1,3,2,4,5,6}\\[-1em]
    &=-\frac{\mathcal{N}_{d}^6}{16}\Big[D_{12}^{2}(D_{13}^{2}+D_{23}^{2})-D_{13}^{2}(D_{12}^{2}+D_{23}^{2})\Big]\nonumber\\
    &\quad\times D_{56}^{2}(D_{45}^{2}+D_{46}^{2}) \nonumber \\
    &=-\frac{\mathcal{N}_d^6}{16}D_{23}^{2}D_{56}^{2}(D_{12}^{2}-D_{13}^{2})(D_{45}^{2}+D_{46}^{2})+\ldots\,,\nonumber
\end{align}
where the ellipsis stands for operators having the commutator $[D_{ij}^2,D_{ik}^2]$ 
at the right-most position, for example, $D_{56}^{2}(D_{45}^{2}+D_{46}^{2})[D_{12}^{2},D_{13}^{2}]$. As discussed 
in eq.~\eqref{vanishing}, such commutators annihilate the six-point $D$-function $D_{d,d,d,d,d,d}$. 
Note that the $D_{23}^{2}D_{56}^{2}$ factor in the left-most position in the last line of eq.~\eqref{eq:NLSMn2} cancels two propagators in the associated trivalent diagram, as expected. It is remarkable that the non-commutativity of the $D_{I}^{2}$ in $\hat{N}$ 
merely leads to commutators at the right-most position of numerator factors and 
annihilate the $D$-function, thus effectively dropping out of the six-point correlator. It would be interesting to confirm that this holds at higher points.

%%%%%%%%%%%%%%%%%%%%%%%%%%%%%%%%%%%%%%%%%%%%%%%%%%%%%

\section{YM in AdS}\label{sec:YMsec}

In this section, we investigate YM theory in AdS space.  Its three- and four-point functions have been discussed in various contexts, both as tests of the AdS/CFT 
correspondence \cite{Freedman:1998tz, Chalmers:1998xr} and as illustrations of the embedding and Mellin space techniques \cite{Paulos:2011ie,Penedones:2010ue,
Kharel:2013mka, Costa:2011mg,Costa:2014kfa}. 
Here we obtain their explicit position-space representation and verify that the four-point correlator satisfies the AdS BCJ relations conjectured in section~\ref{AdS_AmplitudesRelationsGeneral}.  
We also construct differential representations of the off-shell three-point YM correlator, recovering the results of ref. \cite{Mizeraetal} when $\Delta_{i}=d-1$. 

\subsection{The three-point correlator}
\label{sec:YM3pt}

To set the stage for the calculation of the four-point correlator, we begin by reviewing and extending existing constructions  of the position-space three-point correlator for 
YM theory in AdS space. We will give give an explicit position-space representation as well as a differential representation for this correlator. The former will be useful for 
the double copy discussion in section~\ref{towardsadsdoublecopy}.

We computed the three-point YM AdS boundary correlator using the Feynman rules discussed in section \ref{AdSrulesEmbeddingSpace}. 
The structure of the vertex together with the property $E_\Delta^{MA}X_A = 0$ of the vector bulk-boundary propagator  imply that the embedding space projector 
$G^{AB}$ defined in eq.~\eqref{projector} can simply be replaced with $\eta^{AB}$. We will compute the correlator for arbitrary weight $\Delta_{i}$, as 
this form will be useful for the four-point calculation in the next section. The three-point correlator is 
\begin{align}
&\mathcal{A}^{a_1a_2a_3}_{\Delta_1\Delta_2\Delta_3}=-f^{a_{1}a_{2}a_{3}}Z_{1,M_{1}}Z_{2,M_{2}}Z_{3,M_{3}}\int_{\ads}dX \\
&\times \left[
E_{\Delta_1}^{M_{1}A_{1}}(P_{1},X)\eta_{A_{2}A_{3}}\left (\partial_{A_{1}}E_{\Delta_2}^{M_{2}A_{2}}(P_{2},X)E_{\Delta_3}^{M_{3}A_{3}}(P_{3},X) -(2\leftrightarrow 3)\right )+\textrm{cyclic}(1,2,3)
\right].\nonumber
\end{align}
Using the expression for $E_\Delta^{M,A}$ in eq.~\eqref{spin1btobound},  $[{\cal D}^{M_{i}A_{i}}_{\Delta_{i}}, \partial_B]=0$, and
\begin{equation}
\partial_{A}E_{\Delta_i}(P_i,X)=-\frac{\Delta_i P_{i,A}}{(P_i\cdot X)}E_{\Delta_i}(P_i,X)\equiv K_{i,A}E_{\Delta}(P_i,X) \ .    
\end{equation}
The correlator can be organized as
\begin{align}\label{eq:splitym3intopre}
\mathcal{A}^{a_1a_2a_3}_{\Delta_1\Delta_2\Delta_3}&=-f^{a_{1}a_{2}a_{3}} \left ( \prod_{i=1}^{3} Z_{i,M_{i}}{\cal D}^{M_{i}A_{i}}_{\Delta_{i}}\right ) 
\pP_{A_{1} A_{2} A_{3}}^{\Delta_1\Delta_1\Delta_3}(P_1, P_2, P_3)\,, \\
\pP_{A_{1} A_{2} A_{3}}^{\Delta_1\Delta_1\Delta_3}(P_1, P_2, P_3)&=\int_{\ads}dX \big[ \eta_{A_{2},A_{3}}(K_{2}-K_{3})_{A_1} +\textrm{cyclic}(1,2,3) \big]\prod_{i=1}^{3}\frac{\mathcal{N}_{\Delta_{i},1}}{(-2P_{i}\cdot X)^{\Delta_{i}}} \ .
\nonumber
\end{align} 
It is straightforward to recognize the remaining bulk integrals as three-point $D$-functions. See appendix~\ref{contactdiagramsinads} and ref.~\cite{DHoker:1999kzh} for general definitions and 
properties of $D$-functions. Unlike their four-point counterparts that will appear in the next section, embedding space isometries (or, equivalently, AdS isometries) completely fixes their dependence on the boundary points, leaving only the overall numerical factor to be determined. 

Accounting for the bulk point dependence in the vectors $K_{i,{A_j}}$, the tensor $P_{A_{1} A_{2} A_{3}}$ in eq.~\eqref{eq:splitym3intopre} evaluates to
\begingroup
\allowdisplaybreaks
\begin{align}\label{eq:offshellfunction}
\pP_{A_{1}A_{2}A_{3}}^{\Delta_1\Delta_1\Delta_3}(P_1, P_2, P_3)&= 2\Bigg[\prod_{i=1}^3\mathcal{N}_{\Delta_i,1}\Bigg]\eta_{A_{2},A_{3}}
\big( \Delta_2 P_{2,A_{1}} D_{\Delta_{1}, \Delta_{2}+1,\Delta_{3}} - \Delta_3 P_{3,A_{1}} D_{\Delta_{1},\Delta_{2},\Delta_{3}+1}  \big)\nonumber\\
&\quad +\textrm{cyclic}(1,2,3) \nonumber\\
&=\pi^{d/2}\Bigg[\prod_{i=1}^3\frac{\mathcal{N}_{\Delta_i,1}}{\Gamma(\Delta_i)}\Bigg]\Bigg[\eta_{A_2A_3}\Bigg(\frac{P_{2,A_1}P_{13}}{\delta_{13}-\frac{1}{2}}-\frac{P_{3,A_1}P_{12}}{\delta_{12}-\frac{1}{2}}\Bigg)+\text{cyclic}(1,2,3)\Bigg]\nonumber\\
&\quad\times\Gamma\!\left(\frac{\Delta_{1}+\Delta_2+\Delta_3-d+1}{2}\right)\text{PT}(1_{\Delta_1},2_{\Delta_2},3_{\Delta_3})\,,
\end{align}
\endgroup
where $\delta_{ij}$ is defined as
\begin{align}
    \delta_{ij}=\frac{\Delta_i+\Delta_j-\Delta_k}{2} 
\end{align}
for $\{i,j,k\}$ being a permutation of $\{1,2,3\}$. We also define for convenience the ``Parke-Taylor factor'' as
\begin{align}
    \text{PT}(1_{\Delta_1},2_{\Delta_2},3_{\Delta_3})=\frac{\Gamma(\delta_{12}+\frac{1}{2})\Gamma(\delta_{23}+\frac{1}{2})\Gamma(\delta_{13}+\frac{1}{2})}{P_{12}^{\delta_{12}+\frac{1}{2}}P_{23}^{\delta_{23}+\frac{1}{2}}P_{13}^{\delta_{13}+\frac{1}{2}}}\,.
\end{align}
One may verify that the resulting three-point correlator is both transverse and obeys the conformal Ward identity,
\begin{equation}
\mathcal{A}^{a_1a_2a_3}_{\Delta_1\Delta_2\Delta_3}\Big|_{Z_{i}\rightarrow P_{i}}=0 \,, \qquad \sum_{i=1}^{3} D_{i}^{AB}\mathcal{A}^{a_1a_2a_3}_{\Delta_1\Delta_2\Delta_3}=0    \ .
\end{equation}
The former relation may be understood as a consequence of the manifest transversality of the bulk-boundary vector field propagator while the latter implies that 
the formalism manifestly preserves conformal invariance and together they imply that the three-point function can be pulled back from the embedding space to AdS \cite{Paulos:2011ie,Costa:2011mg}. 

The correlator does not obey the current conservation for generic $\Delta_{i}$. This is, of course, to be expected as boundary current conservation is a reflection of 
a bulk gauge symmetry for the vector fields, which fixes $\Delta=d-1$ for spin-1 fields. Other ``massive'' vector fields may be interpreted as higher Kaluza-Klein modes and, while corresponding to BPS currents in a supersymmetric holographic framework, 
do not exhibit gauge invariance.

The on-shell correlator follows from eqs.~\eqref{eq:offshellfunction} and \eqref{eq:splitym3intopre} with $\Delta_{i}=d-1$. It can be put in a compact form in terms of the $V_{i,jk}$ and $H_{ij}$ functions introduced in ref.~\cite{Costa:2011mg}:
\begin{align}
    V_{i,jk}&=\frac{(P_j\cdot Z_i) (P_i\cdot P_k)-(P_k\cdot Z_i)(P_i\cdot P_j)}{P_j\cdot P_k}\,,\nonumber\\
    H_{ij}&=-2\big[(Z_i\cdot Z_j)(P_i\cdot P_j)-(Z_i\cdot P_j)(Z_j\cdot P_i)\big] \ .
    \label{VHdefinitions}
\end{align}
With the notation $V_1\equiv V_{1,23}$, $V_2\equiv V_{2,31}$ and $V_3\equiv V_{3,12}$, the full YM AdS boundary correlator becomes 
\begin{align}
\mathcal{A}^{a_{1}a_{2}a_{3}}_3&=-f^{a_1a_2a_3}\frac{d\, \Gamma(d-2)}{8\pi^d(d-2)}\frac{N_3}{(P_{12}P_{23}P_{13})^{d/2}} \ ,
\end{align}
where
\begin{align}\label{eq:N3YM}
    N_3 &= (4\Lambda_1-V_1V_2V_3)-\frac{6}{d}\Lambda_1
    \\
    \Lambda_1&=V_1V_2V_3+\frac{1}{2}(V_1H_{23}+\text{cyclic})
\end{align}
%
%and $\Lambda_1=V_1V_2V_3+\frac{1}{2}(V_1H_{23}+\text{cyclic})$.  
This form reproduces the result of ref.~\cite{Zhiboedov:2012bm}. In section \ref{towardsadsdoublecopy} 
we will use this form of the correlator and the analogous one corresponding to massive vectors.

In addition to the position space representation, we also construct a differential representation of the off-shell three-point correlator. Its existence is a nontrivial indication for our conjecture that on-shell YM correlators can be written in the form of eq.~(\ref{eq:conjecturedadsBCJrep}). 
With the definitions
\begin{align}\label{calE}
    \mathcal{E}_i^{AB}=P_i^A Z_i^B-P_i^B Z_i^A\,,
    %~~\text{and}~~
    %\delta_{ij}=\frac{1}{2}(\Delta_{i}+\Delta_{j}-\Delta_{k})\; ,\quad i\neq j \neq k     \ ,
\end{align}
the differential form of $\mathcal{A}^{a_1a_2a_3}_{\Delta_1\Delta_2\Delta_3}$ is
\begin{subequations}\label{offshelldifferentiacorrector}
\begin{align}
\mathcal{A}^{a_1a_2a_3}_{\Delta_1\Delta_2\Delta_3}&=\frac{f^{a_{1} a_{2} a_{3}}\Gamma\left(\frac{\Delta_{1}+\Delta_{2}+\Delta_{3}-d+1}{2}\right )}{16\pi^{d}\prod_{i=1}^3\big[\Gamma(\Delta_{i}{-}\frac{d}{2}{+}1)(\Delta_{i}{-}1)\big]}\hat{A}_{\Delta_1\Delta_2\Delta_3}\text{PT}(1_{\Delta_1},2_{\Delta_2},3_{\Delta_3})\,,\\
%\prod_{1\leqslant i<j}^3\frac{\Gamma(\delta_{ij}+\frac{1}{2})}{P_{ij}^{\delta_{ij}+\frac{1}{2}}}    
\hat{A}_{\Delta_1\Delta_2\Delta_3}&=\Big[(2\delta_{12}-1)(\mathcal{E}_{1}\cdot \mathcal{E}_{2})(\mathcal{E}_{3}\cdot D_{1})+2(\Delta_1^2-2\Delta_1\Delta_2+2\Delta_1-1)\text{Tr}(\mathcal{E}_1\mathcal{E}_2\mathcal{E}_3) \nonumber\\
&\quad +\text{cyclic}(1,2,3)\Big]\,,
\end{align}
\end{subequations}
where the dot products are defined in the sense of eq.~\eqref{conformalgener}. 
In the massless limit $\Delta_i = d-1$, the differential representation becomes
\begin{subequations}\label{onshell}
\begin{align}
    \mathcal{A}^{a_1a_2a_3}_3&=f^{a_1a_2a_3}\frac{\Gamma(d-2)}{16\pi^d(d-2)^2} \hat{A}_3 \,\frac{1}{ P_{12}^{d/2}P_{23}^{d/2}P_{13}^{d/2}}\,,
    \\
    \hat{A}_3&=(d-2)\Big[(\mathcal{E}_1\cdot\mathcal{E}_2)(\mathcal{E}_3\cdot D_1)+\text{cyclic}(1,2,3)\Big]-6(d-2)^{2}\text{Tr}(\mathcal{E}_1\mathcal{E}_2\mathcal{E}_3) \ .
\end{align}
\end{subequations}
This reproduces the result of ref.~\cite{Mizeraetal}. As assumed in section~\ref{AdS_AmplitudesRelationsGeneral}, this expression has uniform scaling 
dimension $-1$ for each external state. 

We note that the factor ${\cal E}$ defined in eq.~\eqref{calE} may be identified with the numerator of the bulk-boundary propagator for the linearized vector field strength; it is curious that, unlike in flat space, it is natural to organize the three-point YM correlator in terms of this tensor. We moreover note that the contribution of a $\Tr[F^3]$ interaction to the AdS boundary correlator involves the same kinematic terms as the YM expression, just with different numeric coefficients. We will return to this observation in section~\ref{diffdoublecopy}.

Notably, the off-shell differential correlator in eq.~(\ref{offshelldifferentiacorrector}) depends explicitly on the conformal weight $\Delta_{i}$ of external states, in sharp contrast to the differential form of the NLSM four-point correlator. In particular, it signals that certain manipulations used in the construction of the six- and possibly higher-point NLSM correlators may not have a direct counterpart in AdS YM calculations. For example, one could have derived the results in section~\ref{6pcomputation} using the split representation and how the differential represention of the NLSM four-point correlator is unchanged off-shell. This computation does not generalize to YM since the YM differential representation is not independent of $\Delta_{i}$. Instead, we must directly compute the correlator as a polynomial in $P_{i}$, $Z_{i}$, and $D$-functions.

%%%%%%%%%%%%%%%%%%%%%%%%%%%

\subsection{The four-point correlator}

We now describe a direct evaluation of the four-point on-shell YM correlator and verify that it satisfies the BCJ relations discussed in section~\ref{sec:4pBCJrep}. 
We follow the computation in ref.~\cite{Kharel:2013mka} and extend it to obtain an explicit polynomial of boundary coordinates $P_i$, polarization vectors $Z_{i}$ and 
$D$-functions \cite{DHoker:1999kzh}. There are two topologies of diagrams that contribute -- the exchange graphs and the contact diagram -- and the color-dressed 
correlator has the general form 
\begin{align}\label{eq:colordressedcorrelator}
\mathcal{A}_{4}^{a_{1}a_{2}a_{3}a_{4}}&=\mathcal{A}_{\textrm{contact}}^{a_{1}a_{2}a_{3}a_{4}}+\mathcal{A}_{s}^{a_{1}a_{2}a_{3}a_{4}}+\mathcal{A}_{t}^{a_{1}a_{2}a_{3}a_{4}}+\mathcal{A}_{u}^{a_{1}a_{2}a_{3}a_{4}}  \\
&=\mathcal{A}_{\textrm{contact}}^{a_{1}a_{2}a_{3}a_{4}}+\mathcal{A}_{s}^{a_{1}a_{2}a_{3}a_{4}}+\big(\mathcal{A}_{s}^{a_{1}a_{2}a_{3}a_{4}}\big|_{1\rightarrow 2\rightarrow 3\rightarrow 1}\big)+\big(\mathcal{A}_{s}^{a_{1}a_{2}a_{3}a_{4}}\big|_{1\rightarrow 3\rightarrow 2\rightarrow 1}\big) \,,\nonumber
\end{align}
where on the second line we used the symmetry properties of Witten diagrams.

We start with the contribution from the four-point contact diagram, which can be read-off 
from the four-field term in YM Lagrangian (\ref{eq:YMlagrangian}),
\begin{align}
\mathcal{A}^{a_{1}a_{2}a_{3}a_{4}}_{\textrm{contact}} &= \begin{tikzpicture}[baseline={([yshift=-.5ex]current bounding box.center)},every node/.style={font=\scriptsize}]\pgfmathsetmacro{\r}{0.75}
\draw [] (0,0) circle (\r cm);
\tikzset{decoration={snake,amplitude=.4mm,segment length=1.5mm,post length=0mm,pre length=0mm}}
\filldraw (45:\r) circle (1pt) node[above=0pt]{$3$};
\filldraw (135:\r) circle (1pt) node[above=0pt]{$2$};
\filldraw (225:\r) circle (1pt) node[below=0pt]{$1$};
\filldraw (-45:\r) circle (1pt) node[below=0pt]{$4$};
\filldraw (0,0) circle (1pt);
\draw [thick,decorate] (45:\r) -- (0,0);
\draw [thick,decorate] (-45:\r) -- (0,0);
\draw [thick,decorate] (135:\r) -- (-0,0);
\draw [thick,decorate] (-135:\r) -- (-0,0);
\end{tikzpicture}=\int_{\ads} dX\, \mathcal{I}^{a_{1}a_{2}a_{3}a_{4}}_{A_{1}A_{2}A_{3}A_{4}} \Bigg[ \prod_{j=1}^4Z_{i,M_i}E_{d-1}^{M_{i}A_{i}} \Bigg]  \,, \\
\mathcal{I}^{a_{1} a_{2} a_{3} a_{4}}_{A_{1} A_{2} A_{3} A_{4}}&=g^2 f^{a_{1}a_{2}x}f^{a_{3}a_{4}x}(\eta_{A_{1}A_{3}}\eta_{A_{2}A_{4}}-\eta_{A_{1}A_{4}}\eta_{A_{2}A_{3}})+\textrm{cyclic}(2,3,4) \,.
\end{align}
\noindent Using the expression for the vector bulk-boundary propagator $E_\Delta^{M_{1},A_{1}}$ in eq.~\eqref{spin1btobound} and the definition 
of the $D$-function in eq.~\eqref{orgdef}, it is straightforward to obtain:
\begin{align}
\mathcal{A}^{a_{1}a_{2}a_{3}a_{4}}_{\textrm{contact}}&=g^{2}\Bigg [ \frac{(d-1)\Gamma(d-1)}{2\pi^{d/2}(d-2)\Gamma(d/2)}\Bigg ]^{4} \mathcal{I}^{a_{1} a_{2} a_{3} a_{4}}_{A_{1} A_{2} A_{3} A_{4}}\Bigg[\prod_{i=1}^4 {Z_{i,M_i}\cal D}_{d-1}^{M_{i}A_{i}}\Bigg] D_{d-1,d-1,d-1,d-1}  \ .     
\end{align}
Acting with the derivatives in ${\cal D}_{d-1}^{M_{i},A_{i}}$ generates a significant number of terms, which can be expressed 
in terms of $D$-functions with shifted indices using the identities in appendix~\ref{contactdiagramsinads}.

We now turn to the evaluation of the $s$-channel exchange diagram.
%The main points of the calculation are as follows. 
We use the split representation of the the bulk-bulk propagator \eqref{spin1bulk} to write the exchange graph as a product of two partly off-shell three-point correlators integrated over a 
boundary point $Q$  and over the dimension/mass of the field corresponding to that point. The three-point correlators are written in Mellin space; this makes the integral over the 
boundary point straightforward and converts the product of three-point correlators to a Mellin-space four-point correlator. After the integral over the dimension of the intermediate 
field is evaluated, an inverse Mellin transform yields the desired position-space correlator. 
Although the computation strategy may appear somewhat convoluted compared to direct integration in the bulk points, it ultimately allows us to write the four-point correlator 
as an explicit polynomial of $P_{i}$, $Z_{i}$, and $D$-functions. Furthermore, the above computation strategy can be systematically generalized to $n$-point correlators at tree 
level~\cite{Kharel:2013mka,Paulos:2011ie}.\footnote{As we will see, the main difficulty in going to higher order is explicitly evaluating the contour integrals that appear due to using the split representation. For example, at four-points, the only non-trivial integral that appears is eq.~\eqref{eq:resultintegral}. However, one can show that the $c$-contours that appear are always equivalent to the $c$-contour integrals that appear in evaluating scalar correlators. Such scalar correlators in AdS are trivial to calculate using Mellin space Feynman rules \cite{Fitzpatrick:2011ia}. Therefore, although technically more challenging than flat space, one can algorithmically calculate tree level YM correlators in AdS in terms of $Z_{i}$, $P_{i}$ and $D$-functions at $n$-point without evaluating any integrals.}

Proceeding to the actual computation and using eq.~\reef{bulktoboundYMN}, the $s$-channel contribution to the  correlator written in terms of the pre-correlator is
\begin{align}\label{eq:pre4pointamp}
\mathcal{A}^{a_{1}a_{2}a_{3}a_{4}}_{s}=\begin{tikzpicture}[baseline={([yshift=-.5ex]current bounding box.center)},every node/.style={font=\scriptsize}]\pgfmathsetmacro{\r}{0.75}
\draw [] (0,0) circle (\r cm);
\tikzset{decoration={snake,amplitude=.4mm,segment length=1.5mm,post length=0mm,pre length=0mm}}
\filldraw (45:\r) circle (1pt) node[above=0pt]{$3$};
\filldraw (135:\r) circle (1pt) node[above=0pt]{$2$};
\filldraw (-135:\r) circle (1pt) node[below=0pt]{$1$};
\filldraw (-45:\r) circle (1pt) node[below=0pt]{$4$};
\filldraw (-0.25,0) circle (1pt) (0.25,0) circle (1pt);
\draw [thick,decorate] (-0.25,0) -- (0.25,0);
\draw [thick,decorate] (45:\r) -- (0.25,0);
\draw [thick,decorate] (-45:\r) -- (0.25,0);
\draw [thick,decorate] (135:\r) -- (-0.25,0);
\draw [thick,decorate] (-135:\r) -- (-0.25,0);
\end{tikzpicture}=g^{2}f^{a_{1}a_{2}x}f^{a_{3}a_{4}x}
\Bigg[\prod_{i=1}^4 Z_{i,M_i}\pD_{d-1}^{M_{i}A_{i}} \Bigg] \pP_{A_{1}A_{2}A_{3}A_{4}}^{s} \ ,
\end{align}
where $\pP_{A_{1}A_{2}A_{3}A_{4}}$ is an integral over the locations of the two three-point vertices. The split representation of the massless spin-1 propagator
in eq.~(\ref{bulktoboundYMN}) expresses it as an integral of the product of two off-shell three-point pre-correlators in eq.~(\ref{eq:offshellfunction}):
\begin{align}\label{eq:splitrepexp}
&\pP_{A_{1}A_{2}A_{3}A_{4}}^{s}=\int_{-i\infty}^{i\infty}\frac{dc}{2\pi i}\frac{-2c^2}{c^{2}-(d/2-1)^{2}} \\
&\quad
\times 
\int_{\partial\textrm{AdS}}dQ \; \eta_{NM}
\Big[\pD^{NA_{Q}}_{d/2+c}\pP_{A_{1}A_{2}A_{Q}}^{d-1\, d-1\, d/2+c}(P_1, P_2, Q)\Big] 
\Big[\pD^{MB_{Q}}_{d/2-c}\pP_{A_{3}A_{4}B_{Q}}^{d-1\, d-1\, d/2-c}(P_3, P_4, Q)\Big]  \, .\nonumber
\end{align}
The derivatives of the three-point pre-correlators are obtained by simply evaluating the derivatives with respect to an off-shell leg in eq.~\eqref{eq:offshellfunction}, 
\begin{align}\label{eq:offshellampt}
   \pD^{NA_{Q}}_{d/2+c}&\pP_{A_{1}A_{2}A_{Q}}^{d-1\, d-1\, d/2+c}=\frac{\left(\frac{d-1}{d-2}\right)^2\Gamma\left(\frac{3d/2+c-1}{2}\right)}{4\pi^{d}(d/2+c-1)\Gamma(d/2)^2\Gamma(1+c)}\text{PT}(1_{d-1},2_{d-1},Q_{d/2+c}) \nonumber\\
& \times \Big [(\eta^{A_{1}A_{2}}P_{1}^{N}-2\eta^{A_{1}N}P_{1}^{A_{2}})P_{2Q}
-(\eta^{A_{1}A_{2}}P_{2}^{N}-2\eta^{A_{2}N}P_{2}^{A_{1}})P_{1Q}\Big ] +(\ldots) \ ,    
\end{align}
where $P_{i Q}=-2P_i\cdot Q$ and similarly for $ \pD^{NA_{Q}}_{d/2-c}\pP_{A_{3},A_{4},A_{Q}}^{d-1\, d-1\, d/2-c}$. The terms in $(\ldots)$ will vanish when the leg $P_1$ and $P_2$ are taken on-shell. More specifically, they are removed as the result of the identity~\cite{Kharel:2013mka}
\begin{equation}
\pD^{MA}_\Delta\frac{\partial}{\partial P_{A}}\mathcal{F}_{\Delta-1}(P)=0\,,  
\end{equation}
where $\mathcal{F}_{\Delta-1}(P)$ is any function of weight $\Delta-1$ in $P$. Thus in the following we will neglect the $(\ldots)$ terms in eq.~\eqref{eq:offshellampt}.

The two terms on the second line of eq.~\eqref{eq:offshellampt} are related by the interchange of labels $1$ and $2$; the terms in the analogous factor in 
$ \pD^{NA_{Q}}_{d/2-c}\pP_{A_{3},A_{4},A_{Q}}^{d-1\,d-1\,d/2-c}$ are related by the interchange of labels $3$ and $4$. Thus, replacing these expressions in eq.~\eqref{eq:splitrepexp} yields four terms, three of which can be obtained from the fourth through the transformations $1\leftrightarrow 2$, $3\leftrightarrow 4$ and $(1,3)\leftrightarrow (2, 4)$.
It is not difficult to find that
\begin{align}\label{Qintegral}
&\int_{\partial\ads} dQ \, \eta_{NM}\pD^{NA_{Q}}_{d/2+c}P_{A_{1}A_{2}A_{Q}}^{d-1\, d-1\, d/2+c} \pD^{MB_{Q}}_{d/2-c}P_{A_{3}A_{4}B_{Q}}^{d-1\,d-1\, d/2-c}=
\frac{\Gamma\Big( \frac{3d/2+c-1}{2}\Big)^{2} \Gamma\Big( \frac{3d/2-c-1}{2}\Big )^{2}}{64\pi^{2d}\Gamma(d/2)^4\Gamma(1+c)\Gamma(1-c)}\nonumber\\
&{}~~\times\Big(\frac{d-1}{d-2}\Big)^4\Bigg[\frac{P_{13}\,\mathcal{K}^{(P_1P_2P_3P_4)}_{A_1A_2A_3A_4}}{ 
P_{12}^{\frac{3d/2-c-1}{2}}P_{34}^{\frac{3d/2+c-1}{2}}} I\Big({}^{\,P_1\; P_2\; P_3\; P_4}_{\;\tilde{\Delta}_1\,\tilde{\Delta}_2\,\tilde{\Delta}_3\tilde{\Delta}_4}\Big)
-\frac{P_{23}\,\mathcal{K}^{(P_2P_1P_3P_4)}_{A_2A_1A_3A_4}}{ 
P_{12}^{\frac{3d/2-c-1}{2}}P_{34}^{\frac{3d/2+c-1}{2}}} I\Big({}^{\,P_1\; P_2\; P_3\; P_4}_{\;\tilde{\Delta}_2\,\tilde{\Delta}_1\,\tilde{\Delta}_3\tilde{\Delta}_4}\Big) \nonumber\\
&{}\qquad
-\frac{P_{14}\,\mathcal{K}^{(P_1P_2P_4P_3)}_{A_1A_2A_4A_3}}{ 
P_{12}^{\frac{3d/2-c-1}{2}}P_{34}^{\frac{3d/2+c-1}{2}}} I\Big({}^{\,P_1\; P_2\; P_3\; P_4}_{\;\tilde{\Delta}_1\,\tilde{\Delta}_2\,\tilde{\Delta}_4\tilde{\Delta}_3}\Big)+\frac{P_{24}\,\mathcal{K}^{(P_2P_1P_4P_3)}_{A_2A_1A_4A_3}}{ 
P_{12}^{\frac{3d/2-c-1}{2}}P_{34}^{\frac{3d/2+c-1}{2}}} I\Big({}^{\,P_1\; P_2\; P_3\; P_4}_{\;\tilde{\Delta}_2\,\tilde{\Delta}_1\,\tilde{\Delta}_4\tilde{\Delta}_3}\Big)\Bigg] ,
\end{align}
where ${\cal K}$ and ${\tilde\Delta}_i$ are defined as
\begin{gather}
    \mathcal{K}_{A_{1}A_{2}A_{3}A_{4}}^{(P_1P_2P_3P_4)}=\frac{(\eta_{A_{1}A_{2}}P_{1}^{N}-2\delta^{N}_{A_{1}}P_{1,A_{2}})(\eta_{A_{3}A_{4}}P_{3,N}-2\eta_{A_{3}N}P_{3,A_{4}})}{P_{13}}\,, \\
\tilde{\Delta}_{1}=\frac{d/2{+}c{+}1}{2}\ , \quad \tilde{\Delta}_{2}=\frac{d/2{+}c{-}1}{2} \ , \quad  
\tilde{\Delta}_{3}=\frac{d/2{-}c{+}1}{2}\ , \quad \tilde{\Delta}_{4}=\frac{d/2{-}c{-}1}{2} \,.
\end{gather}
Importantly, $\tilde{\Delta}_i$ satisfy the relation $\sum_{i=1}^4\tilde{\Delta}_i=d$.
The function $I\Big({}^{\,P_1\; P_2\; P_3\; P_4}_{\;\tilde{\Delta}_1\,\tilde{\Delta}_2\,\tilde{\Delta}_3\tilde{\Delta}_4}\Big)$ is the result of converting a certain four-point contact integral over the boundary to its Mellin representation~\cite{Kharel:2013mka},
\begin{align}\label{eq:bdryInt}
I\Big({}^{\,P_1\; P_2\; P_3\; P_4}_{\;\tilde{\Delta}_1\,\tilde{\Delta}_2\,\tilde{\Delta}_3\tilde{\Delta}_4}\Big)&=\int_{\partial \textrm{AdS}}dQ \prod_{i=1}^{4}\frac{\Gamma(\tilde\Delta_{i})}{(-2P_{i}\cdot  Q)^{\tilde{\Delta}_{i}}}\qquad \big(\text{under the constraint $\textstyle\sum_{i=1}^4\tilde{\Delta}_i=d$}\big)\nonumber\\
&=\pi^{d/2}\int_{-i\infty}^{i\infty} \Bigg[\prod_{1\leqslant i<j}^{4}\frac{d\tilde{\delta}_{ij}}{2\pi i}\Gamma(\tilde{\delta}_{ij})P_{ij}^{-\tilde{\delta}_{ij}}\Bigg]\Bigg[ \prod_{k=1}^{4} \delta\Big(\tilde{\Delta}_{k}-\sum_{l=1,\,l\neq k}^4\tilde{\delta}_{lk}\Big)\Bigg ]\,,
\end{align}
where we also assume that the integration variable $\tilde{\delta}_{ij}$ is symmetric in its indices. We note that the four integrals entering eq.~\eqref{Qintegral} differ by 
interchange of $\tilde\Delta_i$ with fixed ordering of $P_i$. Thus, they are different even though $I\Big({}^{\,P_1\; P_2\; P_3\; P_4}_{\;\tilde{\Delta}_1\,\tilde{\Delta}_2\,\tilde{\Delta}_3\tilde{\Delta}_4}\Big)$ is invariant under the interchange of {\em pairs} $(P_i, {\tilde\Delta}_i)$.
%
\iffalse
We note that the permutation sum in eq.~\eqref{Qintegral} is taken over all the labels. For example, the permutation $1\leftrightarrow 2$ means that we should exchange $P_1 \leftrightarrow P_2$, $A_1\leftrightarrow A_2$ and $\tilde{\Delta}_1\leftrightarrow\tilde{\Delta}_2$. In general, the function $I_{\tilde{\Delta}_1\tilde{\Delta}_2\tilde{\Delta}_3\tilde{\Delta}_4}$ is not symmetric under such permutations, namely,  $I_{\tilde{\Delta}_1\tilde{\Delta}_2\tilde{\Delta}_3\tilde{\Delta}_4}\neq I_{\tilde{\Delta}_2\tilde{\Delta}_1\tilde{\Delta}_3\tilde{\Delta}_4}$ if $\tilde{\Delta}_1$ and $\tilde{\Delta}_2$ are different.
\fi

Now that we have converted the integral over the boundary point insertion into Mellin form, we can proceed and perform the contour integral over $c$. We start with the change of variables,
\begin{align}
\delta_{12}&=\tilde{\delta}_{12}+\frac{3d/2-c-1}{2}\,,  ~ & ~ \delta_{34}&=\tilde{\delta}_{34}+\frac{3d/2+c-1}{2}\,, \\
\delta_{13}&=\tilde{\delta}_{13}-1\,,   ~&~  \delta_{ij}&=\tilde{\delta}_{ij} \quad \textrm{for all others}\,, \nonumber
\end{align}
for the integral $I_{\tilde{\Delta}_1\tilde{\Delta}_2\tilde{\Delta}_3\tilde{\Delta}_4}$ given in eq.~\eqref{eq:bdryInt}, together with its images under the specified permutation maps for the other three terms in the sum of eq.~\eqref{Qintegral}, to align the constraints on 
the Mellin integration variables, which now become
\begin{equation}
\prod_{k=1}^{4}\delta\Big(\tilde{\Delta}_{k}-\sum_{l=1,\,,l\neq k}^{4}\tilde{\delta}_{lk}\Big) \rightarrow  \prod_{k=1}^{4}\delta\Big(d-1-\sum_{l=1,\,,l\neq k}^{4}\delta_{lk}\Big)     
\end{equation}
in all four terms in eq.~\eqref{Qintegral}. The pre-correlator then has the rather compact expression:
\begin{align}\label{eq:finalexpresiontoint}
\pP_{A_{1}A_{2}A_{3}A_{4}}^{s}&=-\frac{\left(\frac{d-1}{d-2}\right)^4}{32\pi^{3d/2}\Gamma(d/2)^4}\int_{-i\infty}^{i\infty} \Bigg[ \prod_{1\leqslant i<j}^{4}\frac{d\delta_{ij}}{2\pi i}\frac{\Gamma(\delta_{ij})}{P_{ij}^{\delta_{ij}}}\Bigg]  \Bigg[\prod_{k=1}^{4}\delta\Big(d{-}1{-}\sum_{l=1,l\neq k}^{4}\delta_{lk}\Big)\Bigg]  \\
& \quad
\times \int_{-i\infty}^{i\infty} \frac{dc}{2\pi i} S(\delta_{12},c) \Bigg[\mathcal{K}_{A_{1}A_{2}A_{3}A_{4}} \, \delta_{13}-\big(1\leftrightarrow 2\big)- \big(3\leftrightarrow 4\big) + \Bigg(\begin{array}{c}
1\leftrightarrow 2 \\
3\leftrightarrow 4
\end{array}\Bigg)\Bigg]\,,\nonumber
\end{align}
where the permutation map acts on $P_i$, $A_i$ and the indices of $\delta_{ij}$. For example, under the permutation $1\leftrightarrow 2$ we exchange $P_1\leftrightarrow P_2$, $A_1\leftrightarrow A_2$, $\delta_{13}\leftrightarrow\delta_{23}$ and $\delta_{14}\leftrightarrow\delta_{24}$. The entire $c$ dependence is contained in the function $S(\delta_{12},c)$,
\begin{equation}
S(\delta_{12},c)= \frac{l(\delta_{12},c)l(\delta_{12},-c)}{(d/2-1)^2-c^2}\Bigg|_{\Delta_{12}=\Delta_{34}=2d-1}  \, ,
\end{equation}
where $l(\delta_{12}, c)$ with generic $\Delta_{12}=\Delta_1+\Delta_2$ 
and $\Delta_{34}=\Delta_3+\Delta_4$ is given by
\begin{align}
    l(\delta_{12},c)=\frac{\Gamma\big(\delta_{12}-\frac{\Delta_{12}-c-d/2}{2}\big)\Gamma\big(\frac{\Delta_{12}+c-d/2}{2}\big)\Gamma(\frac{\Delta_{34}+c-d/2}{2})}{\Gamma(\delta_{12})\Gamma(c)}\,.
\end{align}
Choosing the contour such that nonphysical poles do not contribute, the $c$ integral in eq.~\eqref{eq:finalexpresiontoint} yields 
\begin{align}\label{eq:resultintegral}
\int^{i\infty}_{-i\infty}\frac{dc}{2\pi i}&\frac{l(\delta_{12},c)l(\delta_{12},-c)}{(d/2-1)^2-c^2} = \Gamma\left( \frac{\Delta_{12}+\Delta_{34}-d}{2}\right ) \sum_{l=1}^{m} \frac{\left (\frac{\sum_{i} \Delta_{i}-d}{2}\right )_{-l}(\delta_{12})_{-l}}{\left (\frac{ \Delta_{12}-d+1}{2}\right )_{1-l}\left (\frac{ \Delta_{12}-1}{2}\right )_{1-l}} \nonumber\\
&=\frac{2\Gamma\left(\frac{\Delta_{12}+\Delta_{34}-d}{2}\right){}_3F_{2}\left(1,\frac{3{-}\Delta_{12}}{2},\frac{d{+}1{-}\Delta_{12}}{2};2{-}\delta_{12},\frac{d{-}\Delta_{12}{-}\Delta_{34}{+}4}{2};1\right)}{(\delta_{12}-1)(\Delta_{12}+\Delta_{34}-d-2)}\,, 
\end{align}
where $m=\frac{1}{2}(\Delta_{12}-d+1)$ and $(a)_n$ is the Pochhammer symbol~\cite{DHoker:1999mqo,Penedones:2010ue}. Although the expression in the second line is derived assuming $m$ is a positive integer, it holds for more generic parameters as a result of analytic continuation.\footnote{To arrive at the right-hand side of eq.~\eqref{eq:resultintegral}, a specific choice of contour for the $c$ integral is required, which is the same one made in eq.~(133) of~\cite{Penedones:2010ue}.} 
For $d=4$, we find that %
\begin{align}
\label{eq:cintegralfinal}
\int_{-i\infty}^{i\infty} \frac{dc}{2\pi i}S(\delta_{12},c)\Bigg|_{d=4}=12\left[ \frac{\Gamma(\delta_{12}-2)}{3\Gamma(\delta_{12})}+\frac{\Gamma(\delta_{12}-1)}{2\Gamma(\delta_{12})}\right ]=\frac{4}{\delta_{12}-2}+\frac{2}{\delta_{12}-1}\ .
\end{align}
As we have evaluated the $c$ integral, we are left with the evaluation of the Mellin integrals in eq.~\eqref{eq:finalexpresiontoint}. They can be converted into $D$-functions using the identity 
\begingroup
\allowdisplaybreaks
\begin{align}\label{eq:mellinspace}
\textbf{M}^{-1}\Bigg[\prod_{1\leqslant i<j}^{4}\frac{\Gamma(\delta_{ij}+l_{ij})}{\Gamma(\delta_{ij})}\Bigg] &=\frac{2}{\pi^{d/2}}
\frac{\prod_{i=1}^{4}\Gamma(\tilde{\Delta}_{i})}{\Gamma\Big(\frac{\tilde\Sigma-d}{2}\Big)}\Bigg[ \prod_{1\leqslant i<j}^4 P_{ij}^{l_{ij}}\Bigg]D_{\tilde{\Delta}_{1} \tilde{\Delta}_{2} \tilde{\Delta}_{3} \tilde{\Delta}_{4}}  \ , \\
\tilde{\Delta}_{i}&=\Delta_{i}+\sum_{j=1,\,j\neq i}^{4}l_{ij} \,,\qquad \tilde\Sigma=\sum_{i=1}^{4}\tilde\Delta_i\,,\nonumber
\end{align}
\endgroup
which will be proven in appendix~\ref{sec:details4pcom}. Here $\textbf{M}^{-1}$ denotes the inverse Mellin transform,
\begin{equation}\label{derivationofInvM}
\textbf{M}^{-1}\Big[f(\delta_{ij})\Big]=\int_{-i\infty}^{i\infty} \Bigg[ \prod_{1\leqslant i<j}^{4}\frac{d\delta_{ij}}{2\pi i}\frac{\Gamma(\delta_{ij})}{P_{ij}^{\delta_{ij}}}\Bigg] \Bigg[ \prod_{k=1}^{4}\delta\Big(\Delta_{k}-\sum_{l=1}^{4}\delta_{lk} \Big)\Bigg]f(\delta_{ij})  \ .    
\end{equation}
Specialized to $d=4$, eqs.~\eqref{eq:cintegralfinal} and \eqref{eq:mellinspace} together can bring the pre-correlator in eq.~\eqref{eq:finalexpresiontoint} to a linear combination of $D$-functions weighted by polynomial of $P_{i}$, $Z_{i}$. An expression valid for generic boundary dimension $d$ can be obtained by using eq.~\eqref{eq:mellinspace} together with the sum
representation \cite{Penedones:2010ue} of eq.~\eqref{eq:resultintegral}. In the following, we focus on $d=4$. The pre-correlator is given by
\begin{align}\label{eq:finalresult}
    \pP_{A_{1}A_{2}A_{3}A_{4}}^{\rm s}\Big|_{d=4}=-\frac{243}{32\pi^8}&\left[\frac{P_{13}\mathcal{R}_{A_1A_2A_3A_4}}{P_{12}}D_{3,2,4,3}-\frac{P_{14}\mathcal{R}_{A_1A_2A_3A_4}^{(3\leftrightarrow 4)}}{P_{12}}D_{3,2,3,4}\right.\nonumber\\
    &\left.+\frac{P_{13}\mathcal{R}_{A_1A_2A_3A_4}}{P_{12}^2}D_{2,1,4,3}-\frac{P_{14}\mathcal{R}_{A_1A_2A_3A_4}^{(3\leftrightarrow 4)}}{P_{12}^2}D_{2,1,3,4}\right],
\end{align}
where
\begin{align}
    \mathcal{R}_{A_1A_2A_3A_4}=\mathcal{K}_{A_1A_2A_3A_4}+\Bigg(\begin{array}{c}
1\leftrightarrow 2 \\
3\leftrightarrow 4
\end{array}\Bigg)\,,\qquad \mathcal{R}_{A_1A_2A_3A_4}^{(3\leftrightarrow 4)}=\mathcal{R}_{A_1A_2A_3A_4}\Big|_{\substack{P_3\leftrightarrow P_4 \\ A_3\leftrightarrow A_4}}\,.
\end{align}
Finally, we apply the $\pD$-derivatives in eq.~\eqref{eq:pre4pointamp} and express the result in terms of $D$-functions by repeated use of the identity
\begin{align}
    \frac{\partial D_{\Delta_1,\Delta_2,\Delta_3,\Delta_4}}{\partial P_{1,A}}=\frac{4\Delta_1}{\sum_{i=1}^4\Delta_i-d}\Big(&\Delta_2P_2^A D_{\Delta_{1}+1,\Delta_2+1,\Delta_3,\Delta_4}+\Delta_3P_3^AD_{\Delta_1+1,\Delta_2,\Delta_3+1,\Delta_4}\nonumber\\
    &+\Delta_4P_4^AD_{\Delta_1+1,\Delta_2,\Delta_3,\Delta_4+1}\Big)\,.
\end{align}
It is then straightforward, albeit tedious, to find an explicit expression for the $s$-channel correlator $\mathcal{A}_{s}^{a_{1}a_{2}a_{3}a_{4}}$ as a linear combination of $D$-functions, from which the $t$- and $u$-channel correlators can subsequently be obtained 
by the relabelings given in eq.~\eqref{eq:colordressedcorrelator}.

The partial correlators can be extracted  from eq.~\eqref{eq:colordressedcorrelator} in the usual way, either by directly going to a trace basis or by 
using  the Jacobi identity 
\begin{equation}
f^{a_{1}a_{4}x}f^{a_{2}a_{3}x}
+f^{a_{1}a_{2}x}f^{a_{3}a_{4}x}+f^{a_{1}a_{3}x}f^{a_{4}a_{2}x} = 0 \ ,  
\end{equation}
to pass to the DDM basis,
\begin{equation}
\mathcal{A}_{4}^{a_{1}a_{2}a_{3}a_{4}}=f^{a_{1}a_{2}x}f^{a_{3}a_{4}x}A_4(1,2,3,4)+f^{a_{1}a_{3}x}f^{a_{2}a_{4}x}A_4(1,3,2,4) \ .  
\end{equation}
We are now in a position to verify that the AdS BCJ relation (\ref{eq:BCJ4pooint}), 
\begin{equation}\label{BCjrelations}
D_{12}^{2}A_4(1,2,3,4)=D_{13}^{2}A_4(1,3,2,4) \ ,    
\end{equation}
is satisfied. The conformal generators $D_{i}^{AB}$ are defined in eq.~(\ref{conformalgener}). While it is in principle possible, albeit tedious, to do so 
analytically through judicious use of the $D$-function identities in appendix~\ref{Dfunctionidentities}, we have verified eq.~\eqref{BCjrelations} numerically at $d=4$
at random kinematic points with very high precision. The part of the conformal generator that acts on the polarization vectors $Z_i$ is crucial for the AdS BCJ relations to hold.

The fact that the four-point AdS BCJ relation is satisfied 
%together with $D$-function identities 
suggests that it may be possible to put the four-point YM AdS boundary correlator in the form put forth in eq.~(\ref{eq:conjecturedadsBCJrep}). Similar to the three-point YM AdS boundary correlator, we expect that the four-point BCJ representation will match the flat space result up to 
possible additional terms that result from the non-commutativity of factors in the AdS kinematic numerators. Algorithms for efficiently computing such differential representations are left to future investigation. 

\section{Towards a bosonic double copy in AdS space}\label{towardsadsdoublecopy}

In this section, we discuss possible double copy procedures in AdS space. 
We first analyze a ``differential'' double copy that is analogous to the celestial double copy in flat space. 
The differential turns out to yield consistent AdS boundary correlators for $d=2$, in agreement with expectations based on the AdS$_3\times$S$_3$ ambitwistor 
string~\cite{Roehrig:2020kck}, but issues develop in higher dimensions even at three-points.
We then study the double copy in position space and find that the most naive construction holds for three-point correlators only in 
the limit of large AdS dimension,  thus recovering results of ref.~\cite{Mizeraetal}. 
Finally, we discuss limiting cases in which connections between AdS boundary correlators and flat space amplitudes should expose double-copy structures in momentum and Mellin space.

\subsection{A differential double copy} \label{diffdoublecopy}

In section~\ref{AdS_AmplitudesRelationsGeneral} we suggested that the NLSM and YM AdS
boundary correlators can be written as sums of differential operators acting on a 
single contact diagram, 
\begin{equation}\label{bcj2for}
\mathcal{A}=\sum_{\textrm{cubic }g}C(g|\alpha_{g}) \left ( \prod_{I\in g}\frac{1}{D_{I}^{2}} \right )  \hat{N}(g|\alpha_{g}) D_{d,d,d,\ldots}\ ,
\end{equation}
and that, as in flat space, color/kinematics duality identifies the algebraic 
properties of the color factor with those of the kinematic numerators ${\hat N}$
when acting on the contact diagram.
Given such a differential representation, the most natural attempt at an AdS double copy 
procedure is to simply replace the color factors, $C(g|\alpha_{g})$, with their associated 
kinematic numerators, $\hat{N}(g|\alpha_{g})$. 
However, direct counting of the conformal weight for each external state suggests that 
certain modifications are necessary.
Indeed, for YM theory, we assumed in section~\ref{bcj2for} and explicitly demonstrated in 
section~\ref{sec:YM3pt} that the conformal weight of the kinematic numerators $\hat{N}$ 
with respect to every external state is $-1$. Combining this with the $d$ conformal weight
of $D_{d,d,d,\ldots}$ for each of its external points implies that the action of two 
kinematic numerators leads to a $d-2$ overall conformal weight for each external
state of the putative differential double copy. Thus, in addition to replacing 
the color factors with kinematic numerators, to obtain the requisite conformal weight 
$\Delta=d$ it is necessary to also increase the conformal weight of each of the external 
legs of the contact diagram by two units. The full double copy procedure should then amount to the replacements
\begin{equation}\label{gravdoublecopy}
\begin{split}
C(g|\alpha_{g})&\rightarrow \hat{N}(g|\alpha_{g})\,, \\
D_{d,d,d,\ldots}&\rightarrow D_{d+2,d+2,d+2,\ldots}\,, 
\end{split}
\end{equation}
where $\hat{N}$ here might differ from the one in eq.~\eqref{bcj2for} by some operators that annihilate $D_{d,d,d,\ldots}$. Remarkably, the ambitwistor string construction of ref.~\cite{Roehrig:2020kck} strongly suggests that spin-2 AdS${}_{3}\times S^{3}$ boundary correlators can be derived by applying the substitution rules in eq.~(\ref{gravdoublecopy}). Specifically, one would apply eq.~(\ref{gravdoublecopy}) to the differential representation (\ref{bcj2for}) of correlators in a YM-Chern-Simons theory deformed by a specific linear combination of certain higher-dimension operators.
The generalization of this double copy procedure from $d=2$ to arbitrary 
$d$ turns out to be more subtle than one might naively expect.\footnote{There are  subtleties even in $d=2$ related to how gravitons do not obey eq. (\ref{onshellcondition}); the equation of motion for a free graviton 
in AdS is $(-\frac{1}{2}D_X^{2}+2)h^{AB}=0$, rather than $D_X^{2} h^{AB}=0$. This implies that the $1/D_{I}^{2}$ factors in eq. (\ref{bcj2for}) should also be shifted in order to interpret these factors as propagators in the associated Witten diagrams. However, the formulas of ref. \cite{Roehrig:2020kck} seem to suggest that this shift is not necessary.} 

To see this, it suffices to consider the differential double copy at three points. 
We derived the differential form of the three-point AdS YM correlator in 
section~\ref{sec:YM3pt}. The normalization of the bulk-boundary vector-field 
propagator in eq.~\eqref{spin1btobound} is singular for $d=2$; so
to have a smooth analytic continuation in dimension for the purpose of this discussion, 
we will change it by removing the offending factor of $(\Delta-1)^{-1}=(d-2)^{-1}$. 
We will also deform the YM theory with the operator $\textrm{Tr}[F^{3}]$ with an arbitrary (Wilson) coefficient $g_{F^3}$.

With these preparations and up to an overall constant which is finite for all positive 
values of $d$,  the differential form of the YM+$g_{F^3}\Tr[F^3]$ three-point correlator 
is 
\begin{align}
    \hat{N}_{3}^{g_{F^3}}&\propto\Big[1+6g_{F^3}(d-2)^2\Big]\Big[(\mathcal{E}_2\cdot\mathcal{E}_2)(\mathcal{E}_3\cdot D_1)+\text{cyclic}\Big]\nonumber\\
    &\quad +6(d-2)\Big[-1+2g_{F^3}(d-2)(d+2)\Big]\text{Tr}(\mathcal{E}_1\mathcal{E}_2\mathcal{E}_3)
\end{align}
where $\mathcal{E}$ is defined in eq.~\eqref{calE}. We note that our $\Tr[F^3]$ contribution is consistent with that in~\cite{Caron-Huot:2021kjy} at $d=3$.
The differential double-copy proposal then suggests that the 
corresponding AdS double-copy boundary correlator is 
\begin{equation}\label{doublecopygravity}
\mathcal{M}_{3}^{\textrm{DC}}\propto\hat{N}_{3}^{g_{F^{3}}}\hat{N}_{3}^{g_{F^{3}}'}D_{d+2,d+2,d+2} \ ,    
\end{equation}
with independent $g_{F^{3}}$ and $g_{F^{3}}'$ coefficients to allow for a general heterotic double copy~\cite{Chi:2021mio}.

In $d=2$, the double copy works straightforwardly. This is due to additional linear relations between $V_1V_2V_3$ and $\Lambda_1$ in eq.~\eqref{eq:N3YM}. Consequently, in $d=2$ we have
\begin{align}\label{eq:dcd=2}
    \hat{N}_3^{g_{F^3}}D_{2,2,2}\propto\frac{V_1V_2V_3}{P_{12}P_{23}P_{13}}\,,&
    & \hat{N}_{3}^{g_{F^{3}}}\hat{N}_{3}^{g_{F^{3}}'}D_{4,4,4}\propto \frac{(V_1V_2V_3)^2}{(P_{12}P_{23}P_{13})^2}\,.
\end{align}
It is also easy to check that both expressions in eq.~\eqref{eq:dcd=2} satisfy current conservation for $d=2$ and therefore can be interpreted as an AdS three-graviton correlator. This result is a non-trivial generalization of ref. \cite{Roehrig:2020kck}, which 
only studied YM-Chern-Simons theory in AdS deformed by a fixed linear combination of higher-dimension operators while here the Wilson coefficient $g_{F^3}$ is arbitrary. In fact, to give nonzero contribution at $d=2$, it needs to be proportional to ${(d-2)^{-2}}$. We see that, just as in flat space, the AdS${}_{3}$ double copy appears to be compatible 
with pure YM theory deformed by certain higher derivative operators, such as
$\textrm{Tr}[F^{3}]$, with arbitrary Wilson coefficients \cite{Broedel:2012rc}. 

For $d>2$, the current conservation of $\mathcal{M}_3^{\textrm{DC}}$ requires $g_{F^3}$ to take specific values. If we follow eq.~\eqref{doublecopygravity}, there are only two solutions,
\begin{align}
    & g_{F^3}=-\frac{1}{6(d-2)^2}\,, & & g_{F^3}'=\frac{d}{6(d-2)^2(3d-4)}\,; \nonumber\\
    & g_{F^3}=\frac{2d-3}{6(d-2)^2}\,, & & g_{F^3}'=-\frac{1}{6(d-2)^2}\,.
\end{align}
which impose that one of the $\hat{N}_{3}^{g_{F^{3}}}$ is proportional to $\Tr(\mathcal{E}_1\mathcal{E}_2\mathcal{E}_3)$. The resultant gravity correlator is a linear combination of the Einstein-Hilbert term and certain higher derivatives operators. Moreover, we can modify eq.~\eqref{doublecopygravity} to make it symmetric with respect to $g_{F^3}$ and $g_{F^3}'$, $\mathcal{M}_3^{\textrm{DC}}\propto(\hat{N}_{3}^{g_{F^{3}}}\hat{N}_{3}^{g_{F^{3}}'}+\hat{N}_{3}^{g_{F^{3}}'}\hat{N}_{3}^{g_{F^{3}}})D_{d+2,d+2,d+2}$.
Then the current conservation leads to a unique solution with $g_{F^3}=-\frac{1}{6(d-2)^2}$ and $g_{F^3}'=\frac{5d-6}{6(d-2)^2(3d-2)}$.

We have seen that to realize the AdS double-copy construction requires certain
generalizations of the flat space case. 
A possible approach to understanding it may be higher-dimensional 
generalizations of the ambitwistor string theory 
of ref.~\cite{Roehrig:2020kck}. Possible 
obstacles relate to the stringy realization of the massless spectrum in $\ads_{5}\times S^{5}$, see refs.~\cite{Gaberdiel:2021jrv,Gaberdiel:2021iil}. 
In flat space, the interplay between gauge invariance and color/kinematics duality guarantees that the result of the double copy exhibits diffeomorphism invariance. Thus, an alternative approach could rely on a thorough exploration of the analogous interplay for AdS boundary correlators.

%%%%%%%%%%%%%%%%%%%%%

\subsection{Position-space three-point double copy and comments on Mellin-space double copy}

Recent results suggest that the differential double copy (\ref{gravdoublecopy}) may not be the only double copy procedure applicable to AdS boundary correlators. To gain some insight into the possible structure of alternative double copy relations between gauge 
and gravity theories in AdS space, it is useful to examine the simple example of the three-point 
AdS boundary correlator in position space. 

Using the three-point Feynman rule following from the Einstein-Hilbert action in AdS space (with cosmological constant $\Lambda = -d(d-1)/2$) and following the same computational strategy as for the YM AdS boundary correlator, we found that the three-graviton AdS boundary correlator is 
\begin{align}
    \mathcal{M}_3&=\frac{d^2\,\Gamma(d)}{16\pi^d (d+1)^3}
     \frac{M_3}{(P_{12}P_{13}P_{23})^{1+d/2}} \ , \\
    M_3&= f_1\Lambda_1^2+f_2\Lambda_1V_1V_2V_3+f_3(V_1V_2V_3)^2+f_4\Lambda_2+f_5\Lambda_3 \nonumber
\end{align}
where $V_i$, $H_{i,j}$, and $\Lambda_1$ are defined in eqs.~\eqref{VHdefinitions}, $\Lambda_2$ and $\Lambda_3$ are 
\begin{align}
    \Lambda_2 &=H_{1,2}H_{2,3}H_{3,1}\,,\\
    \Lambda_3 &=V_1V_2H_{1,3}H_{2,3}+\text{cyclic}\,.\nonumber
\end{align}
and the functions $f_{1,\dots 5}$ are
\begin{align}
    f_1 &=16-\frac{16}{d}-\frac{8}{d^2}\,,
    &f_2 &=-8-\frac{8}{d}+\frac{24}{d^2}+\frac{16}{d^3}\,, \nonumber\\
    f_3 &=1+\frac{4}{d}-\frac{4}{d^2}-\frac{16}{d^3}\,,
   & f_4& =\frac{8}{d}\,, \qquad \qquad
     f_5 =\frac{4}{d^2}+\frac{8}{d^3}\,.
\end{align}
This expression for $\mathcal{M}_3$ agrees with ref.~\cite{Zhiboedov:2012bm} up to some 
notational translation:
\begingroup
\allowdisplaybreaks
\begin{align}
    \Lambda_1^{(\text{here})}&=-(-2P_1\cdot P_2)(-2P_2\cdot P_3)(-2P_3\cdot P_1)\Lambda_1^{(\text{there})}\,,\nonumber\\
    (V_1V_2V_3)^{(\text{here})}&=-(-2P_1\cdot P_2)(-2P_2\cdot P_3)(-2P_3\cdot P_1)(V_1V_2V_3)^{(\text{there})}\,,\nonumber\\
    \Lambda_2^{(\text{here})}&=(-2P_1\cdot P_2)^2(-2P_2\cdot P_3)^2(-2P_3\cdot P_1)^2\Lambda_2^{(\text{there})}\,,\nonumber\\
    \Lambda_3^{(\text{here})}&=(-2P_1\cdot P_2)^2(-2P_2\cdot P_3)^2(-2P_3\cdot P_1)^2\Lambda_3^{(\text{there})}\,.
\end{align}
\endgroup

It is not difficult to see the numerator $M_3$ above and the analogous quantity $N_3$ in the three-point YM correlator in eq.~\eqref{eq:N3YM}
are related by
\begin{align}
\label{squaring}
\lim_{d\rightarrow\infty} {M}_3 = \lim_{d\rightarrow\infty}  ({N}_3)^2 \ ,
\end{align}
That is, to leading order in the expansion in the large dimension of the AdS space, the three-graviton correlator equals the square of the 
three-gluon correlator in eq~\eqref{eq:N3YM}, in agreement with ref.~\cite{Mizeraetal}.

We note that, for three-point correlators, the position-space factors $ {M}_3$ 
and  $N_3$ coincide (up to possible {\em overall} normalization factors) with the corresponding Mellin-space amplitudes. With this observation, eq.~\eqref{squaring} 
above also implies that simple squaring relations between gauge and gravity three-point amplitudes may hold in Mellin space only in the large-$d$ limit.
In contrast, ref.~\cite{Zhou:2021gnu} reports such a squaring relation for the scalar
components of the super-gluon and super-graviton multiplet at $d=4$. The difference
is presumably due to the action of supercharges which introduces a nontrivial dependence 
on the conformal weight.

\iffalse
We note that, for three-point correlators, the position-space factors $ {M}_3$ and  
$N_3$ coincide (up to possible {\em overall} normalization factors) with 
the corresponding Mellin-space amplitudes. With this observation, the conclusion \eqref{squaring} departs from that of ref.~\cite{Zhou:2021gnu}, which found that
the super-graviton three-point function is the square of the super-gluon three-point function.  While ref.~\cite{Zhou:2021gnu} uses a different 
normalization from ours, the formulae above show that the $d$ dependence of the difference ${M}_3  - ({N}_3)^2 $ prevents 
it from vanishing for any choice of normalization at any finite $d$. 
\fi

We will refrain from conjecturing the generalization of this relation to higher-point correlators or how it might be formulated for the differential 
form of correlators. It is however difficult not to note, as was also noted in \cite{Alday:2021odx}, certain similarities between the large-dimension limit above 
and the relation between flat space S-matrix and 
AdS boundary correlators. Indeed, it was argued in ref.~\cite{Paulos:2016fap, Penedones:2010ue, Fitzpatrick:2011ia} that these two quantities are closely related; a formulation of 
this connection which holds for amplitudes of massive fields is  \cite{Paulos:2016fap}
\begin{align}
m_1^a T(k_i) = 
\lim_{\Delta_i\rightarrow \infty}\frac{\Delta_1^a}{{\cal N}} M\left(\gamma_{ij}=\frac{\Delta_i\Delta_j}{\sum_{k=1}^n\Delta_k}\left(1+\frac{k_i\cdot k_j}{m_i m_j}\right)\right)
\end{align}
where $M$ is a Mellin-space amplitude, $\gamma_{ij}$ are Mellin variables obeying the standard constraints, $T$ is a flat space amplitude, 
$k_i$ are flat-space momenta, $a=\frac{n(d-1)}{2}-d-1$ and
\begin{align}
{\cal N} = \frac{\pi^d}{2}\Gamma\bigg(\frac{\sum_{i=1}^n\Delta_i - d}{2}\bigg)\prod_{i=1}^n \frac{\sqrt{{\cal N}_{\Delta_i}}}{\Gamma(\Delta_i)}\,,
\qquad
{\cal N}_{\Delta} = \frac{\Gamma(\Delta)}{2\pi^{d/2}\Gamma(\Delta-d/2+1)} \ .
\end{align}
It clearly implies that, at least in the limit of large AdS energies, the Mellin-space amplitude exhibits a double-copy structure which is inherited from the corresponding flat 
space S-matrix element. The large-$\Delta$ limit may be realized either by considering very massive particles or, as in the three-point example discussed above, by taking the 
space-time dimension to be large. It would be interesting to understand better in what sense $\ads_{d\rightarrow\infty}$ may be interpreted as flat space.
More involved relations \cite{Penedones:2010ue,Li:2021snj, Fitzpatrick:2011ia} connecting Mellin-space and flat space amplitudes are also suggestive of a double-copy structure in this limit. 

Taking at face value the observation that we may assume the dimension to be large, let us discuss another limit on AdS boundary correlators that points to a double-copy structure in AdS 
momentum space.

\subsection{An argument for double copy at high energies}

The AdS Poincar\'e patch that we have been using exhibits translational invariance -- and thus conserved momentum -- in the directions parallel to the boundary.
It is therefore natural to consider momentum-space AdS boundary correlators -- i.e. Fourier-transforms of AdS boundary correlators along the boundary coordinates. Properties of momentum-space correlation functions of gauge-invariant operators have been discussed from dual gauge theory perspective in refs.~\cite{Engelund:2012re, Bzowski:2019kwd, Bzowski:2020kfw, Jain:2021qcl}.

A hard high energy scattering process (i.e. a scattering process for which the momentum transfer is large) may be expected to be localized in a small region of the space.
Thus, for weakly-curved spaces,  the scattering effectively occurs in flat space. An important point, emphasized in refs.~\cite{Polchinski:2001tt, Brower:2006ea} and used there to provide a connection between the soft high-energy string theory S-matrix elements and the hard S-matrix elements of gauge theories, is that the momenta of particles in the scattering region are not the same as the momenta at infinity/boundary.  Rather than the boundary momentum $p$, it is the momentum ${\tilde p}$ in the local inertial frame, 
\begin{align}
{\tilde p}_a = e_a^\mu p_\mu \ 
\end{align}
with the vielbein $e_\mu^a$ and $p_\mu \sim \partial/\partial x^\mu$, that governs the local scattering process. Moreover, since the propagation from the boundary to the 
interaction region probes a large region of the curved space, the asymptotic states are captured by the curved-space bulk-boundary propagators.

Thus, if the extent of the scattering region is not too large, the correlation function labeled by boundary momenta is schematically
\begin{align}
{\cal M}_{\Delta_1\dots \Delta_n}(p)=\langle {\cal O}_{\Delta_1}(p_1)\dots {\cal O}_{\Delta_n} (p_n) \rangle
= \int_{M}\, {\cal M}_\text{flat}({\tilde p})  \prod_{i=1}^n E_{\Delta_i} \ ,
\label{factorized}
\end{align}
where $E_{\Delta_i}$ are the bulk-boundary propagators labeled by boundary momenta for the fields dual to the operators ${\cal O}_{\Delta_i}$, ${\cal M}_\text{flat}$ is the flat space amplitude for 
these fields and $M$ is the entire space (e.g. AdS$_5\times$X). For tree-level boundary correlators the integration runs over the coordinates that are not Fourier-transformed (e.g. in AdS it is only the 
transverse direction). The asymptotic states for fields with spin naturally carry tangent space indices, so their bulk-boundary propagators are similarly labeled.

The Poincar\'e patch the metric is
\begin{align}
ds^2 = \frac{R^2}{z^2}\left(\eta_{\mu\nu}dx^\mu dx^\nu +  dr^2 \right) \ .
\end{align}
In these coordinates, the boundary is at $z=0$ and the momentum in the local inertial frame is
\be
{\tilde p} = \frac{z}{R} p \ .
\ee
Thus, for any finite boundary momentum $p$, the local momentum is large if the scattering occurs away from the boundary. 
One may extend the range of validity of this approximation by taking the boundary momenta to be parametrically large, but the 
scattering region is required to have a relatively small extent in the  AdS transverse direction. 
Therefore, the double-copy structure of the flat space gravitational amplitudes, formally written as 
${\cal M}_\text{flat}({\tilde p})  = {\rm DC}[{\cal A}_\text{L, flat}({\tilde p}), {\cal A}_\text{L, flat}({\tilde p})]$, 
implies that in the regime eq.~\eqref{factorized} holds the transverse-space integrand of AdS boundary correlators also have certain double-copy properties. For graviton asymptotic states:
\begin{align}
{\cal M}_{\Delta_1\dots \Delta_n}(p) = \int_{M}\, {\rm DC}[{\cal A}^{M_1\dots M_n}_\text{L, flat}({\tilde p}), {\cal A}^{N_1\dots N_n}_\text{L, flat}({\tilde p})] 
\prod_{i=1}^n E^{M_iN_i, A_iB_i}_{\Delta_i} Z_{i, A_i B_i} \ ,
\end{align}
where graviton $Z_i$ are polarization tensors, $E^{M_iN_i, A_iB_i}_{\Delta_i} $ are graviton bulk-boundary propagators and ${\cal A}_{L,R}$ are the left 
and right gauge theory amplitudes entering the flat space double copy.

Inspection of the bulk-boundary propagators reveals that these properties may be further enhanced in the limit of large AdS dimension or large~$\Delta$.  As discussed 
in section~\ref{YMfeynamnrules}, the vector-field propagator may be written as a differential operator acting on the scalar propagator, \emph{cf.} eq.~\eqref{bulktoboundYMN}. 
The graviton propagator has a similar form 
\begin{align}\label{gravprop}
E_{\Delta}^{MN,AB} &= {\cal D}^{MN,AB}E_\Delta\,,\\
{\cal D}^{MN,AB} &= \eta^{MA}\eta^{NB} + \frac{1}{\Delta}\left(\eta^{MA} P^B\frac{\partial}{\partial P_N} + \eta^{NB} P^A\frac{\partial}{\partial P_M}  \right) 
+ \frac{1}{\Delta(\Delta{+}1)} P^AP^B \frac{\partial^2}{\partial P_M \partial P_N} \ .
\nonumber
\end{align}
In the limit of large $\Delta$ or for massless fields in the limit of large AdS dimension, 
this operator may also be written as
\begin{align}
E_{\Delta\rightarrow\infty}^{MN,AB} &= {\cal D}^{MN,AB}E_{\Delta\rightarrow\infty} = {\cal D}^{MA} {\cal D}^{NB}E_{\Delta\rightarrow\infty}  =   {\cal D}^{NB}  {\cal D}^{MA} E_{\Delta\rightarrow\infty} 
\\
{\cal D}^{MA} &= \eta^{MA} +\frac{1}{\Delta}  P^A\frac{\partial}{\partial P_M} \ .
\end{align}
Even though ${\cal D}$ contains terms with a manifest $\Delta^{-1}$ which might seem possible to ignore in the large-$\Delta$ limit, the derivatives with respect to the bulk point  provide
an additional factor ${\cal O}(\Delta)$ which render this term finite in this limit. The factorization of ${\cal D}^{MN,AB} $ relies on dropping various terms ${\cal O}(\Delta^{-1})$ after the derivatives 
are  evaluated. Thus, in the high energy (from the boundary perspective) and large large AdS dimension, the integrand of the momentum-space gravitational AdS boundary correlator can be written 
as the square of a differential operator acting on scalar bulk-boundary propagators:
\begin{align}
\label{doublecopy}
{\cal M}_{\Delta_1\dots \Delta_n}(p) = \int_{M}\, {\rm DC}[{\cal A}^{M_1\dots M_n}_\text{L, flat}({\tilde p}), {\cal A}^{N_1\dots N_n}_\text{L, flat}({\tilde p})] 
\prod_{i=1}^n Z_{i, A_i} {\cal D}^{M_i,A_i} Z_{i, B_i}{\cal D}^{N_i,B_i}\,E_{\Delta_i\rightarrow\infty} \ .
\end{align}
A single power of this differential operator,
\begin{align}
\label{singlecopy}
{\cal A}_{\Delta_1\dots \Delta_n}(p) = \int_{M}\, {\cal A}^{M_1\dots M_n}_\text{flat}({\tilde p})
\prod_{i=1}^n Z_{i, A_i} {\cal D}^{M_i,A_i} E_{\Delta_i\rightarrow\infty} \ 
\end{align}
where we suppressed color indices, is a color-dressed gauge theory AdS boundary correlator.  While in general the weight of vector fields and gravitons is different, 
their differences are subleading in the large dimension or large energy limit so $E_{\Delta_i\rightarrow\infty}$ are the same in both eq.~\eqref{doublecopy} and eq.~\eqref{singlecopy}.

Similar reasoning suggests AdS boundary correlators with other asymptotic states can be double-copied in the same sense as outlined here.  Fourier-transforming the boundary momenta provides
a possible connection to the position-space representation of AdS boundary correlators. It would be very interesting to understand whether a more direct relation can be formulated.

\section{Conclusion}\label{conclusions}

The scattering amplitudes program has uncovered remarkable structures in 
flat space quantum field theories, most of which are not manifest from a 
Lagrangian perspective. 
Their far-reaching consequences make it worthwhile to explore and develop 
their curved space analog. Its maximal symmetry and central role in the
most concrete realization of the holographic principle make AdS space an important 
arena fur such investigations. 
In this paper, we proposed a framework to discuss color/kinematics duality 
and the associated BCJ amplitude relations for AdS boundary correlators in position 
space. 
Other approaches \cite{Albayrak:2020fyp,Armstrong:2020woi,Alday:2021odx,Zhou:2021gnu} focus 
on the AdS momentum space and Mellin space representations of boundary correlators.
While somewhat technical, we argue that the linearly-realized conformal symmetry
of the position embedding space has certain advantageous consequences analogous 
the on-shell properties of flat space scattering amplitudes. 

We conjectured an AdS analog of color/kinematics duality and of the BCJ amplitudes 
relations. We supported our conjecture by showing the AdS$_5$ boundary 
correlators of the NLSM and YM theory obey the AdS BCJ relations through six and four points, respectively.
As in flat space, these properties are obscure from a Lagrangian perspective. 
The AdS amplitudes relations suggest the existence of a differential 
representation of the tree level boundary correlators of the form of eq.~(\ref{eq:conjecturedadsBCJrep}), which we explicitly found for both the four- and 
six-points NLSM correlators.
We also gave a broad overview of several possible double-copy procedures in AdS space, 
focusing on two distinct ones that satisfy certain necessary consistency conditions 
for the $d=2$ and $d\rightarrow \infty$ limits, respectively.

The relations we conjectured from the perspective of weakly-coupled theories 
in AdS space translate, through the holographic duality, to conditions on
strongly-coupled CFT flavor correlation functions. Incorporating them in the CFT 
bootstrap program would help identify theories with a weakly coupled bulk dual. 

Since our analysis covers only (appropriately-defined) massless states, our conclusions 
apply at leading order in the large 't Hooft coupling limit of the boundary theory. 
It would clearly be interesting to extend them to the complete string theory 
in AdS space (suitably extended by a compact space) and thus access (through tree-level 
string calculations) subleading terms in the large 't Hooft coupling limit or $1/N$ 
corrections to correlators through string loops. 
Such calculations may be aided by the differential representation studied in this 
paper, as it makes manifest the analytic structure of correlators. While such calculations
are expected to be IR finite \cite{CALLAN1990366,Aharony:2012jf,Aharony:2015hix}, 
the specific spectrum and interactions of string theory should be necessary to ensure 
their UV consistency.
It would moreover be interesting to understand if the differential representations
discussed here have a supersymmetric generalization analogous to the on-shell 
superspace formalism of scattering amplitudes~\cite{Nair:1988bq,Herderschee:2019ofc, Herderschee:2019dmc, Dennen:2009vk,Boels:2012ie,Craig:2011ws}.

While the discussion of color/kinematics-satisfying representations for boundary correlators and of the relations they satisfy was carried out in AdS space, the results are applicable with minor modifications to computing the wave-function in the Bunch-Davies vacuum de Sitter (dS) space.
To find the Bunch-Davies wave-function, the relevant Witten diagrams correspond to integrating over dS instead of AdS space. 
There are a number of techniques available for computing experimentally-measurable, late-time correlators using the Bunch-Davies wave-function.\footnote{Appendix A of ref.~\cite{Arkani-Hamed:2018kmz} summarizes the precise relation between the Bunch-Davies wave-function, which can be computed using analytically continued Witten diagrams, and late time correlators. See also ref.~\cite{Harlow:2011ke}.} In particular, the relation between dS late time correlators and AdS boundary correlators in momentum-Mellin space amounts to including a simple pre-factor \cite{Sleight:2020obc,Sleight:2021iix,Sleight:2019hfp, Sleight:2019mgd}. 
Studying whether the differential representation is useful for directly computing late-time correlators, not just the Bunch-Davies wave-function, could be very rewarding.\footnote{We thank Sebastian Mizera for discussion on this point.}

Our discussion of the double-copy in AdS space is clearly incomplete; identifying a 
construction that relates gauge and gravitational theories at finite boundary dimension $d$
is clearly an important problem. 
Celestial amplitudes are perhaps the closest flat space analog to our AdS boundary correlators
and their structure \cite{Casali:2020vuy,Casali:2020uvr,Kalyanapuram:2020aya} is analogous to
that of the differential representations discussed here. Notably, they are constrained by 
symmetries \cite{Strominger:2017zoo,Kapec:2014opa,Pasterski:2015tva,Arkani-Hamed:2020gyp} 
exposed by soft-momentum properties of  the corresponding momentum-space amplitudes.
 An interesting question is whether such symmetries could generalize in some form to the 
 AdS boundary correlators, providing an AdS analog of the flat space soft theorems.  
 More conceptually, it is important to understand the meaning of the double copy from 
 the point of view of the boundary theory and in particular the sense in which the product 
 of two operators is local in the quantum theory, as a double-copy construction might suggest.
 
 We briefly discussed in section~\ref{diffdoublecopy} a simple deformation of YM theory 
 by a higher-dimension operator. Extensive work has been carried out in flat space directed at understanding the interplay between higher dimension operators and color/kinematics duality \cite{Broedel:2012rc,Elvang:2020kuj,He:2016iqi}. 
Specifically, ref.~\cite{Chi:2021mio} proposed a bootstrap procedure for identifying higher dimension operators compatible with color/kinematics duality and a BCJ-like construction in the presence of such operators was discussed in refs.~\cite{Carrasco:2021ptp,Carrasco:2019yyn}. It would be interesting to understand their realization in curved space, and study the double copy at higher points.

\acknowledgments
We would like to thank Murat G\"unaydin, Sebastian Mizera, Mukund Rangamani, 
Oliver Schlotterer, Diandian Wang and Leopoldo Pando Zayas for stimulating discussion. We also thank Henrik Johansson and Xinan Zhou for comments on the draft. 
AH would like to especially thank Henriette Elvang for her continued support 
and comments. 
AH is supported in part by the US Department of Energy under Grant No. DE-SC0007859 and 
in part by a Leinweber Center for Theoretical Physics Graduate Fellowship. 
PD, RR and FT are supported by the US Department of Energy under Grant No. DE-SC0013699.

\newpage

\appendix

\section{Embedding space formalism \label{app_embedding_space}}
The $\ads_{d+1}$ background can be realized as a space-like hypersurface in the $(d{+}2)$ dimensional flat spacetime $\mathbb{R}^{d+1,1}$.
Under Cartesian coordinates $X^A=(X^{\mathsf{a}},X^d,X^{d+1})$, the hypersurface is given by the equation
\begin{align}\label{eq:adssurface}
    (X^0)^2+(X^1)^2+\cdots+(X^{d-1})^2+(X^d)^2-(X^{d+1})^2=-R^2\,.
\end{align}
The Poincar\'{e} coordinates of $\ads_{d+1}$ are given by the following parametrization, 
\begin{align}
\label{eq:embedding}
    & X^{\mathsf{a}}=\frac{R}{z}x^{\mathsf{a}}\,,
    && X^d=\frac{R}{z}\frac{1-x^2-z^2}{2}\,,
    & X^{d+1}=\frac{R}{z}\frac{1+x^2+z^2}{2}\,,
\end{align}
such that for $R=\text{constant}$ we have
\begin{align}\label{eq:metricAdS}
    ds^2_{\ads_{d+1}}=g_{\mu\nu}dx^{\mu}dx^{\nu}=\frac{R^2}{z^2}\left(dz^2+dx_{\mathsf{a}} dx^{\mathsf{a}}\right)\,.
\end{align}
We can also view eq.~\eqref{eq:embedding} as a coordinate transformation from the Cartesian coorinates $X^A$ to the coordinates $(R,z,x^{\mathsf{a}})$, under which metric of $\mathbb{R}^{d+1,1}$ becomes
\begin{align}
    ds^2_{\mathbb{R}^{2,d}}=-dR^2+ds^2_{\ads_{d+1}}\,.
\end{align}
In fact, $(R,z,x^{\mathsf{a}})$ are Gaussian normal coordinates adapted to the AdS hypersurface and the $g_{\mu\nu}$ in eq.~\eqref{eq:metricAdS} is the induced metric.

When we study a scalar field $\phi$ on $\ads_{d+1}$, it is convenient to extend its definition to the entire embedding space. However, we need to make sure that the dynamics does not depend on the variation normal to the AdS hypersurface. It is not difficult to show that 
\begin{align}
\eta^{AB} \partial_A \phi \partial_B \phi &=
-(\partial_R\phi)^2 + g^{\mu\nu}\partial_{\mu} \phi\partial_{\nu} \phi\,,\qquad
\frac{1}{R} X^A\partial_A \phi = \partial_R\phi\,.
\end{align}
We can thus define
\begin{align}\label{eq:GAB}
    G_{AB}=\eta_{AB}-\frac{X_A X_B}{X^2}=g^{\mu\nu}\frac{\partial X_A}{\partial x^{\mu}}\frac{\partial X_B}{\partial x^{\nu}}\,,
\end{align}
such that 
\begin{align}
    G^{AB}\partial_A\phi\partial_B\phi=g^{\mu\nu}\partial_{\mu}\phi\partial_{\nu}\phi\,.
\end{align}
One can also prove that the AdS Laplacian is given by
\begin{align}
    \partial_A(G^{AB}\partial_B)=-\frac{1}{2}D_X^2=\nabla^2_{\ads}\,.
\end{align}
The matrix form of $G_{AB}$ under the $(R,z,x^{\mathsf{a}})$ coordinates is
\begin{align}
    G_{AB}\xrightarrow{\text{transformation}~\eqref{eq:embedding}}\begin{pmatrix}
    0 & 0 \\
    0 & g_{\mu\nu}
    \end{pmatrix}\,,
\end{align}
namely, $G^{AB}$ is indeed a projector to $\ads_{d+1}$. One can easily show that $G_{AB}$ is idempotent and transverse to the normal vector $X^A$,
\begin{align}
    G_{A}{}^BG_{B}{}^C=G_{A}{}^C\,,\qquad G_{AB}X^B = 0\,.
\end{align}
Geometrically, $G_{AB}$ is the first fundamental form associated to the hypersurface~\eqref{eq:adssurface}. From now on, we fix $R=1$ for the AdS hypersurface, such that the AdS integration measure becomes 
\begin{align}
    \int_{\ads}dX\equiv\int_{\ads} \frac{\delta(R-1)R^{d+1}dRdzd^dx}{z^{d+1}}=\int_{\ads} \frac{dzdx^d}{z^{d+1}}\,.
\end{align}
Thus we can perform integration-by-parts for $\partial/\partial X^A$ if it appears in the integrand as $\frac{\partial}{\partial X^A}(G^{AB}\cdots)$ or $G^{BA}\frac{\partial}{\partial X^A}(\cdots)$, since the projector $G^{AB}$ removes $\frac{\partial}{\partial R}$ so the measure remains invariant.

A tensor $H_{A_1A_2\ldots A_n}$ in the embedding space defines a tensor on AdS if it is \emph{transverse} to the AdS hypersurface,
\begin{align}
    X^{A_i}H_{A_1A_2\ldots A_n}=0\,,\qquad 1\leqslant i\leqslant n\,.
\end{align}
The AdS tensor can be recovered through the projection
\begin{align}
    h_{\mu_1\mu_2\ldots\mu_n}=\frac{\partial X^{A_1}}{\partial x^{\mu_1}}\frac{\partial X^{A_2}}{\partial x^{\mu_2}}\ldots\frac{\partial X^{A_n}}{\partial x^{\mu_n}}H_{A_1A_2\ldots A_n}\,.
\end{align}
From this equation and the relation~\eqref{eq:GAB} between $g_{\mu\nu}$ and $G_{AB}$, we can show that the contraction between two AdS tensors can be uplifted into the embedding space,
\begin{align}
    h_{\mu_1\mu_2\ldots\mu_n}g^{\mu_i\nu_j}f_{\nu_1\nu_2\ldots\nu_m}=H_{A_1A_2\ldots A_n}G^{A_iB_j}F_{B_1B_2\ldots B_m}=H_{A_1A_2\ldots A_n}\eta^{A_iB_j}F_{B_1B_2\ldots B_m}\,,
\end{align}
where the second equality is due to the transversality. Conditions like being symmetric and traceless of an AdS tensor can be directly imposed on its embedding space uplift. A more formal discussion on this topic can be found in ref.~\cite{Costa:2014kfa}.

The conformal boundary of AdS is located at $z\rightarrow 0$. We can represent a boundary point $x_i^{\mathsf{a}}$ by a projective null vector in the embedding space,
\begin{align}
    P_i = \Big(x_i^{\mathsf{a}},\frac{1-x_i^2}{2},\frac{1+x_i^2}{2}\Big)\,.
\end{align}
Obviously we have $P_i^2=0$ and we identify $P_i\sim \lambda P_k$. A polarization vector $\epsilon_i^{\mathsf{a}}$ on the boundary can also be represented in the embedding space,
\begin{align}
    Z_i=(\epsilon_i^{\mathsf{a}},-\epsilon_i\cdot x_i,\epsilon_i\cdot x_i)\,, 
\end{align}
such that $P_i\cdot Z_i=0$ and $Z_i\cdot Z_i=\epsilon_i^2=0$. A tensor current $F_{A_1A_2\ldots A_n}(P)$ defined on the null cone $P^2=0$ is physical only if it is homogeneous and transverse,
\begin{align}
    F_{A_1A_2\ldots A_n}(\lambda P)=\lambda^{-\Delta}F_{A_1A_2\ldots A_n}(P)\,,\qquad P^{A_i}F_{A_1A_2\ldots A_n}(P)=0\,,\quad 1\leqslant i\leqslant n\,.
\end{align}
Finally, we list a few useful relations that connect the embedding space and physical space expressions,
\begin{align}
-2X\cdot P_i 
&=z+z^{-1}(x-x_i)^2\,,  & -2P_i\cdot P _j  &= (x_i-x_j)^2\,, \nonumber\\
     P_i\cdot Z_j&=\epsilon_j\cdot (x_i-x_j)\,,
   &  Z_i\cdot Z_j&=\epsilon_i\cdot\epsilon_j\,,
\end{align}
where $X$ is a bulk point defined as in eq.~\eqref{eq:embedding}.

\section{Contact diagrams in AdS}\label{contactdiagramsinads}

Contact diagrams of scalar operators play a crucial role in flat space scattering amplitudes. Crucially, the delta function that imposes momentum conservation on an $n$-point scattering amplitudes arises from the contact diagram of an $n$-point scalar vertex
\begin{equation}
\begin{split}
\mathcal{A}_{\textrm{contact}}^{\textrm{flat}}&=\int d^{d}x\prod_{i}e^{ip_{i}x} \\
&=\delta^{d}\Big(\sum_{i}p_{i}^{\mu}\Big)
\end{split}
\end{equation}
The contact diagrams of $n$-point scalar vertex operators play an analogous role in AdS, giving the AdS analog of the flat space delta function. The contact diagram associated with the vertex $\prod_{i}\phi_{i}$ in AdS is

\begin{equation}
\mathcal{A}^{n}_{\textrm{contact}}=\left( \prod_{i=1}^n\frac{\Gamma(\Delta_{i})}{2\pi^{d/2} \Gamma(\Delta_{i}-d/2+1)} \right )D_{\Delta_1\dots \Delta_n}    
\end{equation}
where
\begin{align}\label{orgdef}
D_{\Delta_1\dots \Delta_n} = 
\int_{\ads} dX \prod_{i=1}^n \frac{1}{(-2P_i\cdot X)^{\Delta_i}}
\end{align}
is the $D$-function. Any tree-level position space correlator in AdS will be a polynomial in $D$-functions, $Z$-variables and $P$-variables. Similar to how all flat space scattering amplitudes contain a universal momentum conserving delta function, each term in the polynomial expansion of the position space correlator contains its own $D$-function. A major challenge for evaluating position space correlators in AdS is finding representations of $D$-functions amiable to numeric approximation.

\subsection{Three-point \texorpdfstring{$D$}{D}-functions}

The $D$-functions that appear in three-point correlators can be integrated directly:
\begin{align}\label{3pdfunction}
D_{\Delta_{1}\Delta_{2}\Delta_{3}}&=\int_{\ads} dX \prod_{i=1}^{3}\frac{1}{(-2 P_{i}\cdot X)^{\Delta_{i}}} \nonumber\\
&=\frac{\pi^{d/2}}{2\prod_{i}\Gamma(\Delta_{i})}\Gamma\left(\frac{\Delta_{1}+\Delta_2+\Delta_3-d}{2} \right )\prod_{i<j}\frac{\Gamma(\delta_{ij})}{(-2P_{i}\cdot P_{j})^{\delta_{ij}}}
\\
\delta_{ij}&= \frac{1}{2}(\Delta_{i}+\Delta_{j}-\Delta_{k}) \ .\nonumber
\end{align}
The three-point $D$-function is the one which has a rational dependence on coordinates; higher point $D$-functions are rather non-trivial functions.

\subsection{Exact solution of four-point \texorpdfstring{$D$}{D}-functions}

In this section, we review a derivation of an exact solution to the four-point $D$-function in terms of derivatives of polylogarithms. Advantages (and disadvantages) of alternative methods are discussed at the end.  

To find a numerically tractable representation of eq.~(\ref{orgdef}), we first rewrote the $D$-function into a Feynman parameterization:
\begin{equation}\label{feynamparm}
D_{\Delta_{1},\Delta_{2},\Delta_{3},\Delta_{4}}=\frac{\pi^{d/2}\Gamma(\frac{\Sigma-d}{2})\Gamma(\frac{\Sigma}{2})}{2\prod_{i}\Gamma(\Delta_{i})}\int \frac{\prod_{j}\alpha_{j}^{\Delta_{j}-1}\delta(\sum_{i}\alpha_{i}-1)}{(\sum_{k,l}\alpha_{k}\alpha_{l}P_{k,l})^{\Sigma/2}} \,,
\end{equation}
where $\Sigma=\Delta_1+\Delta_2+\Delta_3+\Delta_4$. Our goal is to find an exact solution for the integral in eq.~(\ref{feynamparm}) that can be used for numerical analysis in Mathematica. We first solve for the integral in eq.~(\ref{feynamparm}) for the simplest case when $\Delta_{i}=1$ for all $i$'s, which we denote as $B(P_{ij})$. It is nothing but the four-mass box integral~\cite{Denner:1991qq,Bern:1993kr},
\begin{equation}\label{4massbox}
\begin{split}
B(P_{ij})&=\int \frac{\prod_j d\alpha_{j}\delta(\sum_{i}\alpha_{i}-1)}{\left ( \sum_{k,l} \alpha_{k}\alpha_{l}P_{k,l}\right )} \\
&=\frac{1}{\sqrt{\Delta'}}\Bigg[\frac{1}{2}\log\left ( \frac{u_{+}u_{-}}{(1-u_{+})^{2}(1-u_{-})^{2}}\right )\log\frac{u_{+}}{u_{-}}\\
&\quad -\textrm{Li}_{2}(1-u_{+})+\textrm{Li}_{2}(1-u_{-})-\textrm{Li}_{2}\Big(1-\frac{1}{u_{-}}\Big)+\textrm{Li}_{2}\Big(1-\frac{1}{u_{+}}\Big)  \Bigg],    
\end{split}
\end{equation}
where
\begin{align}
    & \Delta'=X^{2}+Y^{2}+Z^{2}-2XY-2YZ-2ZX\,, & & X=P_{12}P_{34}\,, & & Y=P_{13}P_{24}\,, \nonumber\\
    & u_{\pm}=\frac{Y+X-Z\pm \sqrt{\Delta'}}{2Y}\,,& & Z=P_{14}P_{23}\,.
\end{align}
Note that eq.~(\ref{4massbox}) is simply the standard four-mass box integral whose solution has been known for 20 odd years.\footnote{The same integral was evaluated in eq.~(39) in ref.~\cite{Bianchi:1998nk}, but their solution appears to have a typo. One can test the relative sign of individual terms in a given convention by checking whether the resulting $D$-function obeys the properties given in appendix~\ref{Dfunctionidentities}.} We then identify a simply relation between derivatives of $B(P_{ij})$ and $D_{\Delta_{1},\Delta_{2},\Delta_3,\Delta_4}$ to find a generic solution for four-point $D$-functions:
\begin{equation}\label{finalsolution}
\begin{split}
&\quad\frac{\pi^{d/2}(-1)^{\sum_{i<j}c_{ij}}\Gamma(\frac{\Sigma'-d}{2})\Gamma(\frac{\Sigma'}{2})}{2\Gamma(2+\sum_{i<j}c_{ij})(\prod_{i} \Gamma(\Delta_{i}'))}\left [ \prod_{i<j}\left ( \frac{\partial}{\partial P_{ij}}\right )^{c_{ij}}\right ]B(P_{ij})\\
&=\frac{\pi^{d/2}\Gamma(\frac{\Sigma'-d}{2})\Gamma(\frac{\Sigma'}{2})}{2(\prod_{i} \Gamma(\Delta_{i}'))} \int \frac{\prod_a d\alpha_{a}\alpha_{a}^{\Delta'_{a}-1}\delta(\sum_{b}\alpha_{b}-1)}{\left ( \sum_{i,j} \alpha_{i}\alpha_{j}P_{ij}\right )^{\Sigma'/2}} \\
&=D_{\Delta_{1}',\Delta_{2}',\Delta_{3}',\Delta_{4}'} \,,
\end{split}
\end{equation}
where $\Delta_{i}'=\sum_{j\neq i}c_{ij}+1$ and $\Sigma'=\sum_{i}\Delta_{i}'$. Equations (\ref{4massbox}) and (\ref{finalsolution}) together provide a closed form expressions for arbitrary four-point $D$-functions in terms of derivatives of polylogarithms. Unfortunately, although this representation is advantageous in that it provides an \textit{exact} representation of the $D$-function in terms of polylogarithms, the $D$-function expressions become cumbersome very quickly, even for symbolic computation programs such as Mathematica. We ultimately used numeric differentiation algorithms in combination with eqs. (\ref{4massbox}) and (\ref{finalsolution}) to solve for the $D$-functions at arbitrary kinematic points. These results were cross-checked with more direct numeric integrations of eq.~(\ref{feynamparm}). 

Finding numerically tractable representations of $D$-functions is generically quite hard. For example, although standard Mellin integral representations of the $D$-function can be integrated numerically with high precision, defining the actual contour for integration is somewhat subtle. While the Mellin representation of Feynman integrals is largely understood, the problem seems to be slightly more technically challenging for the $D$-functions corresponding to AdS contact diagrams. One might hope that numeric integration in the Feynman representation would offer a more realistic approach, as it suffers from no contour ambiguities. Unfortunately, direct numeric integration of Feynman parameterized integrals is often unstable for arbitrary kinematics and $\Delta_{i}$ unless sophisticated weighted Monte-Carlo sampling techniques are applied. For example, see ref.~\cite{Borinsky:2020rqs}.

\subsection{\texorpdfstring{$D$}{D}-function identities}\label{Dfunctionidentities}
When computing NLSM and YM AdS boundary correlators, we have used some $D$-function identities to simplify the results. In this appendix, we derive these identities starting from the conformal Ward identity. We start with the four-point case,
\begin{align}\label{eq:CWI}
    (D_1^{AB}+D_2^{AB}+D_3^{AB}+D_4^{AB})D_{\Delta_1,\Delta_2,\Delta_3,\Delta_4}=0\,.
\end{align}
We can act respectively $(D_1+D_3)^{AB}$ and $(D_2+D_4)^{AB}$ onto this equation, giving
\begin{align}
    &(2D_{13}^2+D_1^2+D_3^2)D_{\Delta_1,\Delta_2,\Delta_3,\Delta_4}=-(D_1+D_3)_{AB}(D_2+D_4)^{AB}D_{\Delta_1,\Delta_2,\Delta_3,\Delta_4}\,,\nonumber\\
    &(2D_{24}^2+D_2^2+D_4^2)D_{\Delta_1,\Delta_2,\Delta_3,\Delta_4}=-(D_2+D_4)_{AB}(D_1+D_3)^{AB}D_{\Delta_1,\Delta_2,\Delta_3,\Delta_4}\,.
\end{align}
Since $(D_1+D_3)$ commutes with $(D_2+D_4)$ and $D_i^2=-2m_i^2=-2\Delta_i(\Delta_i-d)$ when acting on a conformal partial wave, we get
\begin{align}\label{eq:DijId}
    (D_{23}^{2}-m_2^2-m_3^2)D_{\Delta_1,\Delta_2,\Delta_3,\Delta_4}=(D_{14}^{2}-m_1^2-m_4^2)D_{\Delta_1,\Delta_2,\Delta_3,\Delta_4}\,.
\end{align}
In particular, when $\Delta_i=\Delta$, we have
\begin{align}
    D_{13}^{2}D_{\Delta,\Delta,\Delta,\Delta}=D_{24}^{2}D_{\Delta,\Delta,\Delta,\Delta}\,.
\end{align}
The above relation can be easily generalized to $n$-points,
\begin{align}
    \Big(D_{I}^2-\sum_{i\in I}m_i^2\Big) D_{\Delta_1,\Delta_2,\ldots,\Delta_n}=\Big(D_{\bar I}^2-\sum_{i\in \bar I}m_i^2\Big) D_{\Delta_1,\Delta_2,\ldots,\Delta_n}\,,
\end{align}
where $I\cup \bar{I}=\{1,2,\ldots,n\}$ and $I\cap \bar I = \emptyset$. 

We can also carry out the derivatives in eq.~\eqref{eq:DijId} explicitly to obtain relations involving boundary positions. For example, we can use
\begin{align}\label{eq:D13action}
    -\frac{1}{2}D_{13}^{2}D_{\Delta_1,\Delta_2,\Delta_3,\Delta_4}=4\Delta_1\Delta_3(P_1\cdot P_3)D_{\Delta_1+1,\Delta_2,\Delta_3+1,\Delta_4}+\Delta_1\Delta_3D_{\Delta_1,\Delta_2,\Delta_3,\Delta_4}\,,
\end{align}
and a similar equation for $D_{24}^2$ to show that
\begin{align}
    (P_1\cdot P_3)D_{\Delta+1,\Delta,\Delta+1,\Delta}=(P_2\cdot P_4)D_{\Delta,\Delta+1,\Delta,\Delta+1}\quad\text{ for }\Delta_i=\Delta\,.
\end{align}
One can also show that acting $D_{12}^2$ and $D_{13}^2$ consecutively gives
\begin{align}\label{eq:D12D13}
    \frac{1}{4}D_{12}^2 D_{13}^2 D_{\Delta_1,\Delta_2,\Delta_3,\Delta_4} &= 16 \Delta_1(\Delta_1+1)\Delta_2\Delta_3 (P_1\!\cdot\! P_2)(P_1\!\cdot\! P_3) D_{\Delta_1+2,\Delta_2+1,\Delta_3+1,\Delta_4}\\
    &\quad + 4 \Delta_1(\Delta_1+1)\Delta_2\Delta_3\Big[(P_1\!\cdot\! P_2)D_{\Delta_1+1,\Delta_2+1,\Delta_3,\Delta_4} + (2\leftrightarrow 3)\Big]\nonumber\\
    &\quad -4\Delta_1\Delta_2\Delta_3(P_2\!\cdot\! P_3) D_{\Delta_1,\Delta_2+1,\Delta_3+1,\Delta_4}+ \Delta_1^2\Delta_2\Delta_3 D_{\Delta_1,\Delta_2,\Delta_3,\Delta_4}\,,\nonumber
\end{align}
which is symmetric under the $2\leftrightarrow 3$ exchange. This also means that $D_{\Delta_1,\Delta_2,\Delta_3,\Delta_4}$ lives in the kernel of $[D_{12}^2,D_{13}^2]$, namely, $[D_{12}^2,D_{13}^2]D_{\Delta_1,\Delta_2,\Delta_3,\Delta_4}=0$.

Next, we act $D_1^{AB}$ onto eq.~\eqref{eq:CWI} and then use eq.~\eqref{eq:DijId} to eliminate $D_{14}^2$. This gives
\begin{align}
    (D_{12}^2+D_{13}^2+D_{23}^2)D_{\Delta_1,\Delta_2,\Delta_3,\Delta_4}+(m_4^2-m_1^2-m_2^2-m_3^2)D_{\Delta_1,\Delta_2,\Delta_3,\Delta_4}=0\,.
\end{align}
When $\Delta_i=\Delta$, we get
\begin{align}
    (D_{12}^{2}+D_{13}^{2}+D_{23}^{2})D_{\Delta,\Delta,\Delta,\Delta}=2\Delta(\Delta-d)D_{\Delta,\Delta,\Delta,\Delta}\,.
\end{align}
This equation also implies the identity
\begin{align}
    (P_1\cdot P_2)D_{\Delta+1,\Delta+1,\Delta,\Delta}&+(P_1\cdot P_3)D_{\Delta+1,\Delta,\Delta+1,\Delta}\nonumber\\
    &+(P_2\cdot P_3)D_{\Delta,\Delta+1,\Delta+1,\Delta}=-\frac{4\Delta-d}{4\Delta}D_{\Delta,\Delta,\Delta,\Delta}\,.
\end{align}
In the on-shell limit $\Delta=d$, we thus derive the relation
\begin{align}
    (D_{12}^{2}+D_{13}^{2}+D_{23}^{2})D_{d,d,d,d,}=0\,,
\end{align}
which resembles the relation $s+t+u=0$ for flat space Mandelstam variables.

The above identities can also be derived using integration-by-parts relations. Below we show the derivation of eq.~\eqref{eq:DijId}. Our calculation will be given in the embedding space, which is in parallel with the one in the physical space~\cite{DHoker:1999kzh}. To start with, we define
\begin{align}\label{eq:id1}
    D_{\partial\Delta_1,\Delta_2,\partial\Delta_3,\Delta_4}
    &\equiv\int_{\ads}dXG^{AB}\Bigg[\frac{\partial}{\partial X^A}\frac{1}{(-2P_1\cdot X)^{\Delta_1}}\Bigg]\frac{1}{(-2P_2\cdot X)^{\Delta_2}}\nonumber\\
    &\qquad\qquad\quad\quad\!\times\Bigg[\frac{\partial}{\partial X^B}\frac{1}{(-2P_3\cdot X)^{\Delta_3}}\Bigg]\frac{1}{(-2P_4\cdot X)^{\Delta_4}}\nonumber\\
    &=4\Delta_1\Delta_3(P_1\cdot P_3)D_{\Delta_1+1,\Delta_2,\Delta_3+1,\Delta_4}+\Delta_1\Delta_3 D_{\Delta_1,\Delta_2,\Delta_3,\Delta_4}\,.
\end{align}
We then use the identity
\begin{align}
    \partial_AG^{AB}\partial_B\left[\frac{1}{(-2P_1\cdot X)^{\Delta_1}}\frac{1}{(-2P_3\cdot X)^{\Delta_3}}\right]&=2G^{AB}\frac{\partial}{\partial X^A}\frac{1}{(-2P_1\cdot X)^{\Delta_1}}\frac{\partial}{\partial X^B}\frac{1}{(-2P_3\cdot X)^{\Delta_3}}\nonumber\\
    &\quad +\frac{m_1^2+m_3^2}{(-2P_1\cdot X)^{\Delta_1}(-2P_3\cdot X)^{\Delta_3}}
\end{align}
to write eq.~\eqref{eq:id1} as
\begingroup
\allowdisplaybreaks
\begin{align}
    &D_{\partial\Delta_1,\Delta_2,\partial\Delta_3,\Delta_4}+\frac{1}{2}(m_1^2+m_3^2) D_{\Delta_1,\Delta_2,\Delta_3,\Delta_4}\nonumber\\
    &=\frac{1}{2}\int_{\ads}dX(\partial_AG^{AB}\partial_B)\left[\frac{1}{(-2P_1\cdot X)^{\Delta_1}}\frac{1}{(-2P_3\cdot X)^{\Delta_3}}\right]\frac{1}{(-2P_2\cdot X)^{\Delta_2}}\frac{1}{(-2P_4\cdot X)^{\Delta_4}}\nonumber\\
    &=\frac{1}{2}\int_{\ads}dX(\partial_AG^{AB}\partial_B)\left[\frac{1}{(-2P_2\cdot X)^{\Delta_2}}\frac{1}{(-2P_4\cdot X)^{\Delta_4}}\right]\frac{1}{(-2P_1\cdot X)^{\Delta_1}}\frac{1}{(-2P_3\cdot X)^{\Delta_3}}\nonumber\\
    &=D_{\Delta_1,\partial\Delta_2,\Delta_3,\partial\Delta_4}+\frac{1}{2}(m_2^2+m_4^2)D_{\Delta_1,\Delta_2,\Delta_3,\Delta_4}\,,
\end{align}
\endgroup
where integration-by-parts has been used to obtain the second equation. This leads to the identity
\begin{align}
    &\quad 4\Delta_1\Delta_3(P_1\cdot P_3)D_{\Delta_1+1,\Delta_2,\Delta_3+1,\Delta_4}+\frac{1}{2}(2\Delta_1\Delta_3+m_1^2+m_3^2)D_{\Delta_1,\Delta_2,\Delta_3,\Delta_4}\nonumber\\
    &=4\Delta_2\Delta_4(P_2\cdot P_4)D_{\Delta_1,\Delta_2+1,\Delta_3,\Delta_4+1}+\frac{1}{2}(2\Delta_2\Delta_4+m_2^2+m_4^2)D_{\Delta_1,\Delta_2,\Delta_3,\Delta_4}\,.
\end{align}
When $\Delta_i=\Delta$, we can recover $(P_1\cdot P_3)D_{\Delta+1,\Delta,\Delta+1,\Delta}=(P_2\cdot P_4)D_{\Delta,\Delta+1,\Delta,\Delta+1}$. Now using eq.~\eqref{eq:D13action} to replace $P_i\cdot P_j$ by $D_{13}^2$ immediately leads to 
\begin{align}
    (D_{23}^{2}-m_2^2-m_3^2)D_{\Delta_1,\Delta_2,\Delta_3,\Delta_4}=(D_{14}^{2}-m_1^2-m_4^2)D_{\Delta_1,\Delta_2,\Delta_3,\Delta_4}\,.
\end{align}

\section{Derivation of eq.~(\ref{eq:mellinspace})}\label{sec:details4pcom}

In this appendix, we give a derivation of eq.~(\ref{eq:mellinspace}), copied here as
\begin{gather}
\textbf{M}^{-1}\Bigg[\prod_{1\leqslant i<j}^{4}\frac{\Gamma(\delta_{ij}+l_{ij})}{\Gamma(\delta_{ij})}\Bigg] =\frac{2}{\pi^{d/2}}
\frac{\prod_{i=1}^{4}\Gamma(\tilde{\Delta}_{i})}{\Gamma\Big(\frac{\tilde\Sigma-d}{2}\Big)}\Bigg[ \prod_{1\leqslant i<j}^4 P_{ij}^{l_{ij}}\Bigg]D_{\tilde{\Delta}_{1} \tilde{\Delta}_{2} \tilde{\Delta}_{3} \tilde{\Delta}_{4}}  \ .
\end{gather}
It suffices to consider the inverse Mellin transform of
\begin{equation}
\frac{\Gamma(\delta_{12}+l)}{\Gamma(\delta_{12})}    \ .
\end{equation}
We first parameterize the delta functions in the inverse Mellin transform~\eqref{derivationofInvM}
\begin{equation}
\delta(\Delta_{i}-\sum_{j}\delta_{ij})=\int_{0}^{\infty}\frac{dt_{i}}{t_{i}}t_{i}^{\Delta_{i}-\sum_{j}\delta_{ij}}
\end{equation}
and then evaluate the $\delta_{ij}$ contour integral using
\begin{equation}
\int_{-i\infty}^{i\infty} \frac{d\delta_{ij}}{2\pi i}\Gamma(\delta_{ij}+l)(t_{i}t_{j}P_{ij})^{-\delta_{ij}}=(t_{i}t_{j}P_{ij})^{l}e^{-t_{i}t_{j}P_{ij}} \ .   
\end{equation}
This leads to
\begin{equation}
\textbf{M}^{-1}\Bigg[\frac{\Gamma(\delta_{12}+l)}{\Gamma(\delta_{12})}\Bigg]=(P_{12})^{l}\int_{0}^{\infty}\Bigg[\prod_{i=1}^{4}\frac{dt_i}{t_i}\Bigg]t_{1}^{\Delta_{1}+l}t_{2}^{\Delta_{2}+l}t_{3}^{\Delta_{3}}t_{4}^{\Delta_{4}}\prod_{i<j}e^{-t_{i}t_{j}P_{ij}}   \ .    
\end{equation}
To convert this expression to a $D$-function, we first insert the identity element
\begin{equation}
\begin{split}
1&=\frac{1}{\Gamma[(\Delta_{1}+l)/2+(\Delta_{2}+l)/2+\Delta_{3}/2+\Delta_{4}/2-d/2]} \\
&\times \int_{0}^{\infty} \frac{dz}{z}z^{-d/2+(\Delta_{1}+l)/2+(\Delta_{2}+l)/2+\Delta_{3}/2+\Delta_{4}/2}e^{-z}
\end{split}
\end{equation}
and then rescale $t_{i}\rightarrow t_{i}'/\sqrt{z}$. Now we can carry out the $z$ integral as
\begin{align}
\frac{\pi^{d/2}}{2}\int_{0}^{\infty}\frac{dz}{z}z^{-d/2}\exp\Big[-z+\frac{1}{z}\Big(\sum_{i}t'_iP_i\Big)^2\Big]=\int_{\ads}dX\prod_i e^{2t_i'X\cdot P_i}\,.
\end{align}
Finally, we perform the $t'_i$ integral as
\begin{equation}
\int_{0}^{\infty}\frac{dt'_i}{t'_i}(t'_i)^{\Delta'_i}e^{t'(2X\cdot P_i)}=\frac{\Gamma(\Delta'_i)}{(-2X\cdot P_i)^{\Delta'_i}}    
\end{equation}
to reach our desired result
\begin{equation}\label{prefinalresultwithgamma}
\textbf{M}^{-1}\Bigg[\frac{\Gamma(\delta_{12}+l)}{\Gamma(\delta_{12})}\Bigg]=\frac{2}{\pi^{d/2}}P_{1,2}^{l}\frac{\Gamma(\Delta_{1}+l)\Gamma(\Delta_{2}+l)\Gamma(\Delta_{3})\Gamma(\Delta_{4})D_{\Delta_{1}+l_{1},\Delta_{2}+l_{2},\Delta_{3},\Delta_{4}}  }{\Gamma[(\Delta_{1}+l)/2+(\Delta_{2}+l)/2+\Delta_{3}/2+\Delta_{4}/2-d/2]} \ .
\end{equation}
It is trivial to see that eq.~(\ref{prefinalresultwithgamma}) generalizes to eq.~(\ref{eq:mellinspace}).

%%%%%%%%%%%%%%%%%%%%%%%%%%%%
\bibliographystyle{JHEP}
\bibliography{Draft.bib}
%%%%%%%%%%%%%%%%%%%%%%%%%%%%

\end{document}